\documentclass[aps,prb, showpacs, superscriptaddress,onecolumn]{revtex4-1}
\usepackage[utf8]{inputenc}
\usepackage[english]{babel}
\usepackage{amsmath, amssymb}
\usepackage{graphicx}
\usepackage{hyperref}
\usepackage[usenames,dvipsnames,svgnames,table]{xcolor}
\usepackage{calc} 
\numberwithin{equation}{section}

\newcommand{\be}[1]{ \begin{eqnarray} \mbox{$\label{#1}$} }

\newcommand{\ee}{\end{eqnarray}}

\newcommand{\oncite}[1]{Ref. \onlinecite{#1} }

\newcommand\ie {{\it i.e. }}

\newcommand\eg {{\it e.g. }}

\newcommand\etal{{\it et al.} }

\newcommand\half{\frac 1 2 }

\newcommand\ket [1] {|#1 \rangle }

\newcommand{\bracket}[2]  { \left<#1 | #2\right>}
\newcommand{\av}[1]{\langle #1\rangle}

\newcommand\noi{\noindent}

\def\i{\imath}
\newcommand{\rmi}{{\imath}}
\newcommand{\rmd}{{\mathrm d}}
\newcommand{\rme}{e}
\newcommand{\zb}{{\bar z}}
\newcommand{\wb}{{\bar w}}

\def\a{\alpha}
\def\b{\beta}

\def\G{\Gamma}

\def\eps{\epsilon}

\newcommand{\etah}{\eta^{H}}

\newcommand{\im}{\mathrm Im \,}

\newcommand{\mbf}{\mathbf}

\newcommand{\US}{U_{\mathcal S}}
\newcommand{\UT}{U_{\mathcal T}}

\newcommand{\elliptic}[3]{\vartheta_{#1}\!\left(#2\middle|#3\right)}
\newcommand{\genelliptic}[4]{\vartheta\!\left[\begin{array}{c}#1\\#2\end{array}\right]\left(#3\middle|#4\right)}

\newcommand {\red}{\textcolor{red}{Red}}
\newcommand {\blue}{\textcolor{blue}{Blue}}
\newcommand {\green}{\textcolor{green}{Green}}
\newcommand {\cyan}{\textcolor{cyan}{Cyan}}
\newcommand {\purple}{\textcolor{Purple}{Purple}}

\newcommand {\orange}{\textcolor{Orange}{Orange}}

\begin{document}

\title { Hall Viscosity of Hierarchical Quantum Hall States }

\author{M. Fremling}
\affiliation{Department of Physics, Stockholm University, AlbaNova University Center, SE-106 91 Stockholm, Sweden}
\author{T.H. Hansson}
\affiliation{Department of Physics, Stockholm University, AlbaNova University Center, SE-106 91 Stockholm, Sweden}
\author{J. Suorsa}
\affiliation{Nordita, Royal Institute of Technology and Stockholm University, Roslagstullbacken 23, SE-106 91 Stockholm, Sweden}

\date{\today}

\date{\today}

\begin{abstract} 
 Using methods based on conformal field theory,
 we construct model wave functions on a torus with arbitrary flat metric
 for all chiral states in the abelian quantum Hall hierarchy.
 These  functions have no variational parameters,
 and they transform under the modular group in the same way as the multicomponent generalizations of the Laughlin wave functions.
 Assuming the absence of Berry phases upon adiabatic variations of the modular parameter $\tau$,
 we calculate the quantum Hall viscosity and find it to be in agreement with the formula, given by Read,
 which  relates the viscosity to the average orbital spin of the electrons.  
 For the filling factor $\nu=2/5$ Jain state, 
 which is at the second level in the hierarchy,
 we compare our  model wave function with the numerically obtained ground state of the Coulomb interaction Hamiltonian in the lowest Landau level,
 and  find very good agreement in a large region of the complex $\tau$-plane.
 For the same example, we also numerically compute the Hall viscosity  and find good agreement with the analytical result
 for both the model wave function and the numerically obtained Coulomb wave function.
 We argue that this supports the notion of a generalized plasma analogy that would ensure that wave functions obtained using the conformal field theory methods do not acquire  Berry phases upon adiabatic evolution.  \\
\end{abstract}


\maketitle

\section{Introduction} 
Thirty years after Laughlin's proposed his famous  the wave functions for the fractional quantum Hall states at filling fractions $\nu = 1/(2m + 1)$,
we still lack a complete  theoretical understanding of the states that proliferate in the lowest Landau level at odd-denominator filling fractions.
Significant progress has been made, however, including the idea of 
a hierarchy of quantum Hall states \cite{Haldane_83,Halperin_83}, the concept of composite fermions\cite{Jain_07_Book}, and 
a general classification scheme, in terms of effective abelian Chern-Simons gauge theories with abelian gauge group \cite{WenZee_92, Wen_95}. 
More recently,  a microscopic realization of the full hierarchy is given in terms of explicit representative wave function\cite{Hansson_07a,Hansson_07b, Soursa_11b}.   

One important insight, that emerged from the Chern-Simon description, due to Wen and Zee,
is that an abelian quantum Hall liquid in its ground state is not only characterized by its topologically quantized Hall response 
---  determined by the filling fraction  ---  but also by the detailed nature of the quasi-particle excitations,
and the response to changes in geometry.
The topological aspect of the latter is captured by the shift $S$,
which gives the offset between the number of flux quanta $N_\Phi$,
and the number of degenerate lowest Landau level states on a sphere: $N_\Phi = N_e/\nu -S$.
This relation can be understood as originating in the orbital spin associated to the cyclotron motion of the electrons.
The orbital spin couples to the local curvature of the manifold,
as charge does to the magnetic field,
and results in Berry phases which effectively cause the shift $N_\Phi \rightarrow N_\Phi + S$.
In many cases, for example the Laughlin states,
the Moore-Read pfaffian state and the states in the positive Jain sequence,
the shift can be be calculated analytically from the trial wave functions,
but also be directly extracted from numerically determined ground states of some suitable Hamiltonian. 
The shift for an arbitrary hierarchy state can be extracted from the effective Chern-Simons theory\cite{Wen_95} or from the explicit wave functions on the sphere as discussed below.  

The shift on  the torus is zero, since the geometry is flat, but for the 
shift to be a genuine topological characteristic of the bulk state it  must also reveal itself in flat geometries. 
The appropriate bulk characteristic is the Hall viscosity\cite{Read_09};
a non-dissipative viscosity that is odd under time-reversal and hence non-vanishing only in systems without time-reversal symmetry.
Read calculated the Hall viscosity for the Laughlin and Moore-Read states,
generalizing the adiabatic response calculation by Avron, Seiler and Zograf\cite{Avron_95},
and found that both the shift and the Hall viscosity are proportional to the average orbital spin of the electrons in the liquid.
He also argued that the same relation should hold for any quantum Hall state
where the electrons can be represented  by primary fields in a conformal field theory. 

The observation of a multitude of odd-denominator quantum Hall states in high quality samples,
led Haldane and Halperin to propose a picture of hierarchical generation of states by quasi-particle condensation.
In a given parent state, quasi-particles may condense into incompressible Laughlin-like states,
thereby iteratively generating new states.
Many of these states, including all of the prominently observed ones,
can be understood within the framework of composite fermions developed by Jain,
but it appears that the hierarchy picture is potentially more general\cite{Pan_08, Bergholtz_07,Bergholtz_08}.
In the original formulation of the hierarchy,
further analysis of the wave functions was difficult due to their implicit form,
involving successive integrations over sets of quasi-particle coordinates.
Later developments,
which combined ideas from the composite fermion construction with methods of conformal field theory (CFT),
have made it possible to give explicit algebraic wave functions in terms of conformal blocks of CFT for any state in the hierarchy.
The Jain states  are identical to the corresponding conformal block wave functions,
thus showing that the composite fermion states indeed form a subset of the hierarchy.
The original construction of the hierarchical conformal block wave functions assumed planar geometry\cite{Hansson_07a,Hansson_07b},
but in a recent paper Kvorning extends these results to the sphere 
and calculates the shift for all the proposed hierarchy wave functions\cite{Kvorning_2013}.
In view of this, we find it interesting to also calculate the Hall viscosity and check whether it relates to the shift according to Read's formula.
To do this, we must first find proper expressions for the hierarchy wave functions on the torus.
These are the objectives of this paper. 

To appreciate the difficulties in calculating the Hall viscosity for the hierarchy states,
one should recall that the viscosity characterizes the response to a velocity gradient, or a strain rate.
For non-interacting electrons it is feasible to set up such a gradient and calculate the response,
while for interacting systems it is simpler to use the techniques developed by Avron, Seiler and Zograf\cite{Avron_95} in the context of the integer QH states.
The idea is to relate the Hall viscosity to the adiabatic curvature in the space of flat metrics.
One considers the ground state wave function on a flat two-torus with a metric parametrized by the modular parameter $\tau$,
and determines the Berry connection corresponding to a constant adiabatic change of $\tau$, 
which amounts to applying a constant and spatially homogeneous strain rate. 
This calculation requires, as emphasized by Read,
knowledge on the full $\tau$-dependence of the normalized wave functions\cite{Read_09}.  
For wave functions that have the form of single conformal blocks of primary fields,
a plasma mapping, and a generalized screening argument, can be used to argue for the precise dependence on $\tau$.

For the hierarchical wave functions the situation is more complicated.
Firstly, these functions are linear combinations of conformal blocks,   
and in this case no simple plasma analogy exists 
(in the concluding section we shall comment on some current work on this question).
Secondly, in their planar and spherical incarnations,
the conformal blocks used in the wave functions are not correlators of only primary fields,
but also of Virasoro  descendants.
It is  less well understood how to translate this construction to the torus geometry. 
Although such conformal blocks are well-defined in CFT,
they do not satisfy the physical boundary conditions imposed on charged particles moving on a torus in a background magnetic field.
A possible way to obtain wave functions with appropriate boundary condition was offered in Ref. \onlinecite{Hermanns_08},
where  the derivatives were replaced by finite translations in the $x$-direction. 
This gave wave functions that are numerically close to the corresponding Coulomb ground states
for rectangular tori with an aspect ratio $L_y/L_x \ge 1$.
However, for the adiabatic response calculation of the Hall viscosity,
it is required that the wave functions are known for all $\tau$ in the upper half plane.

In this article, we propose torus versions of the wave functions for the chiral  subset of the abelian hierarchy.
This comprises all states that can be obtained from the Laughlin states by successive condensations of quasi-electrons only,
including the positive Jain series. 
In Section \ref{sec:Hier_wfn}, these functions are given for a general $\tau$,
and  are shown to transform covariantly, in a way to be specified,  under the modular group.
In Section \ref{sec:colomb-numerics}, we compare our proposed wave function for the $\nu = 2/5$ Jain state with the numerically obtained Coulomb ground state,
and find a very good agreement in a wide range of  $\tau$. 
In Section \ref{sec:Viscosity}, we  calculate the Hall viscosity  analytically,
assuming the absence of Berry phases.
In Section \ref{sec:Visc_numerics} we again use the example of the $\nu = 2/5$  state to compute the viscosity numerically,
both for our proposed wave functions,
and for the numerical Coulomb wave functions.
Both these computations are in good agreement with the analytical results.
The following Section \ref{Sec:Background} contains necessary background results,
with some technical calculations presented in more detail in the appendices.

\section{Multicomponent Laughlin wave functions on the torus}\label{Sec:Background}

\subsection{Torus geometry and Landau levels}
In this section we reproduce some known material and define our notation and choice of gauge.
More comprehensive accounts can be found in Refs. \onlinecite{Haldane_85b} and \onlinecite{Read_11}.
We consider a family of tori $\mathbb{C}/[L(\mathbb{Z}+\tau\mathbb{Z})]$
parametrized by the modular parameter $\tau=\tau_1 + \rmi\tau_2$ and an overall scale factor $L$.
A torus can be thought of as a parallelogram spanned by the vectors 
$\vec e_1 = L \hat e_1$ and $\vec e_2 = L\vec\tau \equiv L( \tau_1  \hat e_1 + \tau_2  \hat e_2)$,
where $\hat e_1$ and $\hat e_2$ are fixed Cartesian unit vectors. 
Instead of using Cartesian coordinates on this parallelogram,
it is convenient to use the coordinates $x^a \in [0,L]$, which are defined on a square with side $L$ with opposite sides identified. The geometry of the torus is then coded in the distance measure $ds^2 = g_{ab}x^ax^b$,
where the (flat) metric is given by
\begin{equation}\label{eq:g_ij}
  g_{ab} = \begin{pmatrix} 1&\tau_1\\\tau_1&|\tau|^2
  \end{pmatrix}\,.
\end{equation}
We shall also use the complex coordinates $z=(x^1+\tau x^2)$.
To avoid notational clutter, we will often use the notation $(x^1,x^2) = (Lx,Ly)$ where $x,y \in [0,1]$.
The transformations $\mathcal S: \, \tau \rightarrow -1/\tau$ and $\mathcal T: \, \tau \rightarrow \tau + 1$ leave the torus unchanged, and  satisfy the algebraic relations $(\mathcal S\mathcal T)^3 = \mathcal S^2 = 1$,
which are the defining relations of the modular group $PSL(2,\mathbb Z)$.
The modular transformation properties of the QH wave functions are of central importance in our analysis.
In particular we want to study the response of homogeneous QH liquid states with a fixed number of electrons.
This means that the area $A=\tau_2 L^2$ must be kept constant when performing the $\mathcal S$ and $\mathcal T$ transformations,  
which can be ensured by using the coordinates $(x,y)$ for which the integration measure $dx dy$ is independent of geometry.
In these coordinates all dependence on the scale $L$ and the area $A = \sqrt{\det g}\, L^2 =\tau_2 L^2$ is explicit. 
Turning to a torus subject to a constant, perpendicular magnetic field $B=\epsilon^{ab}\partial_a A_b$,
the Hamiltonian for a free electron is 
\begin{equation}\label{eq:H0}
  H = \frac{1}{2m} g^{ab}\Pi_a\Pi_b, 
\end{equation}
where $\Pi_a= p_a-eA_a$,
and $g^{ab}$ is the inverse of the metric $g_{ab}$. 
The kinematical momenta
$\Pi_a$ satisfy 
$[\Pi_a,\Pi_b] = \rmi\tau_2\epsilon_{ab}\hbar^2 \ell^{-2}$,
where $\ell^2 = \hbar/eB$ is the magnetic length (we shall take $B>0$).
The guiding-center coordinates $R^i = \tau_2x^i +\frac{\ell^2}{\hbar}\epsilon^{ij} \Pi_j$ obey
$[R^i,R^j] = -\rmi \tau_2\epsilon^{ij} \ell^{2}$,
and commute with $\Pi_i$ and the Hamiltonian \eqref{eq:H0}.
On the infinite plane, the symmetry group is that of the magnetic translations, 
$t_{\vec l} = e^{i l^a K_a}$, generated by $K_a = -\epsilon_{ab} R^b$.
The magnetic translation operators satisfy the Girvin-MacDonald-Platzmann (GMP) algebra
\begin{equation}
  t_{\vec l}\, t_{\vec l^\prime}=t_{\vec l+\vec l^\prime}\, e^{\rmi\frac{\tau_2}{2\ell^2}(\vec l\wedge\vec l^\prime)}.
\end{equation}
Notice the factor of $\tau_2$ in this relation (and several others).
It appears since the physical area of the torus is $ \tau_2 L^2$. 

The Hamiltonian \eqref{eq:H0} is quadratic and can be diagonalized 
algebraically by introducing the ladder operators $\Pi_\pm$.
These operators are gauge dependent but satisfy $[\Pi_-,\Pi_+]=1$ in any gauge, 
and the Hamiltonian becomes $H = \hbar\omega_c(\Pi_+ \Pi_- + 1/2)$, with $\omega_c = eB/m$. 
The states in the lowest Landau level (LLL) satisfy $ \Pi_- \ket{\psi} = 0$.
This equation imposes a particular analytic structure on the wave functions in the coordinate representation.
Up to a gauge dependent Gaussian factor,
they are analytic functions in the coordinate
$z=x^1+\tau x^2 \equiv L(x + \tau y)$. 

When periodic boundary conditions are imposed as $t_{\vec e_j} \ket{\psi} = e^{2\pi\rmi \phi_j}\ket{\psi}$,
only a discrete subgroup of magnetic translations remains a symmetry.
This symmetry is generated by 
\begin{equation}
  t_{m,n}=e^{\rmi\frac{L}{N_\Phi}(nR^1-mR^2)} 
  \ \ ; \ \ m,n = 1, 2, \dots N_\Phi \, ,   \label{eq:trans_operator}
\end{equation}
where $N_\Phi = BA/(h/e)$ is the number of flux quanta piercing the torus. 
For fixed boundary conditions, the symmetry algebra thus consists of 
$N_\Phi^2-1$ independent operators (excluding the identity), and 
the full magnetic translation algebra reduces to\cite{Haldane_2011} $su(N_\Phi)$.
The periodic boundary conditions are
\begin{equation}
  \label{eq:pbc}
  t_a^{N_\Phi}\ket\psi = \exp\left(\rmi \phi_a \right)\ket\psi \ \ ; \ \ a=1,2\,,
\end{equation}
where we introduced the operators $t_1 = t_{1,0}$, $t_2 = t_{0,1}$ and the notation $\eps = 1/N_\Phi$.
These operators satisfy $t_1t_2 = e^{2\pi \rmi \eps}\,t_2 t_1$,
which is a special case of 
\begin{equation}
  \label{tonetwo}
  t_{m,n}t_{m^\prime,n^\prime} = e^{2\pi \rmi \eps(mn^\prime-nm^\prime)} \, t_{m^\prime,n^\prime}t_{m,n}\,.
\end{equation}
The constants $\phi_a$  in \eqref{eq:pbc} can be thought of as magnetic fluxes through the cycles of the torus. 

To obtain explicit expressions for the operators $t_a$ we must pick a gauge.
For our purpose, the most convenient choice is $(A_1,A_2) = (\tau_2Bx^2,0) =  (2\pi N_\Phi y, 0 )$;
we shall refer to this  as the $\tau$-gauge.\footnote 
{In the Cartesian coordinates $z=\tilde x+\rmi \tilde y$,
this corresponds to the vector potential
$\vec{\tilde A}= \frac{\tilde yB}{\tau_2} (\tau_2,-\tau_1)$,
that is perpendicular to $\vec \tau$.}
Note that when expressed in  the coordinates $(x,y)$, the vector potential in the $\tau$-gauge has
no explicit $\tau$-dependence, which will simplify the calculations of the Berry phases needed to extract the Hall viscosity\cite{Read_11}.  

In the $\tau$-gauge the ladder operators are 
\begin{equation}\label{pimin}
  \Pi_- = \sqrt{2} \left(\partial_{\bar z} + \frac {\tau}2 x^2 \right)\,,
\end{equation}
and the LLL wave functions become
\begin{equation}
  \label{lll}
  \psi_{LLL} = f(z) e^{\rmi\frac{\tau\tau_2}{2}\left(\frac{x^2}{\ell}\right)^2}=f(z) e^{\rmi\pi\tau N_\Phi y^2} \,.
\end{equation}
Note that in the $\tau$-gauge, the Gaussian factor is holomorphic in $\tau$, and 
the elementary translation operators take the following simple form: 
\begin{eqnarray}
  t_1 &= \makebox[\widthof{$e^{ L\eps( \partial_2 + \rmi\tau_2 x^1) }$}][l]{$e^{ L \eps \partial_1}$}
  &=e^{ \eps \partial_x  } \nonumber\\
  t_2 &= e^{ L\eps( \partial_2 + \rmi\tau_2 x^1) }
  &=e^{ \eps \partial_y + 2 \pi \rmi x } \,.
\end{eqnarray}
Turning to the $N$-body problem, the wave functions in the LLL are,
again up to the universal Gaussian factor in \eqref{lll},
analytic functions of the electron coordinates $z_i$, $i = 1, 2 \dots N_e$.
It is useful to introduce the unitary center of mass (CM) translation operators 
\begin{eqnarray}\label{Ti}
  T_1\equiv T_{1,0}&=& \prod_{i=1}^{N_e} t_{1,0}^{(i)} \nonumber\\ 
  T_2\equiv T_{0,1}&=& \prod_{i=1}^{N_e} t_{0,1}^{(i)} \, ,
\end{eqnarray}
in terms of which a complete set of commuting operators is $\{H,T_1,T_2^q\}$ \cite{Haldane_85b}.
As a consequence, all states, including the ground state, are at least $q$-fold degenerate
and distinguished by the eigenvalues of $T_1$, and any state in the LLL can be characterized by the eigenvalues
$e^{\rmi2\pi K_1}$ and $e^{\rmi2\pi qK_2}$ of $T_1$ and $T_2^q$ respectively. 
$N_\Phi K_i$ is the total momentum in the direction $i$.
Taking the eigenvalues of $t_1^{(j)}$ acting on the j$^{th}$ particle  as $e^{\rmi2\pi k_1/{N_\Phi}}$, we have
 $N_\Phi K_1=\sum_{j=1}^{N_e} k_1^{(j)}$ which is simply the sum of the individual electron momenta.
In the following, we shall denote the $q$ ground state wave functions at $\nu = p/q$ with $\psi_s$ where $s = 0,\dots q-1$.

We now discuss the boundary conditions \eqref{eq:pbc} in more detail. 
As will be shown below (in Section \ref{sec:mod_prop}), the modular transformations of the torus
relate wave functions with different boundary conditions, so we cannot just keep to one 
choice. 
It is, however, sufficient to consider only periodic or anti-periodic boundary conditions, which are
parametrized by the phases $\phi_a$ as 
\begin{equation}\label{eq:BC_r_t}
  (\phi_1, \phi_2) = \pi (r,t) \, ,
\end{equation}
where $r,t = 0,1$, such that $r$ and $r+2$ denote the same boundary conditions.
We shall use the notation $\psi^{(r,t)}$ for the corresponding wave functions. 
The generalization to arbitrary fluxes $\phi_i$ is straightforward.

For a general state at  $\nu = p/q$, with $q$ odd,
we have the following relations between  the indices $(r,t)$ and $s$:
\begin{eqnarray}\label{eq:s,t,r_relations}
  \psi_{s+q}^{(r,t)}=\psi_s^{(r+2,t)}&=&(-1)^{2qK_2}\psi_s^{(r,t)}\nonumber\\
  \psi_s^{(r,t+2)}=\psi_s^{(r,qt)}&=&\psi_{s}^{(r,t)}.
\end{eqnarray}
where we recall that the index $s$ labels the $q$ different eigenstates of $T_1$.
For details about the wave functions $\psi_s^{(r,t)}$, see Appendix \ref{app:PB}.
The operator $T_2$ relates states with different $s$ as $T_2\psi_s^{(r,t)}=\psi_{s+1}^{(r,t)}$.
In terms of the parameters $(s,r,t)$, the eigenvalues of $T_1$ and $T_2^q$ are
$K_1=s\frac pq+p\frac{N_\phi+1+r}2$ and $qK_2=p\frac{N_\Phi+1+t}2$,
where both $K_1$ and $qK_2$ are defined $\mathrm{mod}$ 1.
For bosons, the corresponding eigenvalues are $K_1=s\frac pq+r\frac p2$ and $qK_2=0$.
See Appendix \ref{app:PB}.

\subsection{The Laughlin wave functions}
A much studied example of a correlated many body state in the LLL is the family of Laughlin wave functions.
On a torus at  filling fraction $\nu=1/q$,
the Laughlin state is $q$-fold degenerate,
and the corresponding multiplet of  $q$ ground state wave functions,
for the boundary condition $r=t =N_\Phi+1$,
is given by
\begin{eqnarray}\label{eq:laughlin_wf}
  \psi_s &=& \mathcal{N}_0 \left[\sqrt{\tau_2} \eta(\tau)^2\right]^{qN_e/2} 
  \prod_{i<j}^{N_e} \left[\frac {\elliptic 1{\frac {z_i-z_j}{L}}\tau}{\eta(\tau)} \right]^q \nonumber
  \\
  &&\times{\cal F}_s (Z) e^{\rmi\pi\tau N_\Phi\sum_{i=1}^{N_e} y_i^2},
\end{eqnarray}
with $s = 0, 1, \dots q-1$. 
$\mathcal{N}_0$ is a  normalization constant that depends on the area of the torus, 
$\eta(\tau)$ is the Dedekind $\eta$ function and the CM wave functions are
\begin{equation}\label{Laugcm}
  {\cal F}_s(Z) =  \frac 1 {\eta(\tau)} \genelliptic {s/q} 0{qZ}{q\tau},
\end{equation}
with $Z=\sum_{i=1}^N z_i/L$.
(For the reader's convenience, some formulas for special functions are collected in Appendix \ref{app:special-functions}.)
These wave functions are eigenstates of the CM translations $T_1$ and $T_2^q$,
with the respective quantum numbers $(K_1, q K_2) = (s/q, 0)$. 

The wave functions \eqref{eq:laughlin_wf} can be obtained by two conceptually distinct procedures.
In the first, one assumes that the torus generalization of the planar coordinate difference $z_i-z_j$ is proportional to the odd Jacobi theta function $\elliptic 1{(z_i-z_j)/L}\tau$.
Then one determines  the CM functions ${\mathcal F}_s$ that are consistent with 
the quasi-periodic boundary conditions\cite{Haldane_85b}. 
In the second procedure, one uses  that the Laughlin wave functions can be expressed as $u(1)_q$ conformal blocks of a compact chiral boson.
This is done by first associating the electron at position $z$ 
with the normal ordered vertex operators
\begin{equation}
  V(z) = \, :e^{\rmi \sqrt{q} \varphi(z)}:  \,.
\end{equation}
The Laughlin wave functions are then obtained by diagonalizing the magnetic translations $t^{N_\Phi}_a$ in the vector space spanned by the torus conformal blocks,
stemming from the operators $V(z_i)$.
These conformal blocks, $\Psi_{e,m} $ are extracted from the correlator of a string of $V(z,\zb) = V(z)\otimes V(\zb)$'s in the presence of a constant neutralizing background charge described by the operator ${\mathcal O}_{ bg}$,
such that
\begin{equation}\label{eq:cftcorr}
  \langle V(z_1,\zb_1) \cdots V(z_N,\zb_N) {\mathcal O}_{ bg} \rangle = \sum_{e,m\in \mathbb{Z}} \Psi_{e,m} \bar\Psi_{e,-m} \, .
\end{equation}
There are various ways to introduce the neutralizing background charge. 
Reference \onlinecite{Read_09} used a torus version of the flux tube method described in \oncite{Hansson_07a}.
We find it more convenient to use a continuous background,
as proposed in \oncite{Hermanns_08},
but being careful in  keeping all $\tau$-dependence.
It is the background charge that will give rise to the Gaussian factor in the wave function,
after taking a suitable square root of \eqref{eq:cftcorr};
the details of the calculations are given in Appendix \ref{app:hierarchy}.

Two comments on the mathematical status of the objects $\Psi_{e,m} $ are in order.
First, in neither method to introduce the background charge,  $\Psi_{e,m} $  are in a strict sense
conformal blocks of primary operators in the underlying CFT.
Using the flux tube regularization, one must  take the limit of infinitely many flux tubes,\footnote{
One must  also use a different charge lattice at each stage in this limiting procedure}
and using the continuous background, the corresponding operator ${\mathcal O}_{bg}$ is not a standard vertex operator.
Second, in both cases, the procedure of taking a square root implies a phase ambiguity.
Since this phase can depend on the electron coordinates $z_i$,
it should be thought of as the freedom of choosing a gauge. 

In evaluating correlators like \eqref{eq:cftcorr},
it is important to use a correctly normalized torus two-point function $\langle\varphi(z,\bar z)\varphi(0)\rangle=K(z,\bar z)$.
This normalization is determined by demanding that the short distance behavior
on the torus is the same as on the plane\cite{Read_09}.
This gives
\begin{equation}\label{eq:twopoint}
  K(z,\zb) = -\ln \left | \frac{L\vartheta_1(z/L|\tau)} {\vartheta'_1(0|\tau) } e^{\rmi\pi\tau y^2}\right |^2 \, 
\end{equation}
for the torus two-point function.
In this context, it is also important to note that we will always use normal ordered vertex operators.
(We comment on this point in Appendix \ref{app:Norm_ord_and_K}.)
To understand the structure of \eqref{eq:laughlin_wf},
recall that $\vartheta'_1(0|\tau) = 2\pi \eta(\tau)^3$
and that factor of $\tau_2$ is extracted in such a way that the normalization constant ${\mathcal N}_0$ 
only depends on the fixed area $A= \tau_2 L^2 = 2\pi N_\Phi\ell_B^2$ of the torus.
Also note that the phase in \eqref{eq:laughlin_wf},
which corresponds to using the $\tau$-gauge,
is obtained by simply extracting the expression inside the absolute value symbol in \eqref{eq:twopoint} as it stands.
Although this might seem obvious, it does amount to a particular choice of gauge. 


\subsection{Generalization to a multicomponent system}

\begin{figure}
  \includegraphics[width=0.4\textwidth]{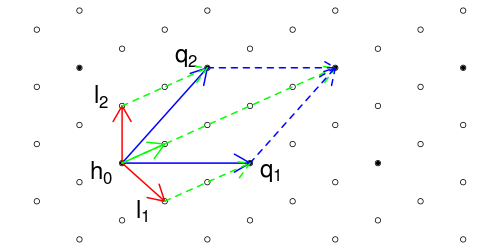} 
  \caption{\label{Lattice_2_5}
    Charge lattice of for $\nu=2/5$.
    Circles ($\circ$) denote $\G^\star$ 
    and dots ($\bullet$) denote $\G$.
    A unit cell of $\G$ is spanned by $\mbf q_\a$ (\blue) 
    and a unit cell of $\G^\star$ is spanned by $\mbf l_\b$ (\red).
    $\G^\star/\G$ is a finite subset of $\G^\star$,
    constructed by equating all points in $\G^\star$ related by a vector in $\G$.
    In the figure, $\G^\star/\G$ is given by the points in the (\blue) parallelogram.
    Note that $\G^\star/\G$ is one dimensional
    and spanned by $\mbf h_0=\mbf l_1+\mbf l_2$ (\green).
    Note that $\mbf l_\a=\mbf q_\a - 2\mbf h_0$.
  }
\end{figure}

The generalization of the Laughlin wave function to multicomponent systems,
\ie systems composed of distinguishable groups of electrons,
is straightforward in the CFT framework.
Instead of a single $u(1)$ theory,
one considers a product theory of multiple $u(1)$ components.
The long-distance behavior of the multi-component theory is characterized by the $K$-matrix in Wen's classification\cite{Wen_95}.
For a given $K$, we choose an electron charge lattice $\G$,
spanned by $\{ \mbf q_\a \}$ so that $K_{\a\b}=\mbf q_\a\cdot\mbf q_\b$.
Two explicit examples  are given below, 
and the charge lattice for $\nu=2/5$ is shown in Fig. \ref{Lattice_2_5}.
the $N_e$ electrons, at filling fraction $\nu=p/q$,
are partitioned into $n$ groups with $N_\a$ electrons in each.
The sizes of the groups are determined by requiring the liquid to be homogeneous,
which implies
\begin{equation}\label{eq:p_alpha_q}
  \frac{N_\a}{N_\Phi}=\frac{p_\alpha}q=\sum_{\b=1}^n K^{-1}_{\a\b} \, ,
\end{equation}
where the integers $p_\alpha$ are relatively prime to $q$, and $p=\sum_\a p_\a$.
In the hierarchy scheme $p_\alpha$ is always odd. 

The correlation functions of the normal ordered vertex operators 
$ V_{\mbf q}(z,\zb)= \, :e^{\rmi {\mbf q}\cdot\vec \varphi (z,\zb)} :$ 
can be calculated using standard methods and have the structure
\begin{equation}
  \langle V_{{\mbf q}_1}(z_1,z_1) \cdots V_{{\mbf q}_{N_e}}(z_{N_e},\zb_{N_e}) {\mathcal O}_{ bg} \rangle = \sum_{\mbf e,\mbf m} 
  \Psi_{{\mbf e},{\mbf m}} \bar\Psi_{{\mbf e},-{\mbf m}}\, , \label{fullcorr}
\end{equation}
where the chiral and anti-chiral sectors are indexed by the integer valued vectors $(\mbf e,\mbf m)$,
consisting of electric and magnetic charges of the Virasoro primary fields in the CFT. 
(For details see Appendix \ref{app:hierarchy}.) 
The physical boundary conditions for the electrons are imposed by diagonalizing the magnetic lattice translations in the space of conformal blocks $\Psi_{{\mbf e},{\mbf m}}$.
We get, again for $r=t =N_\Phi+1$,
the multiplet of wave functions with quantum numbers $(K_1,q K_2) = (s\frac pq, 0)$:
\begin{eqnarray}\label{eq:basis}
  \psi_{\bf h}
  &=&  \mathcal N_0\left[\sqrt{\tau_2}\eta(\tau)^2\right]^{\frac12\sum_{i=1}^{N_e}\mbf{q}_i\cdot \mbf{q}_i}
  \prod_{i<j}^{N_e} \left[\frac{\vartheta_1(z_{ij}/L |\tau)}{\eta(\tau)}\right]^{\mbf{q}_i\cdot \mbf{q}_j}\nonumber \\
  &&\times e^{\rmi \pi \tau N_\Phi\sum_{i=1}^{N_e} y_i^2}
  \mathcal{F}_{\mbf h}(z_1,\ldots, z_{N_e}|\tau).
\end{eqnarray}
Again, $\mathcal N_0$ is a constant that depends on $\tau$ only via the constant area.
The CM functions are labeled by the elements $\mbf h$
of the quotient lattice  $\mbf h \in \G^\star/\G$,
where $\G^\star$ is the charge lattice  of the CFT. 
$\G^\star$ is spanned by the quasi-particle charge vectors $\mbf l_\b$,
and duality means that  $\mbf q_\a\cdot\mbf l_\b=\delta_{\a\b}$ for all  vectors $\mbf q_\a \in \G$ and  $\mbf l_\b \in\G^\star$.
This  guarantees that the electron operators are trivial with respect to all other particles.
Also, the ground state degeneracy, $g$, is given by 
$g = \mathrm{vol} (\G)/ \mathrm{vol} (\G^\star) = \sqrt{\mathrm{ det } K} / \sqrt{\mathrm {det} K^{-1}} = \mathrm {det} K  $,
in agreement with the corresponding result in effective Chern-Simons theories\cite{Wen_95}.

 $\G \subset \G^\star$ and any quasi-particle charge vector $\mbf l  $, can 
 be decomposed as $\mbf l = \mbf q + \mbf h$. See Fig. \ref{Lattice_2_5}. for an example with $\nu=2/5$.
 Explicitly we have, 
\begin{equation}\label{cmwf}
  \mathcal{F}_{\mbf h}(\{z_i\}|\tau) = \frac{1}{\eta(\tau)^n}
  \sum_{\mbf q \in \G} e^{\rmi\pi\tau (\mbf q+\mbf h)^2}
  e^{2\pi\rmi (\mbf q +\mbf h)\sum_{i=1}^{N_e} \mbf q_i z_i},
\end{equation}
where  $n$ is the level of the hierarchy, or the number of groups of electrons.


In Appendix \ref{app:G_star_G} we show that for all chiral states in the hierarchy,
there is a convenient parametrization of the elements $\mbf h \in \G^\star/\G$ 
as $\{ \mbf h = s\mbf h_0, s=1,2\dots q \}$ for filling fraction $\nu = p/q$,
where $\mbf h_0=\frac{\mbf Q}{N_\Phi}=\frac1q\sum_\a p_\a\mbf q_\a=\sum_\alpha\mbf l_\alpha$.
The product of two elements $\mbf h=s\mbf h_0$ and $\mbf h'=s'\mbf h_0$
is readily computed as $\mbf h\cdot\mbf h'=ss'\frac pq$. 
Here $\mbf Q=\sum_{i=1}^{N_e}\mbf q_i$ is the total charge,
which also equals (minus) the charge of the background.
As on the plane, homogeneity of the liquid implies, through \eqref{eq:p_alpha_q}, 
the constraint $\mbf Q\cdot\mbf q_\a=N_\Phi$,
for all $\mbf q_\a$ spanning $\G$.

We illustrate this parametrization with two examples.  First consider a single field with compactification radius $R=\sqrt q$,
corresponding to the $\nu = 1/q$ Laughlin state.
Here the charge lattice is one-dimensional: $\G = \{ n\sqrt{q} ; n\in \mathbb{Z}\}$,
so we directly have $ \G^\star = \{ n/\sqrt{q} ; n\in \mathbb{Z}\}$,
and consequently $\G^\star/\G = \{0,1,\dots,q-1\}/{\sqrt q}$.
The $K$-matrix is simply $K=q$, $\mbf h_0=1/\sqrt{q}$, and the expression \eqref{eq:basis} reduces to \eqref{eq:laughlin_wf}.
The simplest multi-component example is the $\nu=\frac25$ state,
where the $K$-matrix is given by
$K=\begin{pmatrix}3&2\\2&3\end{pmatrix}$.
The inverse is $K^{-1}=\frac15\begin{pmatrix}3&-2\\-2&3\end{pmatrix}$,
giving $p_1=p_2=1$.
In this case there are two charge vectors 
$\mbf q_1=(\frac 3{\sqrt3},0)$ and $\mbf q_2=(\frac 2{\sqrt3},\frac 5{\sqrt{15}})$, depicted Fig. \ref{Lattice_2_5}.
The dual lattice is spanned by $\mbf l_1=(\frac 1{\sqrt3},-\frac 2{\sqrt{15}})$
and $\mbf l_2=(0,\frac 3{\sqrt{15}})$.
Since there are an equal number of particles in the two groups,
$\mbf h_0=(\mbf q_1 +\mbf q_2)/q=(\frac 1{\sqrt3},\frac 1{\sqrt{15}})=\mbf l_1+\mbf l_2$.

In the case of two components,
\eqref{eq:basis} is the torus version of the Halperin $(m,m',n)$  of states, that
were constructed to describe QH bilayers or states that are not fully spin polarized.
The generalization to four components can be used to describe spin-full bilayers,
but there is no obvious experimental realization of states with more components.
This is in sharp contrast to the hierarchy states that proliferate in the lowest Landau level.
These states cannot however be described directly by \eqref{eq:basis},
since these wave functions vanishes identically under anti-symmetrization between the groups. 
On the disk, and on the sphere, it is known how to construct representative hierarchy wave functions from the multicomponent conformal blocks.
There it is done by supplementing differential operators that act differently on the different groups
and thus make them distinguishable.
In the composite fermion language,
the differential operators emerge as polynomials of derivatives when the wave functions in the higher "effective Landau levels" are projected to the lowest Landau level.
In the CFT description, the derivatives appear since the electron operators are in general \emph{not} primary fields.
Instead they are Virasoro descendants,
characterized by a higher conformal spin,
reflecting the orbital spin of the electrons.
As mentioned in the introduction,
this procedure cannot readily be generalized to the torus 
since the correlators of the descendant fields do not satisfy the torus boundary conditions.

\subsection{Modular properties}\label{sec:mod_prop}
The Hall viscosity, and the related average orbital spin of a Hall liquid,
can  be calculated if the full $\tau$-dependence of the wave function is known.
As a guiding principle in constructing hierarchy wave functions,
we shall assume that they  transform in the same way under the modular transformations $\mathcal S$ and $\mathcal T$,
as the associated multicomponent functions \eqref{eq:basis}.
Here the $\mathcal S$ transform denotes the simultaneous change of $\tau\to\tau^\prime=-1/\tau$ and the rotation $z\to z^\prime=\frac {|\tau|} \tau z$,
while the $\mathcal T$ transform only denotes the modular transformation $\tau\to\tau^\prime=1+\tau$.
A direct calculation, given in Appendix \ref{app:mod_trans}, yields
\begin{eqnarray}\label{modtrans}
  \psi_s^{(r,t)}(z)&\stackrel{\mathcal{S}}{\to}&
  \prod_{i=1}^{N_e} \US(z_i)\,\,
  B_{\mathcal S}\sum_{s^\prime=1}^q \mathcal S_{s+\Delta_r,s^\prime+\Delta_t} 
  \psi_{s^\prime}^{(t,r)}(z)\nonumber\\
  \psi_s^{(r,t)}(z)&\stackrel{\mathcal{T}}{\to}&
  \prod_{i=1}^{N_e}\UT(z_i)\,\,
  B_{\mathcal T} \sum_{s^\prime=1}^q\mathcal T_{s+\Delta_r,s^\prime+\Delta_r}\psi_{s^\prime}^{(r,t+r+N_\Phi)}(z)\,.
\end{eqnarray}
Note the changes in boundary conditions.
$\US$ and $\UT$ are gauge transformations, and 
$B_{\mathcal S}$ and $B_{\mathcal T}$ are constant phase factors. 
$S_{s,s^\prime}$ and $T_{s,s^\prime}$ are the modular ${\mathcal S}$
and ${\mathcal T}$ matrices of the underlying CFT and are given by 
\begin{eqnarray}
  \mathcal S_{s,s^\prime} &=& \frac{1}{\sqrt q}e^{-2\pi\rmi ss^\prime\frac pq}\nonumber\\
  \mathcal T_{s,s^\prime} &=& \delta_{s,s^\prime}e^{2\pi\rmi (s^2\frac p{2q}-\frac n{24})} \, .
\end{eqnarray}
Here $n$ is the level of the hierarchy, which coincides with the central charge of the CFT.
The shifts in the arguments of the modular matrices are
$\Delta_r=q(r+N_\phi+1)/2$ and $\Delta_t=q(t+N_\phi+1)/2$.

The occurrence of the gauge transformations $\US $ and $\UT$ can be understood,
since the $\mathcal S$ and $\mathcal T$ transformations induce the changes
$(x,y)\to(-y,x)$ and $(x,y)\to(x+y,y)$ respectively in the invariant coordinates.
So although the vector potential $A_y=2\pi N_\Phi By$
has no explicit $\tau$-dependence in terms of the invariant coordinates $x$ and $y$, 
the gauge is effectively changed by the modular transformations.
These changes are compensated by the gauge transformations
\begin{eqnarray}
  \US(z) &=&\exp(\rmi2\pi N_\Phi xy)\nonumber\\
  \UT(z) &=&\exp(\rmi\pi N_\Phi y^2) \, .\label{eq:US_UT_gauges}
\end{eqnarray}

The modular transformations do not only affect the conformal blocks,
but also change the form of the operators $t_{m,n}$, 
defined in \eqref{eq:trans_operator},
that generate the magnetic symmetry algebra.
To see this, first note that for any function $f(z)$ we have by definition 
$t_{m,n} f(z)\propto f(z+\frac{L}{N_\Phi}(m+n\tau))$, and also  the transformations
\begin{eqnarray*}
  f(z+\frac{L}{N_\Phi}(m+n\tau))
  &\stackrel{\mathcal S}\to&
  f(z+\frac{L}{N_\Phi}(-n+m\tau))\\
  f(z+\frac{L}{N_\Phi}(m+n\tau))
  &\stackrel{\mathcal T}\to&
  f(z+\frac{L}{N_\Phi}(m+n+n\tau)),
\end{eqnarray*}
under $\mathcal S$ and $\mathcal T$.
Taking into account that the area of the torus is $\tau_2L^2$, 
so that a change in $\tau$ implies a rescaling of $L$ to preserve the area, 
the modular group acts on the operators $t_{m,n}$ as
\begin{eqnarray}
  t_{m,n}&\stackrel{\mathcal S}\to& \US t_{-n,m} \US^\dagger\nonumber\\
  t_{m,n}&\stackrel{\mathcal T}\to& \UT t_{m+n,n} \UT^\dagger, \label{ttrans}
\end{eqnarray}
where $\US$ and $\UT$ are the same gauge transformations as in \eqref{eq:US_UT_gauges}.
The detailed derivation is given in Appendix \ref{app:t_transform}. 

\begin{figure}
  \includegraphics[width=0.6\textwidth]{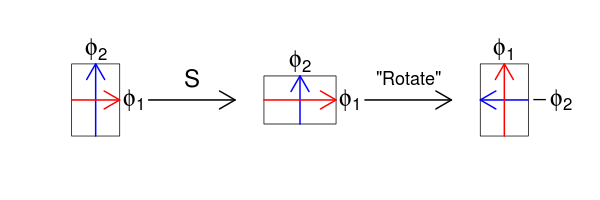} 
  \includegraphics[width=0.6\textwidth]{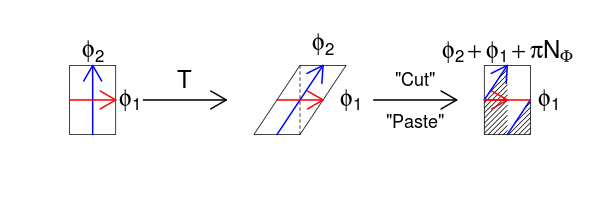} 
  \caption{\label{BC_Change}
    Changes to a path in the $x$-direction (\red) and $y$-direction (\blue) under modular transformations 
    for a rectangular torus. 
    Under $\mathcal S$, the coordinate system is rotated and 
    the $x$- and $y$-directions are effectively interchanged
    and so are the boundary conditions.
    Under $\mathcal T$,
    noting happens with $y$, but $x\to x+y$,
    such that a path in the $y$ direction gets mapped on a path winding in the $x$ direction.
    Also the twisted path encloses $N_\Phi/2$ fluxes, compared to the original path.}
\end{figure}

The changes in boundary conditions $(r,t)$, defined in \eqref{eq:BC_r_t}, under the $\mathcal S$ and $\mathcal T$ transformations in \eqref{modtrans} can be understood as follows.
Since the coordinates $x$ and $y$ are interchanged under $\mathcal S$,
it is natural that the boundary conditions are also interchanged and hence $r\leftrightarrow t$.
The $\mathcal T$ transformation acts as $r\to r$ and $t\to t+r+N_\Phi$. 
It is clear that $t$ will get an additional contribution from $r$,
since as the torus is twisted under $\tau\to\tau+1$, 
any path corresponding to a cycle in the $y$ direction
will also make a cycle in the $x$ direction (See Fig. \ref{BC_Change}).
The extra phase occurs since the two different paths enclose a surface containing $N_\Phi/2$ fluxes.
Formally, this can be seen by noting that $t_2^{N_\Phi}\to e^{\rmi\pi N_\Phi}t_2^{N_\Phi}t_1^{N_\Phi}$ under $\mathcal T$.
From Fig. \ref{BC_Change}, it also follows that a ground state wave function with a good $K_1$ quantum number in the $A_y = 2\pi N_\Phi By $, $\tau$-gauge, should, under $\mathcal S$,
transform into a superposition of states with good $K_2$ quantum number in the $A_y = -2\pi N_\Phi Bx$ gauge.
This is precisely the content of the first line in \eqref{modtrans}.

The constant phase factors $B_{\mathcal S}$ and $B_{\mathcal T}$ in \eqref{modtrans},
are given by
\begin{eqnarray}\label{eq:B_ST}
  B_{\mathcal S}  &=& \left(\frac{\tau}{|\tau|}\right)^{h_{\psi}}
  e^{-\rmi\frac\pi4\mbf Q^2}\nonumber\\
  B_{\mathcal T}  &=& e^{\frac{\pi}{12}\rmi\mbf Q^2}\, ,
\end{eqnarray}
so the only $\tau$-dependence in \eqref{modtrans} is the phase $(\tau/|\tau|)^{h_\psi}$
that occurs in the modular $\mathcal S$ transformation. 
Here $h_\psi$ is the total conformal weight of all the electron operators, \ie 
$h_\psi=\frac12\sum_{j=1}^{N_e}\mbf q_j^2=\frac12\sum_{\alpha=1}^{n} N_\alpha K_{\a\alpha}$, 
where $K_{\a\alpha}$ are the diagonal entries in the $K$-matrix
and $N_\alpha$ is the number of particles in group $\alpha$.
From here on, we shall refer to wave functions that transform according to \eqref{modtrans} 
as being modular covariant.

Since we are considering only holomorphic blocks, the conformal weight equals the conformal spin. 
By making a rotation of the system, we can determine its total (orbital) spin.
Under the assumption of vanishing Berry phases, this equals the total conformal spin.
Thus, for wave functions constructed from primary correlators,
the mean orbital spin $\bar s$ simply equals the average conformal 
spin of the electron operator, $\bar s = h_\psi/N_e$. 
In the special case of the Laughlin states, all of the above results are in agreement 
with those previously found by Read\cite{Read_09, Read_08}. 

\section{Torus hierarchy wave functions}\label{sec:Hier_wfn}

\subsection{Background and preliminary discussion}\label{sec:from-der-to-trans}

As already mentioned,
hierarchy states differ from the Laughlin states and their multi-component generalizations in two crucial ways.
First, they describe a single-component system (that of a single layer of spin polarized electrons) in terms of a multicomponent theory.
Second, in the CFT formalism, the operators associated to electrons are not all Virasoro primaries.
More concretely, the representative wave functions for hierarchy states,
which involves only chiral conformal block,
were constructed in Ref. \onlinecite{Bergholtz_08b} using the chiral vertex operators 
$V_\alpha (z) = \partial_z^{\a-1} \rme^{\rmi\vec q_\a\cdot\vec \varphi (z)}$ to describe the electrons in group $\alpha$. 
As pointed out in the introduction, one cannot directly carry this approach over to the torus.
For $\alpha >1$, $V_\alpha$ are descendant fields,
and the corresponding correlators involve the derivative operators $\partial_{z_i}$,
which are not compatible with the quasi-periodic boundary conditions. 
This is because the magnetic symmetry algebra on the torus differs from that on the plane. 
A consistent torus formulation of the hierarchy wave functions 
requires operators that are compatible with the $SU(N_\Phi)$ symmetry algebra
rather than with the continuous GMP algebra, and which are the torus counterpart
 of the derivative operators $\partial_{z_i}$ on the plane. 
A first attempt to achieve this was made by Hermanns \etal\cite{Hermanns_08}, who noted that the finite magnetic translations, 
$T_1$ and/or $T_2$, preserve the quasi-periodic boundary conditions and 
reduce to the holomorphic derivatives in the limit of large tori. 
Based on this, they proposed
\begin{equation}\label{eq:psi_Khs}
  \tilde\psi_s^{(r,t)} = {\mathcal A} T_1^{s_\a} \,  \psi_s^{(r,t)}
  = {\mathcal A} \prod_{\alpha=1}^n T^{(\alpha)}_{s_\alpha,0} \,  \psi_s^{(r,t)},
\end{equation}
where $\mathcal A$ is an anti-symmetrizer,   $s_\alpha = \alpha - 1$, and where
we introduced the notation
\begin{equation}\label{eq:T_nm-alpha}
  T^{(\alpha)}_{m,n}=\prod_{i_\alpha\in I_\alpha}t_{m,n}^{(i_\alpha)}\,,
\end{equation}
for the many body operator that acts on the electrons in group $I_\alpha$. 
Note that $T_{s_\alpha,0}^{(\alpha)}$ in \eqref{eq:psi_Khs} neither changes 
the boundary conditions nor the $K_1$ quantum number
of the parent multi-component state $\psi_s^{(r,t)}$. 
The relation $s_\alpha = \alpha - 1$ between the descendants level $s_\alpha$,
which enters the conformal spin of the operator $V_\alpha$ and the group index $\alpha$,
is specific for the minimal hierarchy constructed in \oncite{Bergholtz_08b}.
For more general states in the Wen classification\cite{WenZee_92},
$s_\alpha$ is related to the spin vector and thus to the shift on the sphere. 

Note that for $\tilde \psi$ not to vanish because of the anti-symmetrization, 
it is important that $\prod_{\alpha=1}^n T^{(\alpha)}_{s_\alpha,0}$ (or some generalization of it)
has the effect of making the groups distinguishable.
The simplest example of the above construction is $\tilde\psi = \mathcal A\,T_{1,0}^{(2)}\psi$,
which describes the level two $\nu = 2/5$ state in the positive Jain series. 
In \oncite{Hermanns_08} this wave function was successfully tested numerically,
for purely imaginary $\tau$ in the range $1 < \tau_2 < 10$.
However, moving to a smaller $\tau_2$, the wave function becomes worse,
and it fails badly for $\tau_2 < 0.3$, a parameter range that was not considered in \oncite{Hermanns_08}. 
A clue to this failure is that although $\tilde \psi_s$ in \eqref{eq:psi_Khs} is a well defined wave function,
obeying the same boundary conditions as $\psi_s$,
it does not transform in the same way as $\psi_s$ under modular transformations.
This is clear, since according to \eqref{ttrans},
$T^{(\alpha)}_{s_\alpha,0} \stackrel{\mathcal S}\to T^{(\alpha)}_{0, s_\alpha}$.
One can try to improve the ansatz \eqref{eq:psi_Khs} by adding a term $\sim T^{(\alpha)}_{0, s_\alpha}$
to restore the proper transformation under $\mathcal S$,
but then, because of $ T^{(\alpha)}_{0, s_\alpha} \stackrel{\mathcal T}\to T^{(\alpha)}_{s_\alpha, s_\alpha}$,
the transformation under $\mathcal T$ would be ruined.
From this we conclude that since all $T_{m,n}$ with $\gcd(m,n)=s_{\alpha}$ can be reached from $T_{s_\alpha,0}$,
by modular transformations, at least all such terms have to be included to construct an ansatz wave function
which transforms according to \eqref{modtrans}. 

\subsection{Modular covariant hierarchy wave functions}

Motivated by the above, we shall adopt as a guiding principle that the hierarchy wave functions $\tilde \psi_s$
should be modular covariant, by which we mean that they should transform in the same way under $\mathcal S$ and $\mathcal T$
as their corresponding primary states $\psi_s $.
In other words, \eqref{modtrans} shall hold for both $\tilde \psi_s$ and $\psi_s$,
albeit with different conformal weights $h_\psi$.
Below we shall present both analytical and numerical evidence that supports this assumption.

As pointed out above, we expect that any modular covariant ansatz, 
at $\nu=p/q$, must involve sums of terms $ T_{m,n}^{(\alpha)}$,
where $m, n = 1,2\dots N_\Phi$. 
However, while this is required, it is not sufficient,
since acting with these operators in general changes the values of the quantum numbers $K_a$.
This can be compensated for by an appropriate CM translation using the operator $T_{m,n}=\prod_{\alpha=1}^nT_{m,n}^{(\alpha)}$,
as shown in Appendix \ref{app:D_qnums}.
A more detailed analysis of the translation operators, found in Appendix \ref{app:Sum_m_n}, 
shows that the $n,m$ sums has to be taken over $2q$ copies of the lattice. 
We refer the reader to  Appendix \ref{app:D_operator} for detailed derivations
and the explicit expressions for general boundary conditions.
Here we only quote the result for the fermionic wave functions
with periodic boundary conditions along both cycles of the torus,
which is
\begin{equation}\label{finwf}
  \tilde\psi_s = \mathcal{A}\prod_{\alpha=1}^n \mathbb D_{(\alpha)}^{s_\alpha}\psi_s \,.
\end{equation}
and where the $\mathbb D_{(\alpha)}$ operator is defined as 
\begin{equation}\label{micked}
  \mathbb{D}_{(\alpha)}=\sum_{m,n=0}^{2qN_\Phi} \lambda_{m,n}^{N_\alpha} \xi_{m,n}T_{m,n}^{\left(\alpha\right)}T_{r_{\alpha}m,r_{\alpha}n} \, .
\end{equation}
The integer parameter $r_\a$ is defined as $r_\a=-p_\a p^{-1}\mod q$,
where the modulo $q$ appears since $[T_1^q,T_2]=[T_2^q,T_1]=0$.
Note that $p^{-1}$ is an \emph{integer},
defined such that $p^{-1} p = 1$ mod q.
The complex coefficients $\lambda_{m,n}^{N_\alpha} $, 
that will be discussed in detail below,
are the same for all boundary conditions,
while the sign factors 
\begin{equation} \label{signfactor}
  \xi_{m,n}=(-1)^{\Lambda (N_\Phi mn+rm+tn)}
  (-1)^{l(m + n +mn)} \ \ ; \ \ l=0,1 \, ,
\end{equation}
do depend on the boundary conditions, 
and have two contributions, of different origin.
The first is related to the boundary conditions $(r,t)$,
where $\Lambda=pp^{-1}+1$ is an integer modulo $2$. See Appendix \ref{ap:Ferm_Bos_D}.
\footnote{
  As $r_\alpha$ and $\Lambda$ are defined modulo different integers, 
  a change $p^{-1}\to p^{-1}+q$ may change the value of $\Lambda$ but not $r_\a$.
  In the fermionic sector, $p^{-1}$ can always be chosen such that $pp^{-1}$ is an even number.
  In the bosonic sector $\Lambda$ is invariant under changes in $p^{-1}$.
}
The second piece encodes the freedom to include a factor $(-1)^{m + n +mn}$ in $\xi_{m,n}$.
This factor is invariant under both $\mathcal S$ and $\mathcal T$ and can not be determined from modular covariance.
The choice of $l$ is the only freedom left in our ansatz.

In Appendix \ref{app:hierarchy_op} we prove the important property
\begin{equation}\label{mickedcom}
  [\mathbb D_{(\alpha)},\mathbb D_{(\beta)}]=0 \, ,
\end{equation}
which ensures that there are no ordering ambiguities in \eqref{finwf}.
The proof that   $T_{r_{\alpha}m,r_{\alpha}n}$ is such that $\mathbb D_{(\a)}$ does not change the 
quantum numbers of $\psi_s$ in \eqref{finwf},
is given in Appendix \ref{app:D_qnums}.

For $\tilde\psi_s$ to transform covariantly under modular transformations,
$\mathbb D_{(\a)}$ must satisfy certain conditions that are detailed in 
 Appendix \ref{ap:Ferm_Bos_D} and \ref{ap:Modtrans_D}.
In the $r=t=N_\Phi$ sector, they are
\begin{eqnarray}\label{eq:mod_ST_D}
  \mathcal S\ \ :\ \ {\mathbb D}_{(\alpha)} &\to& 
  \left(\frac\tau{|\tau|}\right)^{N_\alpha}  \US {\mathbb D}_{(\alpha)}\US^\dagger
  \nonumber\\
  \mathcal T \ \ : \ \  {\mathbb D}_{(\alpha)} &\to& 
  \UT{\mathbb D}_{(\alpha)} \UT^\dagger \,.
\end{eqnarray}
The phase $(\tau/|\tau|)^{N_\a}$ in the $\mathcal S$ transformation
ensures that the conformal dimension of the $s^{\text{th}}$ descendant differs from that of the primary field by $s$. 
Another way of interpreting this phase is to note that $(\tau/|\tau|)^{N_\a}$ should transform the same way as the derivative $\partial_z$ under rotations.
In fact, in Appendix \ref{sec:thermo_limit} we show that in thermodynamic limit, \ie large area limit,
effectively $\mathbb D_{(\alpha)} \to \prod_{j\in I_\a} \partial_{z_j} $.
Thus we recover the result in the plane. 

We do not know whether or not the conditions \eqref{eq:mod_ST_D} determine the coefficients $\lambda_{m,n}$ uniquely
(up to a $\tau$-independent constant).
However, in Appendix \ref{ap:Modtrans_D}, we show that
\begin{equation}\label{weight_app1}
  \lambda_{m,n}=\sqrt{\tau_2}\eta^3(\tau)
  \frac{e^{-\rmi\pi\tau n^2\epsilon^2}e^{-\rmi\pi nm\epsilon^2}}
       {\elliptic 1{m\epsilon+n\epsilon\tau}{\tau}}\,,
\end{equation}
indeed has the correct transformation properties.
As mentioned above, it possible to include a factor $(-1)^{m+n+mn}$ in $\lambda_{m,n}$, without changing the modular properties.
This sign is here incorporated in  the definition \eqref{signfactor} of the factor $\xi_{m,n}$ above.
We again stress that we have not {\em derived} \eqref{weight_app1},
but rather obtained it by a combination of physics reasoning,
and by imposing modular covariance. 
(The physics argument is outlined in the next section which can be omitted by readers only interested in the results.)
We can thus not exclude the possibility of having other modular covariant 
solutions, and we have also not found any theoretical argument for choosing $l$ as 0 or 1. 

An important example, 
that will be tested numerically in the Sections \ref{sec:colomb-numerics} and \ref{sec:Visc_numerics},
is the $\nu=\frac25$ Jain state.
In this case there is only one group of derivatives that acts on the second of the two equally large groups,
and the wave function is
\begin{equation}\label{eq:2/5-wfn}
  \tilde\psi_s^{2/5} = \mathcal A\,\mathbb D_{(2)}\psi_s^{2/5}=
  \mathcal A\,\sum_{m,n=0}^{2qN_\Phi}D_{m,n}^{(2)}\psi_s^{2/5}\,,
\end{equation}
where we introduced the operator 
\begin{equation}\label{D_mn_small}
  D_{m,n}^{(\alpha)}=\lambda_{m,n}^{N_\alpha} \xi_{m,n}T_{m,n}^{\left(\alpha\right)}T_{r_{\alpha}m,r_{\alpha}n},
\end{equation}
to label the different terms.
This notation will be used in the following sections.

\section{Hierarchy construction of the coefficients $\lambda_{m,n}$ }\label{sec:Hierachy_for_lambda}

Rather than using a trial and error approach to find  coefficients $\lambda_{m,n}$
that ensure that $\mathbb D_{(\a)}$ transforms as \eqref{eq:mod_ST_D},
we shall determine them using a physics argument based on the hierarchical construction of QH liquid states.
This approach has the additional advantage of showing why we would expect the hierarchy wave function to have similar modular properties as the Laughlin states.
We will consider the construction of the $\nu = 2/(2q-1)$ state by quasi-electron condensation in the $\nu = 1/q$ Laughlin state, using the methods developed in Refs. \onlinecite{Hansson_07a,Hansson_07b}.
As already mentioned; the Laughlin state can be expressed as a correlator of electron operators $V_1 = e^{\rmi\sqrt q \varphi (z)}$.
A quasi-electron at the position $w$ is described by $H^\star (w) = e^{-\rmi\varphi(w)/{\sqrt q}}$.
A correlator containing both $V_1$ and $H^\star$ will have simple poles $\sim 1/(z_i - w_j)$,
which need to be regularized in order to obtain a proper electronic wave function.
It is possible to remove these short-distance singularities while preserving the good infrared properties of the ansatz,
by letting $w_j$ approach a corresponding set of the coordinates $z_i$.
The leading singularity is subtracted by a normal ordering prescription $V_1(z) H^\star(w) \rightarrow 
N[V_1(z) H^\star(w)] = \partial_z e^{\rmi(\sqrt q - 1/{\sqrt q})\varphi(z)} \equiv \partial_z \tilde V_2$,
where $\tilde V_2 (z)$ is a primary field.
This procedure singles out $M$ electrons at positions $z_i$,
that are fused with the quasi-electrons.
To get a good wave function one must be careful to keep the wave function single valued and also anti-symmetrize over all possible ways of selecting the $M$ $z_i$:s among the $2M$ electrons.
As explained in detail in Refs. \onlinecite{Hansson_07a,Hansson_07b},
this precisely yields the $\nu = 2/(2q-1)$ composite fermion wave function. 

A key property of the normal-ordering regulator is that it preserves the conformal dimension of the constituent fields.
Its torus analogue needs to preserve the modular properties of the primary field correlators. 
Indeed, on the torus, we shall {\em not} take the limit $w \rightarrow z$.
Rather we keep a fixed finite difference,
such that $w_i = z_i + \delta$ and make sure that modular covariance is preserved.
Hence, we propose to regularize according to
\begin{eqnarray}
  N[V_1(z) H^\star(w)] &=& e^{K(\delta)}t_\delta e^{\rmi(\sqrt q - 1/{\sqrt q})\varphi(z)}\nonumber \\
  &\equiv& e^{K(\delta)}t_\delta\tilde V_2(z).\label{eq:normal}
\end{eqnarray}
More specifically, consider the correlator 
\be{fincorr}
\av{ \prod_{i=1}^{2M} V_1(z_i, \zb_i) \prod_{j=M+1}^{2M} H^\star(w_j , \wb_j ) {\cal O}_{bg}   }
\ee
with $N_e=2M$ electrons. 
With a properly chosen background charge,
this is a perfectly well defined correlator of primary fields,
and it can be calculated for a general modular parameter $\tau$
using the CFT machinery described earlier.
Taking the fixed-difference limit $w_j\to z_j$, for each $i>M$,
this will give a factor $\sim 1/\delta$,
but there will be unphysical singularities at $z_i = z_j + \delta$ for $i \le M$ and $j>M$ or \emph{vice versa}.
Here is the crucial idea:
Since $\delta = L /N_\Phi \sim L/N_e \sim \ell_B/\sqrt{N_e} $,
which for a large system is much less than the average distance between the electrons $d \sim \ell_B /\nu $, 
the singular terms $ {|z_i - z_j|^{2q} } / {|z_i - z_j - \delta|^2} $ can be expanded in powers of $\delta/\ell_B$.
Since the original correlation function involves only primary fields, we know from the previous discussion that it will transform
covariantly under modular transformations taking the conformal spin of the quasi-particle operators properly into account. 
It follows that each term in the expansion in $\delta$ must also transform properly, so truncating this expansion will not
ruin the good modular properties. The expansion converges except  when $|z_i - z_j| \sim \delta$, so for large values of $N_e$,
it will be a very good approximation for almost all configurations to keep only the leading term in the expansion
\be{exp}
\left| \frac {(z_i - z_j)^q } {(z_i - w_j )} \right|^2 
= |z_i - w_j|^{2(q-1)} + {\cal O}(\delta / \ell)\,,
\ee 
Although this approximation is not valid for $|z_i - z_j| \sim \delta$, it yields perfectly well defined expressions for all configurations.
Put differently, it allows us to keep the correct long distance behavior of the 
wave function, and at the same time to obtain a regular short distance behavior simply by analytical continuation
of the approximate expression. 
In the full expression \eqref{fincorr} this amounts to the replacements
$qK(z_i - z_j) -K(z_i - w_j) \to (q-1) K(z_i - w_j) $ for $i\le M$, $j>M$ and 
$qK(z_j - z_k) -K(z_j - w_k) -K(z_k- w_j) \to (q-2) K(w_j - w_k)$ for $j,k>M$, 
where $K(z) \equiv K(z,\zb)$. 
Recalling the definition of $\tilde V_2$,
we conclude that \eqref{fincorr} should be replaced with
\be{fincorr2}
\left|e^{M K(\delta,\bar\delta )}\right|^2\prod_{j=M+1}^{2M} t^{(j)}_\delta
\av{ \prod_{i=1}^{M} V_1(z_i, \zb_i) \prod_{j=M+1}^{2M} \tilde V_2(w_j , \wb_j ) {\cal O}_{bg}}\,.
\ee
The first, $z$-independent, factor is the contribution from the $i=j >M$ term
$\av{ V_1(z_j,\zb_j) H^\star (w_j, \bar w_j)}$,
which diverges for $\delta\rightarrow 0$ and cannot be neglected in the large $N_e$ limit.
The first product is essentially the operator $T_{m,n}^{(\alpha)}$ in \eqref{micked},
assuming $\delta=\delta_{m,n}=L\epsilon(m+n\tau)$, where $\epsilon=1/N_\Phi$.
The correlator of the primary operators $V_1$ and $\tilde V_2$ gives $\psi_s$ in \eqref{finwf}. 
This suggests that we should identify the coefficients  $\lambda_{m,n}$ as a suitably taken square root of the first factor,
with $\delta=\delta_{m,n}=L\epsilon(m+n\tau)$.
From \eqref{eq:twopoint} we obtain 
\begin{equation}\label{eq:sqrt_tow_point}
  \lambda_{m,n}=\zeta_{m,n}e^{K(\delta_{m,n})}
  =\zeta_{m,n}\frac{\vartheta_1^\prime(0|\tau)}
  {L\elliptic1{\delta_{m,n}/L}\tau}e^{-\rmi\pi\tau n^2 \epsilon^2}\,,
\end{equation}
where $\zeta_{m,n}$ is an undetermined phase factor that we set to $\zeta_{m,n} =  e^{-\rmi\pi nm\epsilon^2}$
in order to ensure simple modular properties for $\lambda_{m,n}$.
(Details are given in Appendix \ref{ap:Modtrans_D}.) 
Again note the possibility to include a factor $(-1)^{m+n+mn}$ in $\zeta_{m,n}$,
which has been incorporated in $\xi_{m,n}$.
With this choice of the coefficients $\lambda_{m,n}$
and ignoring a constant scale factor $\sqrt{2\pi/N_\Phi}$,
we obtain the final expression \eqref{weight_app1} quoted in the previous section.

\section{Numerical test of the $\nu = 2/5$ torus wave function } \label{sec:colomb-numerics}

In this section we shall test our $\nu = 2/5$ wave function \eqref{eq:2/5-wfn}
by comparing it to the one obtained by numerically diagonalizing the unscreened Coulomb potential in the LLL. 
Of the two possibilities for the integer $l$, we find good overlaps only for $l=N_\Phi + 1 \mod 2$, which will be assumed in this section.
Details on how to perform the diagonalization can be found in \eg Refs. \onlinecite{Yoshioka_83,Yoshioka_02_Book}.
Since the sum in \eqref{eq:2/5-wfn} has $2q N_\Phi^2$ terms,
the expression will be useful only if it yields good approximations when only a few terms of low order are included.
This is expected to be the case, since in the large $L$ limit,
 the lowest non-vanishing terms dominate, and reduces to the planar result.
As we shall now demonstrate, a few low order terms also give very good results for the small systems we can study numerically.
We will however also demonstrate that it is crucial to keep more than one term to get good agreement with the numerical solutions.
We shall thus compare the numerical wave functions for different $\tau$,
with various combinations of the terms $\mathcal A D_{m,n}\psi_s$ appearing in \eqref{eq:2/5-wfn},
and for simplicity we will often write just $D_{m,n}$,
instead of $\mathcal A\,D_{m,n}\psi_s$.

\begin{figure} 
  \includegraphics[width=0.45\textwidth]{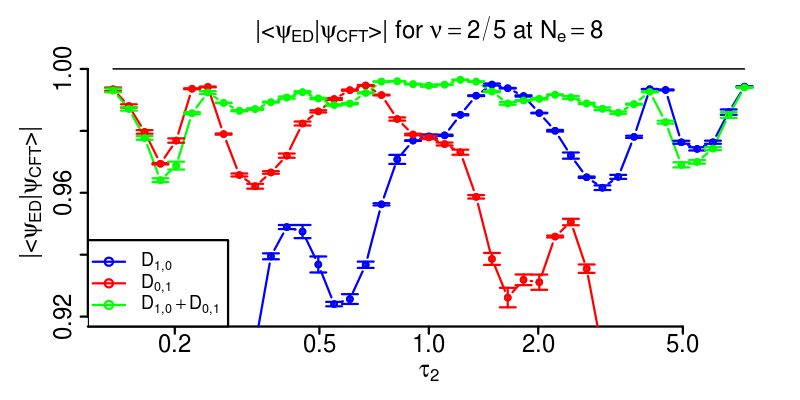} 
  \caption{\label{Overlap_Im_tau}
    Overlap of the terms $D_{1,0}$ (\blue), $D_{0,1}$ (\red) and 
    $D_{1,0} + D_{0,1}$ (\green) in \eqref{eq:2/5-wfn} with the exact Coulomb state,
    for $\tau_1=0$ and $0.1<\tau_2<10$.
    Note the logarithmic scale of $\tau_2$ which makes the figure symmetric around $\tau_2=1$.
    Note that in the limit of a thin torus, $\tau_2\rightarrow \infty$ or $\tau_2\to0$,
    the overlap goes to 1, and that the
     combination $D_{1,0}+D_{0,1}$  is  good over the full range $0<\tau_2<\infty$.
    At $\tau_2\approx5\,(\tau_2\approx0.2)$, there is a dip in the overlap,
    that is improved by adding the term $D_{2,0}$ ($D_{0,2}$).
  }
\end{figure}

In \oncite{Hermanns_08} a similar analysis was carried out for the term $D_{k,0}$ at $\tau=\rmi\tau_2$ with $\tau_2\geq1$.
It was found that already the term $D_{1,0}$ gives a good overlap with the exact Coulomb ground state in the investigated region.
Extending the region to $\tau_2\lesssim1$, the overlaps become much worse,
and the obvious minimal way to improve the wave function is to also include the contribution $D_{0,1}$. 
The resulting overlaps are shown in Fig. \ref{Overlap_Im_tau} for rectangular tori, $\tau=\i\tau_2$,
with aspect ratios in the range $0.1<\tau_2<10$.
As expected, the terms $D_{1,0}$ and $D_{0,1}$ dominate in the large and small $\tau_2$ 
region respectively,
Also as expected, the sum of the two gives a much higher overlap than either of the individual terms in the region $\tau_2 \approx 1$, 
\ie for almost quadratic tori.
Further, note that the two contributions have a reflection symmetry about $\tau_2=1$;
this follows from $D_{1,0}$ and $D_{0,1}$ transforming into each other under the modular $\mathcal S$ transformation.
In all cases we have investigated, 
the contributions $D_{m,n}$ and $D_{-m,-n}$ are identical within numerical accuracy,
presumably due to a reflection symmetry.
We have not attempted to  prove this analytically, but note that 
 this symmetry was pointed out in \oncite{Hermanns_08} in the special case of translations in only one direction.

\begin{figure} 
  \begin{tabular}{cc}
    \includegraphics[width=0.45\textwidth]{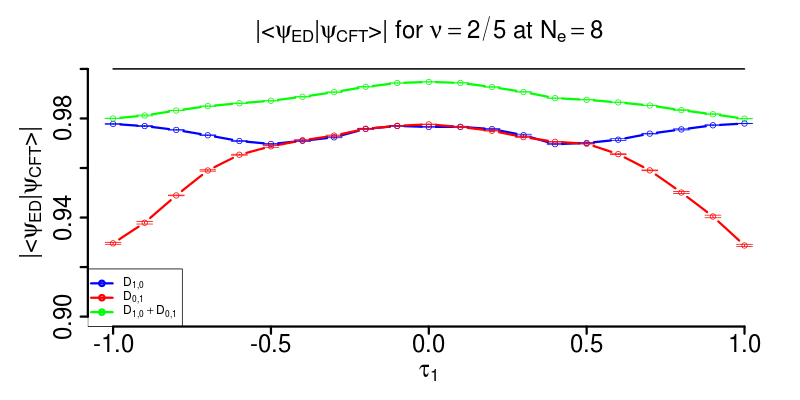} 
    &\includegraphics[width=0.45\textwidth]{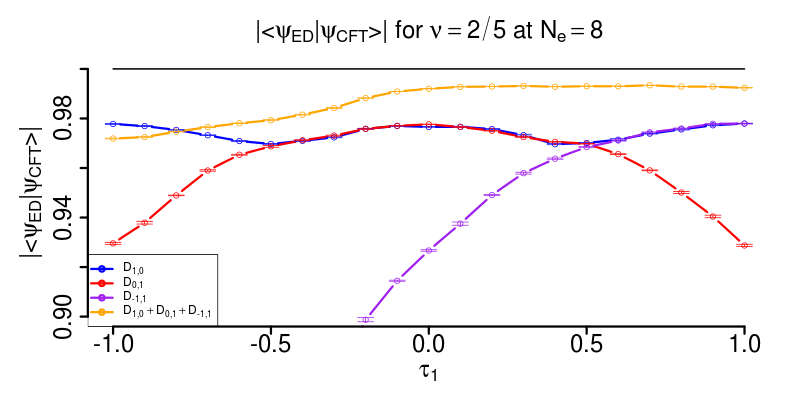}  \\
    $a)$ & $b)$
  \end{tabular}
  \caption{\label{Overlap_Re_tau}
    $a)$ Overlap of the terms $D_{1,0}$ (\blue), $D_{1,0}$ (\red) and 
    $D_{1,0} + D_{0,1}$    (\green)  \eqref{eq:2/5-wfn} with exact Coulomb state,
    for $\tau_2=1$ and $-1<\tau_1<1$.
    $b)$ The same as $a)$ except that the term $D_{-1,1}$ (\purple) has been added to the linear combination $(D_{1,0}+D_{0,1}+D_{-1,1})$ (\orange).
    For highly skew tori $\tau\approx1+\rmi$,
    the term $D_{-1,1}$ plays a similar role as $D_{0,1}$ at $\tau\approx\rmi$.
    At $\tau\approx\frac12+\rmi$, the two terms have equal weight.
    Note that the overlap is boosted in the region around $\tau=1+\rmi$ when the term $D_{-1,1}$ accounted for.
    We emphasize that keeping only \emph{two} terms,
    the overlap with Coulomb is good for all $\tau_1$,
    but which are the terms that dominate depends on $\tau$.
  }
\end{figure}

Before discussing the importance of higher order terms in \eqref{eq:2/5-wfn},
we shall consider skew tori, \ie taking $\tau_1 \neq 0$.
In this case we expect terms where both $n$ and $m$ are different from zero to be important,
the simplest ones being $D_{1,-1}$ and $D_{1,1}$.
This is illustrated in Fig. \ref{Overlap_Re_tau}.
As expected from the above, and shown in the left panel,
the two terms $D_{1,0}$ and $D_{0,1}$ describe the almost quadratic tori quite well.
However, it performs worse for tori of more pronounced rhombic shape.
In the right panel, the contribution $D_{1,-1}$ is added,
and this gives a considerable improvement for $0.5\lesssim\tau_1\lesssim 1.5$.
Adding $D_{1,1}$ would give a similar improvement for negative $\tau_1$. 

In both the above examples,
the numerical improvements for large aspect ratios and large skewness respectively, is quite remarkable.
Since the leading terms already give overlaps ranging between 95\% and 98\%,
and we are adding complex numbers of comparable magnitudes,
the phases must be very precise to yield an improved overlap.
Using \eqref{eq:2/5-wfn}, we achieve such an improvement \emph{without any adjustable parameters}. 

\begin{figure} 
  \includegraphics[width=0.45\textwidth]{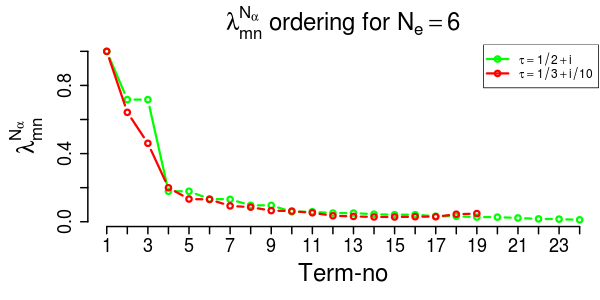}
  \caption{\label{fig:weight_order}
    Absolute value of the coefficients $\lambda_{m,n}$ as a function of increasing $|\delta_{m,n}|$,
    for $\tau=\frac12+\rmi$ (\green) and $\tau=\frac13+\frac\rmi{10}$ (\red).
    In order, the first six terms are:
    $(m,n)=(1,0),\,(0,1),\,(-1,1)\,(-2,1),\,(1,1),\,(2,0)$ and 
    $(m,n)=(-1,3),\,(0,1),\,(-1,2)\,(-2,4),\,(-2,6),\,(2,0)$.
    All terms $(-m,-n)$ have been omitted, as discussed in the text.
    The normalization is such that the weight of the first term is $\lambda_{m,n}=1$.
  } 
\end{figure}

Let us now turn to the higher order terms 
and the precise meaning of our claim that the sums in \eqref{eq:2/5-wfn} are dominated by terms of low order.
First, notice that the coefficients $\lambda_{m,n}$ decrease with increasing $|\delta_{m,n}|=\frac{L}{N_\Phi}\sqrt{(m+n\tau_1)^2+n^2\tau_2^2}$,
as illustrated in Fig. \ref{fig:weight_order} .
(For small $\delta_{m,n}$, $\lambda_{m,n}\propto 1/{\delta_{m,n}}$.)
For a given $\tau$,
it is thus reasonable to add the terms in order of decreasing $|\delta_{m,n}|$.
This is illustrated by considering the special point $\tau_1=\frac12$ in Fig. \ref{Overlap_Re_tau}.
Here $|\delta_{0,1}|=|\delta_{-1,1}|$,
which implies that $D_{-1,1}$ is expected to replace $D_{0,1}$ as the second most important term in the expansion \eqref{eq:2/5-wfn}.
This is illustrated in the figure. 

\begin{figure} 
  \includegraphics[width=0.45\textwidth]{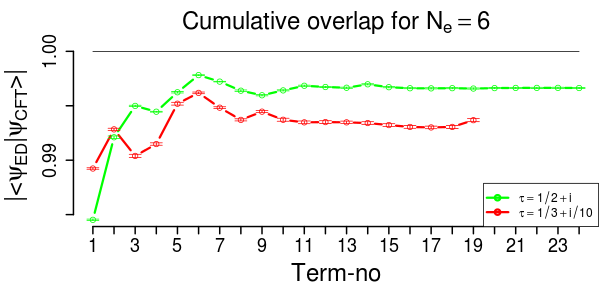}
  \caption{\label{Overlap_Aditional_Terms}
    The cumulative overlap between \eqref{eq:2/5-wfn} and the exact Coulomb result as terms are added,
    in order of increasing $|\delta_{m,n}|$ for $N_e=6$ at $\tau=\frac12+\rmi$ (\green) and $\tau=\frac13+\frac\rmi{10}$ (\red).
    The terms in the sum are ordered as the same way in Fig. \ref{fig:weight_order}.
    For the first few terms added,
    the overlap is improved, but the curve is not monotonic.
    Note that since there are no adjustable parameters there is no reason to expect that monotonicity.
    The overlap is converging with increasing number of terms, 
    but however not at the maximum value of the overlap,
    so there is a point where adding more terms does not necessarily make the result better.
    In fact, for all values of $\tau$, it is sufficient to keep as little as only tow terms, provided these are chosen appropriately.
  }
\end{figure}

\begin{figure} 
  \includegraphics[width=0.45\textwidth]{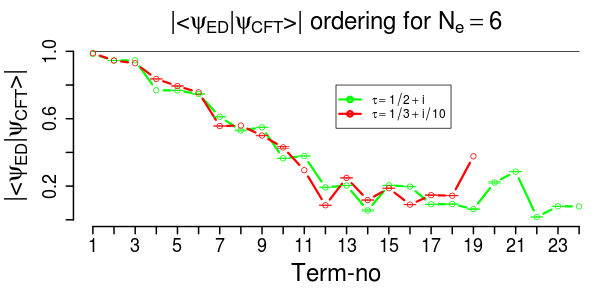}
  \caption{\label{Overlap_ordering}
    The Overlap with exact Coulomb as a function of increasing $|\delta_{m,n}|$.
    The setup is the same as in Fig. \ref{fig:weight_order} and \ref{Overlap_Aditional_Terms}.
    Larger values of $|\delta_{m,n}|$ has smaller overlap,
    which is in agreement with $\delta_{m,n}$ representing a derivative.
    From this image it is clear that when truncating the sum in \eqref{eq:2/5-wfn} to only a few terms, 
    the most important ones have small $\delta_{m,n}$.    
  }
\end{figure}

Additional evidence for our hypothesis that the expansion in \eqref{eq:2/5-wfn}
is dominated by terms with small values for $\delta_{m,n}$
is provided in Fig. \ref{Overlap_Aditional_Terms}, where we show the cumulative overlap for two different tori,
as more and more terms are added in order of increasing $|\delta_{m,n}|$.
In both cases the sums converge towards a very high, stable, overlap.
Again, this is nontrivial,
since the cumulative effect to the higher order terms,
could easily destroy the overlap, given that  the magnitude of $\lambda_{m,n}$ does not fall off very rapidly.
(The magnitudes of the first six or seven terms are 10 -- 80 \% of the largest one.)
Thus the phases of those terms must be such that they effectively cancel in \eqref{eq:2/5-wfn},
and also have smaller overlap with the Coulomb wave function. 
The latter is illustrated in Fig \ref{Overlap_ordering}.
Note that although the saturation in Fig. \ref{Overlap_Aditional_Terms} does no occur at the maximum value of the overlap,
for $N_e=6$ particles the saturated overlap
is $\av {\psi_{\mathrm{Coulomb}}|\tilde\psi}\approx0.996$ for $\tau=\frac12+\rmi$,
and $\av{ \psi_{\mathrm{Coulomb}}|\tilde\psi}\approx0.994$ for $\tau=\frac13+\frac\rmi{10}$.
This is still higher than for any of the individual terms.

\begin{figure}
  \begin{tabular}{ccc}
   \includegraphics[width=0.45\textwidth]{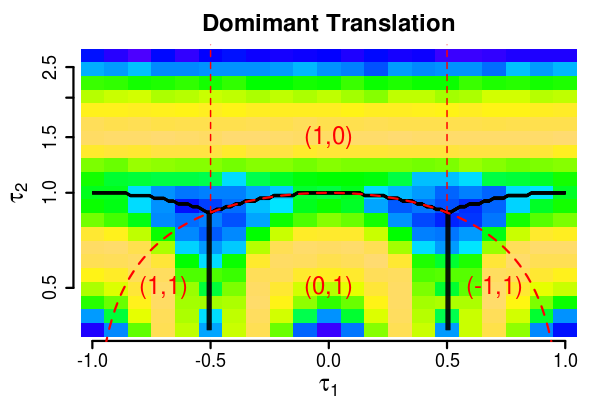} 
   &\includegraphics[width=0.075\textwidth]{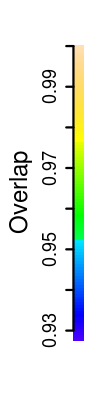} 
   &   \includegraphics[width=0.45\textwidth]{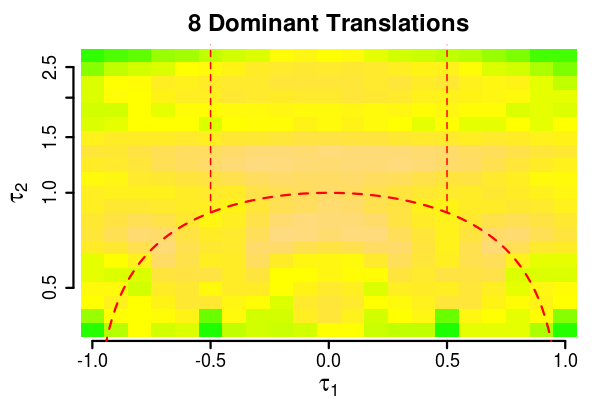}\\
   $a)$&&$b)$
 \end{tabular}
 \caption{\label{Fig:2D-overlap}
   Overlap with the exact Coulomb ground state in a region $-1<\tau_1<1$ and $0.37<\tau_2<2.72$ of the $\tau$-plane,
   for $N_e=8$ particles.
   The thick black lines mark the boundaries of regions with different minimal translation steps $\delta_{m,n}$.
   Note the logarithmic scale of $\tau_2$ for a more symmetric plot.
   $a)$ Only the dominant term is included in the sum \eqref{eq:2/5-wfn}.
   The overlap with the exact Coulomb state is good everywhere.
   $b)$ The eight most dominant terms at $\tau=\rmi$ ( ($m$,$n$)=(1,0), (1,0), (2,0), (0,2), (-1,1), (1,1), (-2,2), (2,2) ) are included in the sum \eqref{eq:2/5-wfn}.
   Overlap with the Coulomb state is better or equal in most parts of the $\tau$-plane;
   at the edges, other terms than the eight used here are dominant (see \eg Fig. \ref{Overlap_Re_tau}).
   This shows that our ansatz is valid in the entire $\tau$-plane, 
   and that the graphs in Figures \ref{Overlap_Im_tau} to \ref{Overlap_ordering} are indeed representative. }
\end{figure}

 Fig. \ref{Fig:2D-overlap} shows that the results presented above are representative, and that we indeed, as claimed in the abstract and introduction, have a very good overlap between our $\nu=2/5$  wave function and the corresponding Coulomb state, in a large part of the $\tau$-plane. 
  In the left panel only the dominant term has been used, \ie the term $D_{m,n}$ for with $m$,$n$ minimizes $\delta_{m,n}$.
  The different boundaries of these regions have been marked with  thick black lines.
  As $\tau_2\to0$ these lines will further split in a fractal pattern.
  The fundamental domain is marked with a dashed red line. 
  
In the right panel of Fig. \ref{Fig:2D-overlap} we include  the eight terms that  dominate  \eqref{eq:2/5-wfn} in the region close to $\tau=\rmi$.
Here the overlap with exact Coulomb state is systematically improved compared to  using one term only.
  This behavior has already been noted in Fig. \ref{Overlap_Aditional_Terms}.
At the boundaries of the plotted $\tau$-plane  the overlap decreases 
  which is to be expected as the dominant terms at these values of $\tau$ differ from those at $\tau=\rmi$.
  Had we also included those terms, the overlap would have improved in these regions as well.

Again we emphasize that there are \emph{no variational parameters} in these fits. 
Although we cannot make any precise statement about the convergence of the sums in \eqref{eq:2/5-wfn}, 
we  believe that our numerical results 
strongly supports that modular covariance is a crucial property of good hierarchy wave functions on the torus.

\section{Hall viscosity}\label{sec:Viscosity}
In Ref. \onlinecite{Avron_95} Avron, Seiler and Zograf showed that the antisymmetric component of the viscosity, 
or the Hall viscosity, $\etah$, can be calculated as the adiabatic curvature on the space of ground state wave functions parametrized by the modular parameter $\tau$.
Physically a constant change of $\tau$ amounts to setting up a constant strain {\em rate} in the system.
The Hall viscosity is a component of the viscosity tensor that describes the response to such a strain rate.
The relation between the Hall viscosity and the adiabatic curvature is in fact very similar to the more commonly known TKNN formula,
which relates the electric conductivity of a Hall liquid
to the response to an adiabatic insertion of magnetic flux into the holes of a torus\cite{Thouless_82}. 
Later, Read calculated the Hall viscosity for the Laughlin and  the Moore-Read state.
He also argued that for any QH state,
which can be expressed as a conformal block of a CFT,
the Hall viscosity is related to the average orbital spin $\bar s$, 
by $\etah=\frac12\bar s\bar n\hbar$, where $\bar n$ is the electron density. 
Also, since the shift on the sphere 
 -- which is a topological invariant --
is related  to the orbital spin by $S = 2\bar s$, it follows that,
at least for a clean system, 
the Hall viscosity has a topological meaning. 
In this section we shall extend these results to the hierarchy states \eqref{finwf}. 

First we recall the formula derived in  Ref. \onlinecite{Avron_95} that relates the Hall viscosity to a Berry phase.
The Berry potential related to the modular parameter $\tau$ is defined by
${\mathcal A}_\tau = \rmi \bracket{\Psi(\tau)}{\partial_\tau \Psi(\tau)}$, and 
the Hall viscosity depends on the corresponding field strength 
${\cal F}_{\tau \bar\tau} = \rmi \partial_{\bar\tau} {\mathcal A}_{\tau}- \rmi \partial_\tau {\mathcal A}_{\bar\tau}$, as
\begin{equation}\label{hvisk}
  \etah = -\frac {2\tau_2^2} A {\cal F}_{\tau\bar\tau},
\end{equation}
where $A$ is the area of the system. 

To get a sense for the difficulty associated with calculating ${\cal F}_{\tau\bar\tau}$ for our trial wave functions,
we first recapitulate Read's argument for the Laughlin state described by the function \eqref{eq:laughlin_wf}.
These wave functions are not normalized,
but using the Laughlin plasma analogy,
we can infer that the partition function,
which is essentially the norm of the wave function squared,
is independent of $\tau$.
Read argues that the two-point function is normalized so that at short distance,
the screening forces are independent of the geometry. 
Recall that in the final expression \eqref{eq:laughlin_wf},
the total normalization constant,
which we assume to be $\tau$-independent by the plasma analogy,
is written as a product of a constant ${\mathcal N}_0$,
which depends only on the area $A$
and an explicit power of $\tau_2$.
Thus, we can write \eqref{eq:laughlin_wf} as
\begin{equation}\label{wfpar}
  \bracket{z_1, z_2,\dots z_N} {\Psi (\tau)}  = {\mathcal N}_0 \tau_2^P  \hat \psi(z_1, z_2,\dots z_N; \tau ) \, ,
\end{equation}
where $\bracket {\Psi(\tau)} {\Psi(\tau)} = 1$.
Since $\hat\psi$ is holomorphic in $\tau$, it follows that 
\begin{equation}\label{laugconn}
  A_\tau = ( \rmi\partial_\tau - \rmi \frac P {\tau_2} \frac {\partial \tau_2} {\partial \tau} ) \bracket \Psi \Psi = - \frac P {2\tau_2}.
\end{equation}
A similar calculation demonstrates that $A_\tau = A_{\bar\tau}$.
This yields ${\cal F}_{\tau\bar\tau} = -P/(2\tau_2^2)$.
Combining these results finally gives the Hall viscosity
\be{finhall}
\etah = \frac PA = \frac{eB \nu} {2\pi} \frac P{N_e} \, .
\ee
For the Laughlin state $P = N_e/(4\nu)$, such that $\etah_{1/q} = {\bar n}/ {(4\nu)}$,
where $\bar n$ is the density.
This is in agreement with \oncite{Read_09}. 

From the earlier sections, we know that the hierarchy states \eqref{finwf} are also on the form \eqref{wfpar}.
However, this does not imply that the Hall viscosity can be directly extracted from the power of $\tau_2$ in the prefactor.
For this, we need the extra assumption 
that some generalized form of the plasma analogy holds for the hierarchy states. 
Recall that if the plasma analogy holds, then the full $\tau$-dependence is given by \eqref{eq:basis}.
We shall return to this question in the last section, 
but for now we will simply assume it to be true. 

By using \eqref{eq:basis}, \eqref{finwf} and \eqref{eq:mod_ST_D}, in \eqref{finhall} we then get,
\be{genhv}
\etah_{\bf K} &=& \frac 1A \left[ \half \sum_\a (\a-1)N_\a + \frac 1 4 \sum_\a N_\a \mbf q_\a^2 \right]  \nonumber\\
&=& \half \sum_\a n_\a \left(\a-1 +\half K_{\a\a} \right) \, .
\ee
We recognize the expression in the last parenthesis in \eqref{genhv} as the conformal spin of the operator $V_\a$.
Thus, we have arrived at the relation $\etah_{\bf K} = \bar n\bar s/2$,
where $\bar s$ is the average conformal spin of the electrons.
This is precisely the relation given by Read. 
From Kvornings work\cite{Kvorning_2013},
we also know that the shift for the states \eqref{finwf} are given by $S_{{\bf K}} = 2 \bar s$,
and thus we have $\etah_{\bf K} = \bar n{ S}_{{\bf K}}/4 $ as expected.
For the case of the Jain series $\nu=\frac p{2p+1}$,
$K_{\alpha\alpha}=3$ and $p_\alpha=1$, so it follows that $\bar s=1+\frac p2$.

\section{Numerical studies of the Hall viscosity at $\nu = 2/5$}\label{sec:Visc_numerics}

In this section we numerically compute the Hall viscosity for the $\nu = 2/5$ Jain state, 
using both the hierarchy wave function \eqref{eq:2/5-wfn}
and the numerically evaluated ground state for the Coulomb potential.
The result for the viscosity will provide the first numerical test of the relation $\etah = \bar nS/4$ for a hierarchy state.
The comparison between the two calculations will test the assumption that the plasma analogy is applicable also to hierarchy states.
This point will be stressed in the concluding section. 

We use the numerical methods developed in Ref. \onlinecite{Read_11}
to compute the Berry field strength ${\mathcal F}$.
The idea  is to calculate the Berry flux through a small circle $\Omega$, of radius $r$,
centered around $\tilde\tau=\tilde\tau_1+\rmi\tilde\tau_2$, and extract the corresponding average
field strength,  $\bar{\mathcal F}(\tilde\tau,r)$,  by dividing this flux with the area $A_\Omega$ of the circle.
For small values of $r$ this should give a good numerical estimate of the value ${\mathcal F} (\tilde\tau)$.

Discretizing the circle into $n$ steps, we get the approximate expression
\begin{equation}\label{berryapprox}
  e^{\rmi A_\Omega \bar{\mathcal F}}=e^{i \oint A_\mu (\lambda) d\lambda_\mu}\approx \prod_{j=0}^{n-1} \bracket{\psi_{j+1}}{\psi_j}.
\end{equation}
For each step $j$ along the curve in the $\tau$-plane,
$\ket{\psi_j}$ labels the ground state at the point $\tau_j$.
To extract $\bar{\mathcal F}$, we also need 
the $SL(2,\mathbb Z)$ invariant area which is,
\begin{equation}
  A_{\Omega}=\int_{\Omega}\frac{d\tau_1\, d\tau_2}{\tau_2^2}
  =2\pi\left[\frac1{\sqrt{1-(r/\tilde\tau_2)^2}}-1\right]\approx\pi(r/\tilde\tau_2)^2.\label{eq:Modular-invariant-area}
\end{equation}

\subsection{Exact Diagonalization of the Coulomb potential}
For the Coulomb ground state,
we evaluate $\psi_j$ in \eqref{berryapprox} numerically in a Fock basis $\{\varphi_n\}$
(\ie a basis of non-interacting electrons in the LLL) 
as $\psi = \sum_n \alpha_n \varphi_n$.
There is a universal contribution to the Berry curvature from the basis states,
while the correlation effects are simply calculated from 
the coefficients $\a_n$.
As explained in section IIB of \oncite{Read_11},
this allows for an efficient numerical evaluation of $\bar{\mathcal F}$. 
In the following, and in all the figures, we shall present our results for the viscosity in terms
of the related average spin $\bar s$.

\begin{figure} 
 \begin{tabular}{cc}
  \includegraphics[width=0.45\textwidth]{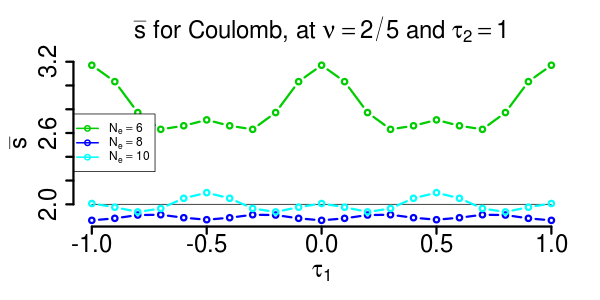} 
  &\includegraphics[width=0.45\textwidth]{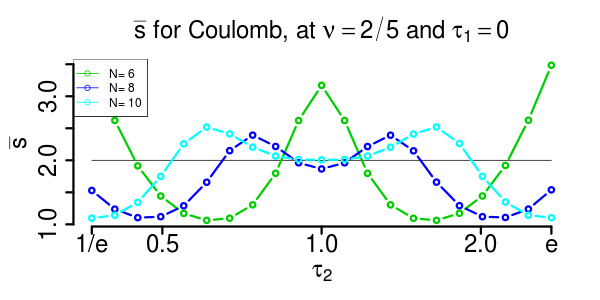}  \\
    $a)$ & $b)$
  \end{tabular}
  \caption{
    Viscosity for exact the Coulomb ground state extracted from the Berry phase
     around a circle with radius $r=0.005$ in the $\tau$-plane, which is discretized into
      $n=200$ points.
    The calculation  is for $N_e=6$ (\green), $N_e=8$ (\blue) and $N_e=10$ (\cyan) particles.
    $a)$ Scan over $-1<\tilde\tau_1<1$ with $\tilde\tau_2=1$ fixed.
    $b)$ Scan over $0.368<\tilde\tau_2<2.72$ with $\tilde\tau_1=0$ fixed.
    Note the logarithmic scale of $\tau_2$.
    The viscosity is not constant over the $\tau$-plane, 
    but seems to converge on $\bar s=2$ for larger system sizes.
    Note that the plateau with $\bar s\approx2$ becomes wider as the system size grows, 
    which indicates that the deviations from $\bar s=2$ are finite-size effects.
  }
  \label{Viscosity_ED}
\end{figure}

Figure \ref{Viscosity_ED} shows $\bar s$ calculated for the exact Coulomb state with $N_e=6,8,10$ 
at $\tilde\tau_2=1$ and $-1\leq\tilde\tau_1\leq1$,
as well as at $\tilde\tau_1=0$ and $0.368\leq\tilde\tau_2\leq 2.72$.
We present the result for a radius of $r = 0.005$ and for 200 steps.
To double check our numerics,
we have also reproduced the results of \oncite{Read_11} for the Laughlin state. 
The mirror symmetry about $\tilde\tau_2=1$ is to be expected,
since $\tau$ and $-\frac1\tau$ represent the same geometry.

The Berry curvature $\bar{\mathcal{F}}$ is \emph{not} constant in the $\tau$-plane.
In fact, there are large finite size deviations from the expected value of $\bar s=2$. 
The dependence on $\tilde\tau_1$ is  much weaker than for $\tilde\tau_2$,
but there are still finite size effects that make $\bar s\neq2$ for the smallest system sizes.
For values of $\tilde\tau_2$ deviating from 1,
the value of $\bar s$ significantly deviates from 2.
This effect however becomes less pronounced as the system size increases.
At $N_e=6$ and $N_e=8$,
the mean orbital spin has stabilized at $\bar s=2$,
at least in the quadratic case.
Indeed it can be inferred from Fig. \ref{Viscosity_ED} that the region with $\bar s\approx 2$ is wider for $N_e=8$ than for $N_e=6$,
indicating that in the 	large torus limit, the viscosity is well defined and $\bar s=2$.

Although not clearly visible in Figure \ref{Viscosity_ED},
the mean orbital spin $\bar s$ does drop to $\bar s=\frac12$ as $\tau_2\to0$ or $\tau_2\to\infty$.
At these extreme aspect ratios,
the ground state reduces to the Tau-Thouless state\cite{Bergholtz_08}. 
It would be interesting to find out to what extent the viscosity can be defined and evaluated in this limit.

\subsection{Hierarchy wave functions and the Monte Carlo algorithm}

The hierarchy trial functions are not given in a Fock basis, 
and therefore, the overlaps $\bracket{\psi_{j+1}}{\psi_j}$ have to be calculated directly using a Monte Carlo algorithm.
This  introduces an extra source of error apart from the one due to the discretization of the path.
The stochastic errors are estimated by diving each Monte Carlo set into several groups,
and calculate $\bar s$ separately for each group.
The statistical error is taken as the standard error of the mean viscosity value of all the groups.
\begin{figure} 
  \includegraphics[width=0.45\textwidth]{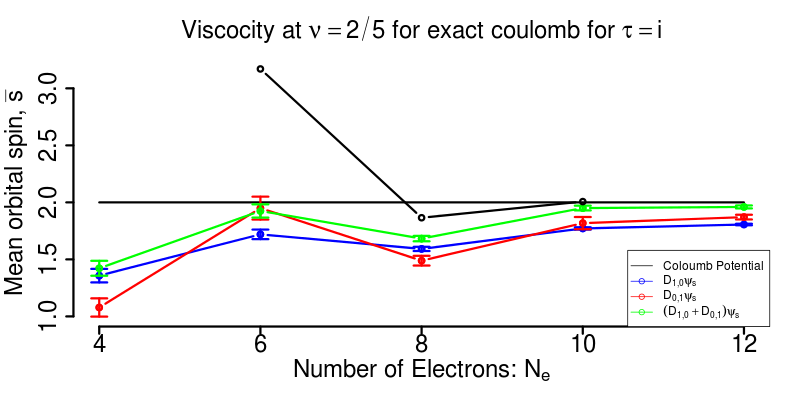} 
  \caption{
    Viscosity for CFT trial wave functions at $\tilde\tau=\rmi$.
    Evaluation is for $D_{1,0}\Psi_s $ (\blue), $D_{0,1}\Psi_s$ (\red) and $(D_{1,0}+D_{0,1})\Psi_s$ (\green) for $N_e=4,6,8,10$.
    As $N_e\to\infty$, the viscosity approaches $\bar s=2$.
    Note the logarithmic scale of $\tau_2$.
    The circle size and number of steps are the same as in Fig. \ref{Viscosity_ED}
  }
  \label{Viscosity_CFT_I=0,R=0}
\end{figure}

In Figure \ref{Viscosity_CFT_I=0,R=0}, the viscosity for $\tilde\tau_1=0$ is shown for $0.223\leq\tilde\tau_2\leq 4.48$. 
It differs between $D_{1,0}\Psi_s$ and $D_{0,1}\Psi_s$
but the values are roughly matched as $\tau\rightarrow-\frac 1 \tau$.
For $\tilde\tau_2$ far from 1,
the viscosity deviates substantially from the expected $\bar s=2$,
but for $\tilde\tau_2$ in a neighborhood of 1 the mean orbital spin is almost $\bar s=2$.

Although it is again not clear from the plot in Fig \ref{Viscosity_CFT_I=0,R=0},
the mean orbital spin approaches $\bar s=\frac12$ in the TT-limit, 
just  as in the case of the Coulomb potential.

\begin{figure} 
  \includegraphics[width=0.45\textwidth]{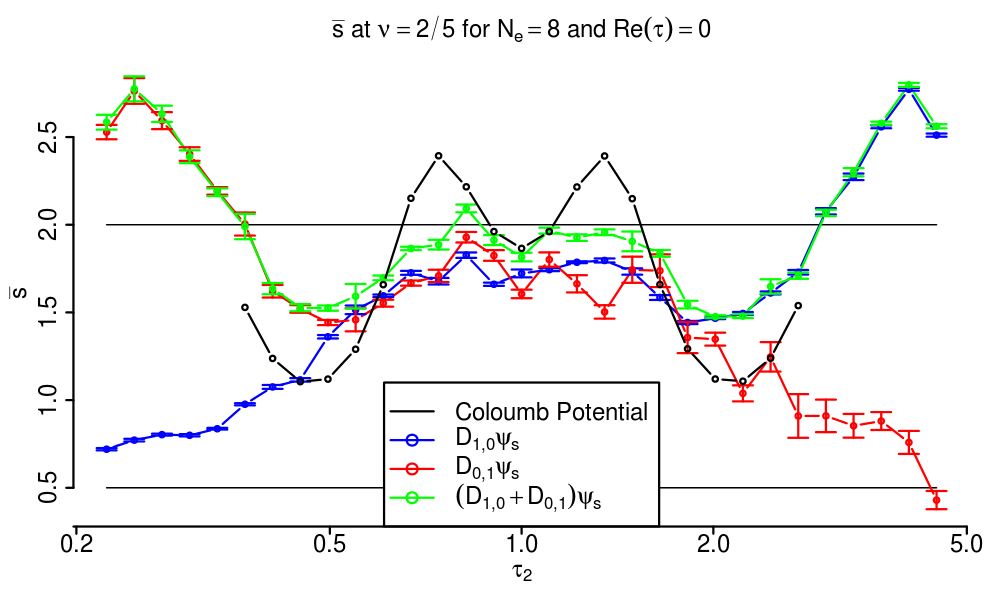} 
  \caption{\label{Viscosity_CFT}
    Viscosity for CFT trial wave functions calculated for $N_e=8$ at $\tilde\tau_1=0$ and $-0.22<\tilde\tau_2<4.5$.
    Evaluation is for $D_{1,0}\Psi_s $ (\blue), $D_{0,1}\Psi_s$ (\red) and $D_{1,0}\Psi_s + D_{0,1}\Psi_s$ (\green).
    At extremal geometries $\tilde\tau_2\neq1$ the viscosity differs substantially from $\bar s=2$ but in the rectangular geometry it fits rather well.
    The discretization of the path in the $\tau$-plane is the same as in Fig. \ref{Viscosity_ED}. 
  }
\end{figure}

Two technical comments are in order. 
First, in our expressions for the wave functions,
we used the $\tau$-gauge, in which the vector potential, when expressed in the invariant coordinates $(x,y)$, is $\tau$ independent.
We can thus directly calculate overlaps,
while in a general gauge we would have had to compensate for the change in gauge between the states along the path.
The second issue concerns the sampling strategy in the Monte Carlo algorithm, where we investigated two alternative methods.
In the first one, we used an independent set of Monte Carlo points for each matrix element $\bracket{\varphi_{j+1}}{\varphi_j}$.
In the other, we used the same coordinate points for the different matrix elements.
Although the first method clearly gives an unbiased estimate of each element,
the cumulative phase displayed large fluctuations that we were unable to control.
The second method resulted in a smooth cumulative phase and was the one used to generate the graphs in Fig. \ref{Viscosity_CFT_I=0,R=0} and Fig. \ref{Viscosity_CFT}.
Although we cannot see any particular reason for why this method should introduce systematic errors, 
we can not exclude that possibility.

 \section{Summary, remarks and  outlook}

Within the context of the CFT approach to the QH hierarchy, we have shown how to construct torus versions of 
all the fully chiral hierarchy states. Our guiding principle was to demand modular covariance, \ie that the hierarchy
wave functions should transform in the same way as the corresponding multicomponent states under  combined modular and
coordinate transformations. In addition to  changing both the modular parameter $\tau$ and the coordinates $z_i$, these  
transformations also involve changes of the gauge, and of the boundary conditions. 

Assuming that no extra Berry phases appear under adiabatic changes of $\tau$, we calculated the Hall viscosity
for a general chiral state, and found that it was related to the average orbital spin, and thus the shift on the sphere, 
by the formula derived by Read in a more restricted context.   

For the simple case of the level two state at $\nu=2/5$, which is in the positive Jain series, we  compared our proposed wave function with a numerical solution of the Coulomb problem, and found, without  fitting any parameters, a very good agreement over a large region in the complex $\tau$-plane. We also calculated the viscosity numerically, both for the Coulomb wave function, and our trial state.  In both cases we found good agreement with the analytical result. 

As  just mentioned, the analytical result for the Hall viscosity 
was derived under the assumption that our wave functions are normalized such that no extra Berry phases are incurred  during the 
adiabatic change of $\tau$, \ie that all relevant $\tau$-dependence is explicit in the wave functions. 
For the Laughlin states, this can be convincingly argued using the exact mapping onto  a classical plasma. 
The corresponding plasma analogy for the hierarchy states\cite{Read_90, Bonderson_2012} is based on weaker arguments,
 and the theoretical status of the conjecture is, in our opinion, unclear. 

A closely related problem arises when one attempts to calculate the braiding statistics of the anyonic quasi-particles
in the hierarchy states on the plane. Assuming that the Moore-Read conjecture regarding the absence of Berry phases for 
wave functions constructed from conformal blocks also applies to the hierarchy states, the statistical phases are easily obtained
from the monodromies of the wave functions.
However, there is very scant numerical support for abelian fractional statistics\cite{Jeon_03} except for the Laughlin states\cite{Kjonsberg_98} so there is little independent supporting evidence for the theoretical result, and thus for the absence of Berry phases. 

In this context we think that our result on the viscosity are important in that they  imply that the Berry phases in question
do vanish, or at least are very small, for the hierarchy wave functions obtained in the CFT approach. This gives independent  
support for  the existence of  a plasma mapping although presumably to a more complicated multicomponent plasma 
with background electric and magnetic charges\cite{Bonderson_2012}. The Hall viscosity is easier to compute 
than the fractional statistics phases where both the size of the system, and the necessity to subtract a large Aharonov-Bohm phases is numerically  very challenging. 

There are at least two other possible ways to construct chiral hierarchy wave functions on the torus. 
For the states in the positive Jain series, it would appear that one should, with suitable modification, be able to carry over the 
composite fermion construction that is so successful on the sphere and on the plane. 
In practice this turns out to be quite difficult, since the various LLL projection methods that have been developed for composite fermions on a sphere cannot be used in the torus geometry. However, in a recent paper Hermanns\cite{Hermanns_2013} has 
derived expression for a class of re-coupling coefficients needed to project the products of single torus wave functions onto the LLL. 
Using these, one can derive expressions for the composite fermion wave functions in the positive Jain series for both fermions and bosons.
Although these wave functions are likely to be modular covariant,
this has still not been proven since the explicit expressions are quite involved.
 The method is however computationally very demanding and has so far only be implemented for  $N_e=6$  at $\nu=2/5$, and for $N=10$
at the bosonic $\nu = 2/3$ state. 

A related method  to deal with the derivatives,
is to use a coherent state basis for the lowest Landau levels single particle wave functions,
and then directly project the derivatives onto this basis.
Doing this at the level of single particle wave functions,
one finds that the finite translations emerge\cite{Fremling_13},
but the resulting expressions are not very transparent,
and it is not clear how to generalize this method to  many-particle states. 

An obvious generalization of this work is to the full hierarchy where the wave functions are built from combinations of chiral and anti-chiral blocks\cite{Soursa_11a, Soursa_11b}.
It should also be possible to generalize to the non-abelian hierarchies  discussed in \oncite{Bonderson_08} and \oncite{Hermanns_10}.  
Several numerical studies could also be of interest. Extension to level three states as well 
as to level two states which are not in the Jain series, such 
as the one observed at $\nu = 4/11$, are obvious possibilities. On the sphere, and on the plane, the composite fermion wave functions are identical to the ones extracted using the CFT approach. A detailed numerical comparison of our wave functions with those of Hermanns, would thus be quite interesting since it could reveal if this holds true also in the torus geometry.  
Finally, referring to the discussion above, a more thorough study of the Hall viscosity for different states might be used to further strengthen the case for a generalized plasma mapping.

\vskip 4mm \noi      {\bf Acknowledgements: }
We thank Nick Read and Stefan Sjörs for  discussions, and Eddy Ardonne and Maria Hermanns for discussions and helpful comments on the manuscript.


\appendix

\section{Multicomponent wave functions of primary correlators}\label{app:hierarchy}
In this appendix we give sufficient details to allow the reader derive the important formulate \eqref{fullcorr}, \eqref{eq:basis} and the general expression for the modular transformation \eqref{modtrans}.

\subsection{Conformal blocks of non-compact  bosons}\label{app:Norm_ord_and_K}

In the main text we have used the normal ordered two-point function,
with equal short distance behavior as on the plane
and given by \eqref{eq:twopoint}.
Normal ordering amounts to removing all $i=j$ contributions in the correlator 
\begin{eqnarray}
  \langle \prod_{i=1}^N e^{\rmi \alpha_i \varphi(z_i,\zb_i)}\rangle
  &=& \prod_{i<j}e^{-\alpha_i\alpha_j K(z_{ij},\zb_{ij})}.
\end{eqnarray}
Using \eqref{eq:twopoint} and $2\pi\eta^3(\tau)=\vartheta_1^\prime(0|\tau)$ the correlator is rewritten as
\begin{eqnarray}
  \langle \prod_{i=1}^N e^{\rmi \alpha_i \varphi(z_i,\zb_i)}\rangle
  &=& \prod_{i<j}\left|\frac{\elliptic 1{z_{ij}/L}\tau}{2\pi L^{-1}\eta^3(\tau)} e^{-\pi\tau_2 (y_i-y_j)^2}\right|^{2\alpha_i\alpha_j}\\ 
  &=& \left|L^{-1}2\pi\eta^2(\tau)\right|^{\sum_i\a_i^2}
  \prod_{i<j}\left|\frac{\elliptic 1{z_{ij}/L}\tau}{\eta(\tau)} e^{-\pi\tau_2 (y_i-y_j)^2}\right|^{2\alpha_i\alpha_j}\, ,\label{eq:corr1}
\end{eqnarray}
where in the last equality we used charge neutrality which implies $\sum_i \a_i^2 = - 2\sum_{i<j} \a_i\a_j$.

A potentially confusing point concerning the choice of \eqref{eq:twopoint} as the torus two-point function $K(z,\zb)$,
is that any normalization constant would naively cancel in a charge neutral correlator;
that is $\sum_i \alpha_i = 0$ where $\alpha_i$ are the different charges. 
In a correlator of normal ordered operators this does not happen since the diagonal terms $\a_i^2$ are not present, and we recall
that the normalization constant $\ln|L^{-1}\vartheta'_1(0|\tau)|$ was crucial for getting the correct result. It is thus a 
legitimate question to ask what would happen if we did not use normal ordering but rather regularized the singular 
self interaction terms. This is the natural approach in a path integral framework, and following \oncite {Polchinski_98_Book}
we write the two-point function of a doubly periodic massless scalar on the torus 
$\mathbb{C}/(L(\mathbb{Z}+\tau\mathbb{Z}))$ as
\begin{equation}\label{eq:twopoint-2}
  \tilde K(z,\zb) = -\ln \big |\elliptic 1{z/L}\tau e^{\rmi\pi\tau y^2}\big |^2
  +k(\tau,\bar\tau)\, , 
\end{equation}
where we recall the parametrization $z=L(x+\tau y)$.
The function $k(\tau,\bar\tau)$ is independent of $z$ and chosen such that the two-point function is orthogonal to the constant zero-mode of the torus Laplacian. 
By orthogonality of the two-point function to the zero mode is meant
\begin{equation}\label{eq:ortho}
  \int \rmd^2 z\,\tilde K(z-z',\zb-\zb') = 0\, ,
\end{equation}
with the integration over the torus.
From this condition,  $k(\tau,\bar\tau)$ in \eqref{eq:twopoint-2} 
can be determined by straightforward computation. 
It is convenient to consider first the integral
\begin{equation}\label{eq:integral}
  I_1 = \int \rmd^2 z \, \ln |\elliptic 1{(z-z')/L}\tau /\eta(\tau)|^2\,.
\end{equation}
The integral can be calculated using the product formulas
\begin{eqnarray}\label{eq:thetatriple}
  \elliptic 1z\tau &=&2 q^{1/4} \sin(\pi z) \prod_{k=1}^\infty \times\nonumber\\
  &&\times (1-q^{2k})(1-q^{2k}e^{2\pi\rmi z})(1-q^{2k}e^{-2\pi\rmi z})\,,\\
  \eta(\tau)&=& q^{1/12}\prod_{k=1}^\infty (1-q^{2k})\,,
\end{eqnarray}
where $q= e^{\rmi \pi \tau}$.
The factors of $(1-q^{2k}e^{\pm2\pi\rmi z})$ vanish and \eqref{eq:integral} reduces to
\begin{equation}
  I_1 = \int\rmd^2 z\, \left(\ln |q^{1/6}|^2 +\ln |2\sin(\tfrac{\pi}{L}(z-z'))|^2\right)=L^2 \tau_2\left[\frac{\pi}{6}\tau_2 +2\pi \tau_2 {y'}^2\right].
  \label{eq:I1result}
\end{equation}
The integral over $\sin z$ gives a contribution of $L^2\tau_2^2\pi/2$ that compensates the $-L^2\tau_2^2\pi/3$ from $q^{1/6}$.
The integral of the Gaussian piece in the two-point function gives 
a contribution that is opposite in sign:
\begin{eqnarray}
  I_2 &=& \int\rmd^2 x\, \ln |e^{\i\pi\tau y^2}|^2\nonumber\\
  &=& L^2\tau_2\left[-2\pi \tau_2 \int_{-1/2}^{1/2}\rmd y\,(y-y')^2\right] \nonumber  \\
  &=& L^2\tau_2\left[-2\pi \tau_2 ({y'}^2+\frac{1}{12})\right]  =-I_1. 
\end{eqnarray}
This implies that the two-point function \eqref{eq:twopoint-2} satisfies the
orthogonality condition \eqref{eq:ortho}, for a symmetric choice of the unit cell, with $k(\tau,\bar\tau) = \ln |\eta(\tau)|^2$.

In calculating the expectation value of a charge-neutral string of 
vertex operators $V =e^{\rmi \a \varphi}$,
we will encounter factors $e^{\frac {\a^2} 2 \tilde K(0)}$ 
that have to be regularized in a manner consistent with the underlying geometry.
The method for doing this is described in \oncite {Polchinski_98_Book} and amounts to the replacement
\begin{equation}
  e^{\frac {\a^2} 2 \tilde K(0)} \to  e^{\frac {\a^2} 2 :\tilde K(0):} 
  = |L^{-1}\vartheta'_1(0|\tau)/\eta(\tau)|^{\a^2}
  = |L^{-1}2\pi\eta^2(\tau)|^{\a^2}
\end{equation}
With this, the correlator is readily evaluated to be
\begin{eqnarray}\label{eq:corr-zero-mode}
  \langle \prod_{i=1}^N e^{\rmi \alpha \varphi(z_i,\zb_i)}\rangle
  &=& \prod_{i,j}^N e^{-\alpha_i\alpha_j \tilde K(z_{ij},\zb_{ij})/2}
  = \prod_i^N e^{-\frac{\alpha_i^2}{2} \colon \tilde K(0)\colon}
  \prod_{i<j}e^{-\alpha_i\alpha_j \tilde K(z_{ij},\zb_{ij})} \nonumber\\
  &=& \left|L^{-1}2\pi\eta^2(\tau)\right|^{2\sum_{i=1}^N\frac{\alpha_i^2}{2}}
  \prod_{i<j}\left|\frac{\elliptic 1{z_{ij}/L}\tau}{\eta(\tau)} e^{-\pi\tau_2 (y_i-y_j)^2}\right|^{2\alpha_i\alpha_j}\, .
\end{eqnarray}
The charge neutrality ensures that the factors of $|\eta(\tau)|$ coming from $k(\tau,\bar\tau)$ cancel,
although we keep them in the above expression.
Notice that the last expression is precisely what we get using the normal ordered vertex operators and the two point function normalized as in \eqref{eq:twopoint}.

\subsection{Conformal blocks of compact bosons}

For a compactified boson the analogue of \eqref{eq:corr1} is more involved,
as sectors of nontrivial winding need to be included.
This can be done by isolating a background contribution in each winding sector by decomposing the field $\varphi_{n,n'}$ as
$\varphi_{n,n'} =\varphi_{n,n'}^{cl}+\tilde \varphi$ where $\tilde\varphi$ is a doubly periodic fluctuating field and $\varphi_{n,n'}^{cl}(z) = \tfrac{2\pi R}{\tau_2} \im \left(z(n'-n \bar\tau)\right)$.
To avoid notational clutter, we set $L=1$ and recover the general $L$ whenever needed.
The background contribution to the partition function is given by
\begin{equation}
  {\mathcal Z}_{n,n'} = \frac{R/\sqrt{2}}{\sqrt{\tau_2}|\eta(\tau)|^2}
  e^{-\frac{\pi R^2}{2\tau_2}|n\tau-n'|^2}.
  \end{equation}
The analogue of the correlator \eqref{eq:corr1} for a compactified boson is therefore given by
\begin{eqnarray}
  {\mathcal Z} \langle \prod_i e^{\rmi \alpha_i \varphi(z_i)}\rangle
  &=& \sum_{n,n'}\mathcal Z_{n,n'}\langle \prod_i e^{\rmi \alpha_i \varphi(z_i)}\rangle_{n,n'}\\
  &=& \langle \prod_i e^{\rmi \alpha_i\tilde \varphi(z_i)}\rangle
  \sum_{n,n'}\mathcal Z_{n,n'}e^{2\pi\rmi\frac{R}{\tau_2}\sum_i \im \left[\alpha_i
      z_i (n'-n\bar\tau)\right]}, \nonumber
\end{eqnarray}
where the correlator in the doubly periodic sector is given in \eqref{eq:corr1} and $\mathcal Z$ is the partition function.
A more useful expression is obtained by Poisson re-summation with respect to $n$:
\begin{eqnarray}\label{eq:compcorr}
  {\mathcal Z} \langle \prod_i e^{\rmi \alpha_i \varphi(z_i)}\rangle
  &=&  \left|L^{-1}2\pi\eta^2(\tau)\right|^{\sum_i \alpha_i^2}
  e^{-2\pi\tau_2 Q\sum_i\a_iy_i^2}\nonumber\\
  &&\times \prod_{i<j}^N \bigg | \frac{\vartheta_1(z_{ij}|\tau)}{\eta(\tau)}\bigg |^{2\alpha_i\alpha_j} \sum_{e,m} \mathcal{F}_{e,m}(Z|\tau)
  \bar{\mathcal{F}}_{e,-m}(\bar Z|\bar\tau),
\end{eqnarray}
where $Z=\sum_i \alpha_i z_i$ is the center of mass (CM) coordinate.
The exponentials of $\im z_{ij}$ present in the two-point function combine with  similar factors in the re-summed correlator and gives 
$e^{-2\pi\tau_2 Q\sum_i\a_iy_i^2}$. 
Here $Q=\sum_i\alpha_i$, which vanishes due to charge neutrality,
but will be important when a neutralizing background is present.
The CM factors are given by
\begin{equation}\label{app:CM-sing-comp}
  \mathcal{F}_{e,m}(Z|\tau) = \frac{1}{\eta(\tau)} e^{\rmi \pi\tau (e/R+mR/2)^2}
  e^{2\pi\rmi (e/R+mR/2) Z}.
\end{equation}
The sum over $e$ and $m$ can be interpreted as a sum over the primary operator
content of the model, consisting of the integer-spin scaling fields which have the left and right $U(1)$ 
charges
\begin{align}
  l_L &= e/R + mR/2 & l_R &= e/R - mR/2 \, .
\end{align}

For a rational compactification radius $R$,
the sum over electric and magnetic charges $e$ and $m$ can be further manipulated into a \emph{finite}
sum over products of Jacobi theta functions. 
Assume that $R^2 =2p/p'$ and write
\begin{eqnarray}
  (e/R+mR/2) &=& R^{-1}( e+m p /p')=R^{-1}( e+n p +\bar m p/p')\nonumber\\
  &=& R^{-1}(2 p n_1 + \bar e +\bar m p/p') = (R p'n_1+ \bar e/R +\bar m R /2),\\
  (e/R-mR/2) &=& R^{-1}( e-m p /p')= R^{-1}( e-n p -\bar m p/p')\nonumber\\
  &=& R^{-1}( 2 p n_2 + \bar e -\bar m p/p')= (R p' n_2 +\bar e/R -\bar m R /2),
\end{eqnarray}
where we write $m = \bar m + n p'$, with $\bar m = 0,\ldots, p'-1$ and $\bar e = 0,\ldots, 2p-1$. Hence, the sum over the charges $e$ and $m$ can equivalently
be written as
\begin{equation}
  \sum_{e,m=-\infty}^{\infty}\mathcal F_{e,m}(Z|\tau) \bar{\mathcal F}_{e,-m}(\bar Z|\bar \tau)
  =
  \sum_{\bar e=1}^{p^\prime}\sum_{\bar m=1}^{2p} 
  \theta_{\bar e,\bar m}(Z|\tau) \bar{\theta}_{\bar e,-\bar m}(\bar Z|\bar \tau)
\end{equation}
where 
\begin{equation}
  \theta_{\bar e,\bar m}(Z|\tau) 
    = \frac{1}{\eta(\tau)}\sum_{n=-\infty}^\infty
  e^{\rmi \pi \tau ( R p' n+\bar e/R + \bar m R/2 )^2}e^{2\pi\rmi( R p' n+\bar e/R + \bar m R/2 ) Z}.
\end{equation}
The re-summed expression reflects the existence of an extending chiral algebra which is generated by the
field $e^{\rmi R p' \varphi(z)}$,
which is local with respect to all chiral primaries of the form $e^{\rmi e/R \varphi}$ and also compatible with $\varphi$ being compact with radius $R$.
For $p'=1$, the algebra is naturally bosonic.
For $p'=2$ it is natural to consider a reduction in terms of an extending super-algebra
and in this case the Hamiltonian form of the partition function will contain an additional sum over possible spin-structures.

\subsection{The background charge}
In applications to the quantum Hall problem, it is necessary to include background charge to ensure charge-neutrality.
On the torus, a convenient choice is the continuous background.
We split the set of charges into a discrete set corresponding to vertex operator insertions with charges $ \alpha_i$,
and a continuous piece corresponding to the background with charge 
$Q = \sum_i \alpha_i = \int \rmd^2 z\, Q(z)$ to guarantee charge-neutrality.
The background operator is taken as 
\begin{equation}
  \mathcal O_{\mathrm{bg}}=e^{-\rmi\int \rmd^2 z\, Q(z)\varphi(z,\bar z)},
\end{equation}
were $Q(z) = \sum_i \a_i/A \equiv Q/(L^2\tau_2)$ is a constant.
In this case, \eqref{eq:compcorr} becomes
\begin{eqnarray}\label{app:BG_Corr}
  \mathcal{Z} \langle \prod_i e^{\rmi \a_i\varphi(z_i)}
  \mathcal O_{\mathrm{bg}}\rangle
  &=& e^{-2\pi\tau_2Q\sum_i\a_iy_i^2}
  \prod_{i<j}^N \left| \frac{\vartheta_1(z_{ij}|\tau)}{L^{-1}\vartheta_1^\prime(0|\tau)}\right|^{2 \a_i\a_j}
  \sum_{ e, m} \mathcal{F}_{ e, m}( Z|\tau)\bar{\mathcal{F}}_{ e,- m}(\bar{ Z}|\tau)\nonumber\\ 
  && \times e^{\sum_i \a_i \int \rmd^2 z Q(z) K(z-z_i)}
  e^{-\frac{1}{2}\int \rmd^2 z\rmd^2z'\, Q(z) Q(z') K(z-z^\prime)},
\end{eqnarray}
where the CM piece has already been extracted.
The background charge does not contribute to the CM piece,
and due to \eqref{eq:ortho} the only contribution is an overall factor of
$\left|L^{-1}2\pi\eta^2(\tau)\right|^{Q^2}$.
Again, charge neutrality ensures that $Q^2=\sum_i\a_i^2+2\sum_{i<j}\a_i\a_j$,
such that the exponent of $|L^{-1}2\pi\eta^2(\tau)|$ is $\sum_i\a_i^2$.
The expression in \eqref{app:BG_Corr} then simplifies to
\begin{eqnarray}\label{eq:sing_comp_vertexstringbg}
  \mathcal{Z} \langle \prod_i e^{\rmi \a_i\varphi(z_i)}
  \mathcal O_{\mathrm{bg}}\rangle
  &=&  \left|L^{-1}2\pi\eta^2(\tau)\right|^{\sum_i\a_i^2}
  \prod_{i<j}^N \bigg | \frac{\vartheta_1(z_{ij}|\tau)}{\eta(\tau)}\bigg |^{2 \a_i \a_j}\nonumber\\
  && \times e^{-2\pi\tau_2 Q\sum_i\a_i y_i^2}
  \sum_{ e, m} \mathcal{F}_{ e, m}( Z|\tau)
  \bar{\mathcal{F}}_{ e,- m}(\bar{ Z}|\tau),
\end{eqnarray}
with the appearance of the Gaussian factor $e^{2\pi\tau_2 Q\sum_i\a_i y_i^2}$.
At this stage, setting $\alpha_i=\sqrt{q}$ the Gaussian and Jastrow factor of the Laughlin wave function \eqref{eq:laughlin_wf} is recovered.
The full wave function is obtained by factoring $\mathcal F$ and $\bar{\mathcal F}$ followed by finding linear combinations that satisfy the bc:s.

\subsection{Multicomponent QH wave functions}
\subsubsection{Multicomponent conformal blocks}
For higher level hierarchy states, a single component correlator does not suffice.
The generalization to more components is straightforward as the correlator of a product theory factorizes.
The single compactified boson $\varphi$ is replaced by $\vec\varphi=\{\varphi_k\}$
and the charges $\a\to\mbf q=\{q_k\}$.
Similarly, the background charge is $\mbf Q = \sum_{i=1}^{N_e} \mbf q_i$ and 
the multi-component generalization of \eqref{eq:sing_comp_vertexstringbg} is
\begin{eqnarray}\label{eq:vertexstringbg}
  \mathcal{Z} \langle \prod_i e^{\rmi \mbf q_i\cdot\vec\varphi(z_i)}
  \mathcal O_{\mathrm{bg}}\rangle
  &=& \left|L^{-1}2\pi\eta^2(\tau)\right|^{\sum_i \mbf q_i^2}
  \prod_{i<j}^N \bigg | \frac{\vartheta_1(z_{ij}|\tau)}{\eta(\tau)}\bigg |^{2\mbf q_i \cdot \mbf q_j}\nonumber\\
  &&\times e^{2\pi\tau_2 \mbf Q\cdot\sum_i\mbf q_i y_i^2}
  \sum_{\mbf e,\mbf m} \mathcal{F}_{\mbf e,\mbf m}(\mbf Z|\tau)
  \bar{\mathcal{F}}_{\mbf e,-\mbf m}(\bar{\mbf Z}|\tau),
\end{eqnarray}
where $\mbf Z = \sum_i \mbf q_i z_i/L$ again is the CM coordinate.
To ensure that all electrons have the same charge,
$\mbf Q\cdot\mbf q_i=N_\Phi$ for all $i$.
The CM piece from \eqref{app:CM-sing-comp} can now be written compactly as 
\begin{equation}\label{eq:COM_block}
  \mathcal F_{\mbf F} = \frac 1{\eta(\tau)^n}e^{\rmi\pi\tau\mbf F^2} e^{2\pi\rmi\mbf F\cdot\mbf Z},
\end{equation}
where $\mbf F = \sum_{k=1}^n(e_k/R_k+m_kR_k/2)\hat{\mbf e}_k$ and $\hat{\mbf e}_k$ is the unit vector in direction $k$.  
This is an exact result and no overall factors have been omitted.
Restoring the factors of $L=\sqrt{A/\tau_2}$,
we can extract the conformal blocks $ \Psi_{{\mbf e},{\mbf m}}$ given in the expression \eqref{fullcorr} in the main text.

\subsubsection{Periodic boundary conditions}\label{app:PB}
In order to diagonalize the magnetic algebra,
the correlator $\mathcal{Z} \langle \prod_i e^{\rmi \alpha_i \varphi(z_i)}\rangle$
must be split into chiral and anti-chiral parts.
For the CM part, the factorization is obvious and so it is for the real Gaussian factor. For the remaining part we
simply take the functions inside the absolute value signs in  \eqref{eq:vertexstringbg}   to get the LLL 
wave function,
\begin{equation}
  \Psi_{\mbf F}=\mathcal N(\tau)e^{\rmi\pi\tau N_\Phi\sum_iy_i^2}
  \prod_{i<j}\left(\frac{\elliptic 1{z_{ij}}\tau}{\eta(\tau)}\right)^{\mbf {q}_{i}\cdot\mbf {q}_{j}} \label{chiralblocks}
  \mathcal{F}_{\mbf F}\left(\mbf Z |\tau\right) \, .
\end{equation}
The prefactor is 
$\mathcal N(\tau)=\mathcal N_0  \left(\sqrt{\tau_2}\eta^2(\tau)\right)^{\frac12\sum_i \mbf q_i^2}$
where $\mathcal N_0$ is a $\tau$-independent constant.
Physical wave functions must diagonalize $t^{(k)}_{N_\Phi,0}$ and $t^{(k)}_{0,N_\Phi}$,
for all particles labeled by $k$.
Acting with the translation operators we obtain
\begin{eqnarray}\label{eq:t_1_2_period_psi}
  t^{(k)}_{N_\Phi,0}\Psi_{\mbf F} & = & 
  (-1)^{N_\Phi+\delta}e^{2\pi\rmi\mbf F\cdot\mbf q_k}\Psi_{\mbf F}\nonumber\\
  t^{(k)}_{0,N_\Phi}\Psi_{\mbf F} & = & 
  (-1)^{N_\Phi+\delta}\Psi_{\mbf F + \mbf q_k},
\end{eqnarray}
where $\delta$ is introduced to distinguish fermions ($\delta=1$) from bosons ($\delta=0$).
The difference arises as the $K$-matrix entries can be written $K_{\a\b}=\delta_{a,\b}+\delta+1\mod 2$.
Combing that information with $pq=p\delta$, as $q$ is odd for fermions and $pq$ is even for bosons, 
leads to \eqref{eq:t_1_2_period_psi}.
Acting with the many body operators $T_1$ and $T_2$ yields
\begin{eqnarray}\label{eq:T_1_2_period_psi}
  T_{1}\Psi_{\mbf F} & = & 
  e^{2\pi\rmi\mbf F\cdot\frac{\mbf {Q}}{N_\Phi}}
  \Psi_{\mbf F}\nonumber\\
  T_{2}\Psi_{\mbf F} & = & \Psi_{\mbf F+\frac{\mbf {Q}}{N_\Phi}}.
\end{eqnarray}

To impose the single particle boundary conditions we use \eqref{eq:t_1_2_period_psi} and consider the linear combination 
\begin{equation}\label{eq:Lin_comb}
  \psi_{\mbf h,\mbf {t}} = 
  \sum_{\mbf q \in \G}e^{\rmi2\pi\mbf t\cdot \mbf q}
  \Psi_{\mbf h+\mbf {q}}.
\end{equation}
Combining \eqref{eq:Lin_comb} and \eqref{eq:COM_block},
the CM functions become
\begin{equation}\label{cmwf-full}
  \mathcal{F}_{\mbf h,\mbf t}(\mbf Z|\tau) = \frac{e^{-\rmi2\pi\mbf h\cdot\mbf t}}{\eta(\tau)^n}
  \sum_{\mbf q \in \G} e^{\rmi\pi\tau (\mbf q+\mbf h)^2}
  e^{2\pi\rmi (\mbf q +\mbf h)\cdot(\mbf Z+\mbf t)}.
\end{equation}
Substituting \eqref{eq:Lin_comb} in \eqref{eq:t_1_2_period_psi} gives the bc:s
\begin{eqnarray}\label{eq:Physical_BC}
  t^{(k)}_{N_\Phi,0}\psi_{\mbf h,\mbf {t}} & = & 
  (-1)^{N_\Phi+\delta}e^{\rmi2\pi\mbf h\cdot\mbf q_k}\psi_{\mbf h,\mbf {t}}\nonumber\\
  t^{(k)}_{0,N_\Phi}\psi_{\mbf h,\mbf {t}} & = & 
  (-1)^{N_\Phi+\delta}e^{-\rmi2\pi\mbf t\cdot\mbf q_k}\psi_{\mbf h,\mbf {t}},
\end{eqnarray}
where $\delta$ is the same as in \eqref{eq:t_1_2_period_psi}.
Combining \eqref{eq:Lin_comb} and \eqref{eq:T_1_2_period_psi} yields the $K_j$ quantum numbers
\begin{eqnarray}
  T_{1}\psi_{\mbf h,\mbf t} & = &
  e^{2\pi\rmi\mbf h\cdot\frac{\mbf Q}{N_\Phi}}\psi_{\mbf h,\mbf t}\nonumber\\
  T_{2}\psi_{\mbf h,\mbf t} & = & 
  \psi_{\mbf h+\frac{\mbf {Q}}{N_\Phi},\mbf {t}}.
  \label{eq:Physical_T1_T2}
\end{eqnarray}
We may now take the $q$:th power of \eqref{eq:Physical_T1_T2} to get 
$T^q_{1}\psi_{\mbf h,\mbf t} = e^{2\pi\rmi q \mbf h\cdot \mbf h_0}\psi_{\mbf h,\mbf t}$ and
$T^q_{2}\psi_{\mbf {r},\mbf {t}} = e^{-2\pi\rmi q \mbf t\cdot \mbf h_0}\psi_{\mbf h,\mbf t}$,
which shows that $\mbf h$ and $\mbf t$ enter on equal footing.
The wave functions \eqref{eq:Lin_comb} thus satisfy the index relations 
$\psi_{\mbf h,\mbf t+\mbf h_0}=\psi_{\mbf h,\mbf t}$
and $\psi_{\mbf h+q\mbf h_0,\mbf t}=e^{-2\pi\rmi q \mbf t\cdot \mbf h_0}\psi_{\mbf h,\mbf t}$.
The many-body momentum quantum numbers are defined as $K_1=\mbf h\cdot\mbf h_0$ and $qK_2=q\mbf h\cdot\mbf h_0$, and both are defined modulo 1.

In summary: the complete expression for the $\nu=\frac pq$ wave function is
\begin{equation}\label{app:complete-wfn}
  \psi_{\mbf h,\mbf t}=
  \mathcal N_0\left(\sqrt{\tau_2}\eta^2(\tau)\right)^{\frac12\sum_i \mbf q_i^2}
  e^{\rmi\pi\tau N_\Phi\sum_iy_i^2}
  \prod_{i<j}\left(\frac{\elliptic 1{z_{ij}}\tau}{\eta(\tau)}\right)^{\mbf {q}_{i}\cdot\mbf {q}_{j}}
  \frac{e^{-\rmi2\pi\mbf h\cdot\mbf t}}{\eta(\tau)^n}
  \sum_{\mbf q \in \G} e^{\rmi\pi\tau (\mbf q+\mbf h)^2}
  e^{2\pi\rmi (\mbf q +\mbf h)\cdot(\mbf Z+\mbf t)}.
\end{equation}

When all particles obey the same (anti-)periodic boundary conditions,
we must require $2\mbf t\cdot\mbf q_k=\tilde t$ and $2\mbf h\cdot\mbf q_k=\tilde h$,
for all $k$, where $\tilde h$ and $\tilde t$ are integers.
We solve for $\mbf h$ and $\mbf t$ as
$\mbf t=\tilde t\frac{\mbf h_0}{2}$ and $\mbf h=\tilde h\frac{\mbf h_0}{2}$.
Given this parametrization the bc:s reduce to $\phi_1/\pi=N_\Phi-\delta-\tilde t$ and $\phi_2/\pi=N_\Phi-\delta-\tilde h$.
The $K_j$ and $qK_j$ quantum numbers are
$K_1=\tilde h\frac{p}{2q}$, $qK_1=\tilde h\frac{p}{2}$ and $qK_2=\tilde t\frac{p}{2}$.
We note that even integer changes in $\tilde h$ and $\tilde t$ do not change the bc:s,
nor $qK_j$, but only the $K_1$ eigenvalue.
This means that the parity of $\tilde h$ and $\tilde t$ carry information about the boundary conditions,
whereas the rest is only the $K_1$ momentum.

For odd $q$, the boundary condition can be separated from the $K_1$ momentum,
by writing $\tilde h=2s+q\tilde r$,
where $s$ labels the $q$-fold degenerate states, and $\tilde r$ the boundary condition.
In terms of $s$ and $\tilde r$,
$K_1=s\frac pq+\tilde r\frac p2$ and $qK_1=\tilde r\frac p2$.
In terms of the boundary conditions $r=\phi_1/\pi$ and $t=\phi_2/\pi$,
the eigenvalues are, for odd $q$, 
$K_1=s\frac pq+r\frac p2+(N_\phi + 1)\frac {p\delta}2$,
$qK_1=r\frac {\delta p}2$ and $qK_2=t\frac {\delta p}2$.
In the above relations we used that $pq=p\delta$.

Using $s$, $r$ and $t$,
\eqref{app:complete-wfn} can be parametrized as 
\begin{equation}\label{eq_s_r_t_param}
  \psi_s^{(r,t)}=\psi_{s\mbf h_0+\frac q2\tilde r\mbf h_0,\frac q2\tilde t\mbf h_0}=\psi_{s\mbf h_0+\frac q2(r+N_\Phi+\delta)\mbf h_0,\frac q2(t+N_\Phi+\delta)\mbf h_0},
\end{equation}
with the axillary index relations
$\psi_{s+q}^{(r,t)}=\psi_s^{(r+2,t)}=(-1)^{tp}\psi_s^{(r,t)}$
and $\psi_s^{(r,t+2)}=\psi_s^{(r,qt)}=\psi_{s}^{(r,t)}$.
Putting $\tilde t=\tilde r=0$, such that $\mbf h\in\G^{\star}/\G$,
reproduces equation \eqref{eq:basis} and \eqref{cmwf} in the main text.

\section{Parametrization of $\G^\star/\G$ }\label{app:G_star_G}

\begin{figure}
  \begin{tabular}{ccc}
    \includegraphics[width=0.4\textwidth]{nu=2_5_lattice.png} 
    &\quad\quad\quad& \includegraphics[width=0.2\textwidth]{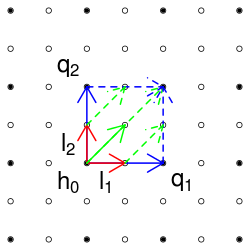}\\
    $a)$ && $b)$
  \end{tabular}
  \caption{\label{Charge_Lattice}
    Charge lattice of 
    $a)$ $K=\begin{pmatrix}3&2\\2&3\end{pmatrix}$ for $\nu=2/5$,
    which satisfies \eqref{eq:GstarG},
    and $b)$ $K=\begin{pmatrix}2&0\\0&2\end{pmatrix}$ for $\nu=1$,
    which does not.
    In both panels open circles ($\circ$) denote $\G^\star$ 
    and closed circles ($\bullet$) denote $\G$.
    A unit cell of $\G$ is spanned by $\mbf q_\a$ (\blue) 
    and a unit cell of $\G^\star$ is spanned by $\mbf l_\b$ (\red).
    $\mbf h_0=\mbf l_1+\mbf l_2$ (\green) is also marked in the figure.
    In $a)$ it is clearly seen that $\mbf l_\a=\mbf q_\a - 2\mbf h_0$
    whereas no such relation exists for $b)$.
    Also in $a)$: $\mbf q_1+\mbf q_2=\det K\cdot\mbf h_0$,
    whereas for $b)$: $\mbf q_1+\mbf q_2\neq\det K\cdot\mbf h_0$.
    The left panel is the same as in Fig. \ref{Lattice_2_5}.
  }
\end{figure}

Here we show that for a class of $K$-matrices that includes all the chiral hierarchy states, 
the lattices $\G$ and $\G^\star$ are related as 
\begin{equation}\label{eq:GstarG}
  \G^\star=\{\mbf q+s\mbf h_0;\mbf q\in \G; s\in \mathbb Z\},
\end{equation}
with $\mbf h_0=\sum_\alpha\mbf l_\alpha$,
and where the quasi-particle charge vectors $\mbf l_\alpha$ span $\G^\star$.
In words: for any vector $\mbf l\in\G^\star$,
it is always possible to find a vector $\mbf q\in\G$ and an integer $s$,
such that $\mbf l=\mbf q+s\mbf h_0$.
See Fig. \ref{Charge_Lattice} for an example where this is possible
and not possible.

The statement in \eqref{eq:GstarG} can be recast as a statement about $K$-matrices,
using that, by definition, $\mbf l_\alpha\cdot\mbf q_\beta=\delta_{\alpha,\beta}$
and $\mbf q_\alpha\cdot\mbf q_\beta=K_{\alpha,\beta}$.
Inserting the expansions $\mbf l=\sum_\alpha n_\alpha\mbf l_\alpha$
and $\mbf q=\sum_\alpha m_\alpha\mbf q_\alpha$, in $\mbf l=\mbf q+s\mbf h_0$
and taking the scalar product with $\mbf q_\beta$,
leads to the equation $\sum_{\a}K_{\b\a}m_{\a}=n_{\b}-s$.
This criteria can be recast in matrix form as
\begin{equation}\label{eq:Indirect-first-m}
  \mathbf{K}\mathbf{m}=\mathbf{n}-\mathbf{t}s,
\end{equation}
where $\mathbf{K}$ is the $K$-matrix,
$\mathbf{m}$ and $\mathbf{n}$ are integer vectors
and $\mathbf{t}=\left(1,1,\cdots,1,1\right)^{T}$ is Wen's $t$-vector in the symmetric basis\cite{WenZee_92}.
Thus, if for any integer vector $\mbf n$, it is possible to find another integer vector $\mbf m$ and an integer $s$,
such that \eqref{eq:Indirect-first-m} is fulfilled,
then the statement in \eqref{eq:GstarG} is also true.

We note that since \eqref{eq:Indirect-first-m} is linear,
it is enough to find solutions for $\mbf n=\mbf e^\beta$,
where $\mbf e^\beta$ are unit vectors in direction $\beta$.
The solution for a general $\mbf n=\sum_\beta n_\beta\mbf e^\beta$,
can the be built from the solutions for $\mbf e^\beta$.

First, assume that $K$ can be written as $\mbf K=\mbf M+r\mbf C$,
where $C_{\alpha\beta}=1$, \ie the matrix where all entries equal 1, 
and $r$ is an integer.
(Here $\mbf C$ is Wen's pseudo-identity matrix.)
Then, \eqref{eq:Indirect-first-m} becomes
\begin{equation}\label{eq:C-matrix reduction}
  \mathbf{M}\mathbf{m}=\mathbf{n}-\mathbf{t}s^\prime,
\end{equation}
with the same $\mathbf{m}$ and $\mbf n$ as in \eqref{eq:Indirect-first-m},
and with $s^{\prime}=s+r\sum_\alpha m_\alpha$.
Thus, if \eqref{eq:C-matrix reduction} has a solution, then so has \eqref{eq:Indirect-first-m}.
This is already is enough to show that \eqref{eq:GstarG} is true for all states in the positive Jain 
series $\nu=p/((q-1)p+1)$, with the $K$-matrices
\begin{equation*}
  \mbf K_{\mathrm{J+}}=\left(q-1\right)\mathbf{C}+\mathbf{1},
\end{equation*}
where $q$ labels the parent Laughlin state at $\nu= 1/q$.
With this $K$-matrix,
\eqref{eq:C-matrix reduction} becomes $\mathbf{m}=\mathbf{n}-\mathbf{t}s^\prime$,
which has the trivial solution $\mbf m=\mbf n$ and $s^\prime=0$.
Reverting back to $\mbf K_{\mathrm{J+}}$,
we can express the vectors $\mbf l_\alpha$,
as $\mathbf{l}_{\alpha}=\mathbf{q}_{\alpha}-\left(q-1\right)\mathbf{h}_{0}$.

Using \eqref{eq:C-matrix reduction},
we can also handle a general level two state, 
which is described by $\mbf K=(q-1)\mathbf{C}+\mbf M$, 
with $\mbf M = \mathrm{diag}(1,r+1)$.
Again, $q$ labels the parent Laughlin state,
and the (even) integer $r$ is the density of the quasi-electron condensate. 
It is simple to solve equation \eqref{eq:Indirect-first-m} for the two unit basis vectors $\mbf e^1 = (1,0)^T$ and $\mbf e^2 = (0,1)^T$,
to get 
\begin{equation*}
  \mathbf{m}_{1}=
  \begin{pmatrix}1-s_{1}^\prime\\\frac{-s_{1}^\prime}{r+1}\end{pmatrix}
  \qquad\mathbf{m}_{2}=
  \begin{pmatrix}-s_{2}^\prime\\\frac{1-s_{2}^\prime}{r+1}\end{pmatrix},
\end{equation*}
with the integer solutions $\mbf m_1 = \mbf e_1$ and $\mbf m_2 = - \mbf e_1$ for 
$s^\prime_\alpha=\delta_{\alpha2}$. 
Transforming back to $s_\alpha$, we have $s_1=1-q$ and $s_2=q$,
such that 
\begin{equation*}
  \mbf l_1=\mbf q_1+(1-q)\mbf h_0 \quad\quad \mbf l_2=-\mbf q_1+q\mbf h_0,
\end{equation*}
independent of $r$.
In general, there is an infinite number of solutions to \eqref{eq:Indirect-first-m} (provided a single solution can be found).
For the two layer example above, the general solution is
$s_\alpha^\prime=\delta_{2\alpha}-k_\alpha(r+1)$ and 
$\mbf m_\alpha=(k_\alpha(r+1)-(-1)^\alpha,k_\alpha)^T$ which implies 
the additional contributions $q(r+2)-1=\det K$ in $s_\alpha$,
and $-(r+1,1)^T=-(p_1 ,p_2)^T$ in $\mbf m_\alpha$.
This is a special example of the more general statement that $s$ can be restricted to 
$s=1,2,\ldots,\det K$, and that $\mbf h_0=\mbf Q/N_\Phi=\sum_\alpha ({p_\alpha} /{\det K})\mbf q_\alpha$.

To handle a general chiral hierarchy state, we need an axillary relation.
Assume that $\mbf K$ is of the block diagonal form
\begin{equation}
  \mathbf{K}=\begin{pmatrix}1 & 0\\0 & \mathbf{M}\end{pmatrix}\label{eq:K_block}
\end{equation}
where $\mathbf{M}$ is any matrix.
As above, it is sufficient to consider an arbitrary basis vector at $\mbf n=\mbf e^\alpha$,
with $e^\alpha_\beta = \delta_{\a,\beta}$.
The two blocks of \eqref{eq:Indirect-first-m} then becomes
\begin{equation*}
  \begin{pmatrix}m_{1\alpha}\\\mathbf{M}\tilde{\mathbf{m}}_\alpha\end{pmatrix}
    =\begin{pmatrix}\delta_{\alpha1}\\\tilde{\mathbf{e}}^\alpha\end{pmatrix}
    -\begin{pmatrix}s_\alpha\\\tilde{\mathbf{t}}s_\alpha\end{pmatrix} \, ,
\end{equation*}
where $\tilde{\mathbf{A}}$ is $\mathbf{A}$, save for the first element.
We consider the two cases $\alpha=1$ and $\alpha\neq1$ separately.
For $\alpha=1$, we have
\begin{eqnarray*}
  m_{11} & = & 1-s_{1}\\
  \mathbf{M}\tilde{\mathbf{m}}_{1} & = & 0-\tilde{\mathbf{T}}s_{1} \, ,
\end{eqnarray*}
with the he simplest solution $\mathbf{m}_{1}=\mathbf{e}_{1}$ and $s_{1}=0$.
For $\alpha\neq1$ the two equations are
\begin{eqnarray*}
  m_{1\alpha} & = & 0-s_{\alpha}\\
  \mathbf{M}\tilde{\mathbf{m}}_{\alpha} & = & \mathbf{e}^{\alpha}-\tilde{\mathbf{T}}s_{\alpha}\, ,
\end{eqnarray*}
which is just a reduced version of \eqref{eq:Indirect-first-m},
with $\mathbf{m}_{\alpha}\rightarrow\tilde{\mathbf{m}}_{\alpha}$ and the extra
constraint that $m_{1\alpha}=-s_{\alpha}$.
Thus if $\mbf K$ can, by subtracting an $ m \mbf C$-matrix with $m$ integer, 
be reduced to a block diagonal form of type \eqref{eq:K_block},
and this solves \eqref{eq:Indirect-first-m}, then so does $\mbf K$.

Returning to a general chiral hierarchy state,
the $K$-matrix can always be written as 
$K_{\alpha\beta}=K_{\alpha\alpha}-(1-\delta_{\alpha,\beta})$ for $\alpha\leq\beta$ and
$K_{\alpha\beta}=K_{\beta\alpha}$\cite{Soursa_11a}.
The diagonal elements of the $\mbf K$ are $K_{\alpha\alpha}=\sum_{j=1}^{\alpha}t_j$ where $t_1$ is odd and $t_j$ are even numbers, greater or equal to zero. 
This $K$-matrix can be written as 
$\mbf K=\mbf K^{(1)}=(t_1-1)\mbf C+\mbf M^{(1)}$, where 
\[
\mathbf{M}^{(1)}=\begin{pmatrix}1 & 0\\0 & \mathbf{K}^{(2)}\end{pmatrix}.
\]
$\mbf K^{(\alpha)}$ can in turn be written as $\mbf K^{(\alpha)}=t_\alpha\mbf C+\mbf M^{(\alpha)}$, such that
\[
\mathbf{M}^{(\alpha)}=\begin{pmatrix}1 & 0\\0 & \mathbf{K}^{(\alpha+1)}\end{pmatrix},
\]
{\em etc}.
By induction, we infer that all chiral hierarchy states,
by a suitable combination of $\mbf C$-matrix shifts and dimensional reductions, 
can be reduced to $\mbf K^{(n)}=t_n+1$,
which explicitly solves \eqref{eq:Indirect-first-m}.
Note that the parity of $t_\a$ was never used,
so the construction is also valid for the bosonic hierarchy.

Finally we note that the parametrization \eqref{eq:GstarG} is valid for a larger class of $K$-matrices than that
of the hierarchy. 
An example is  $\mbf K=\mathrm{diag}(1,2)$. 
On the other hand, $\mbf K=\mathrm{diag}(2,2)$ does not satisfy \eqref{eq:Indirect-first-m},
so \eqref{eq:GstarG} does not apply to all $K$-matrices.
For instance, it can be shown for a diagonal $\mbf K$,
that \eqref{eq:Indirect-first-m} is satisfied if, and only if,
all entries $K_{\a\a}$ are relatively prime to each other.
This is illustrated by two examples in Fig. \ref{Charge_Lattice}.

\section{Modular transformations of conformal blocks}\label{app:mod_trans}

In this appendix we shall derive how the $\nu=p/q$ wave function $\psi_{\mbf h,\mbf t}$ in \eqref{app:complete-wfn},
and the translation operators $t_{m,n}$ in \eqref{eq:trans_operator},
transform under modular transformations performed at a constant area.

\subsection{Conformal blocks -- the $\mathcal S$ transformation}\label{app:S-transform}

Under $\tau \to -1/\tau$ we have
\begin{eqnarray}
  \sqrt{\tau_2}&\to& \sqrt{\im (-1/\tau)} = \frac{\sqrt{\tau_2}}{|\tau|}\\
  \eta(\tau)&\to& \eta(-1/\tau) = \sqrt{-\rmi \tau}\eta(\tau)
\end{eqnarray}
and hence
\begin{equation}
  \sqrt{\tau_2}\eta^2(\tau)\to \left(\frac{\tau}{\rmi |\tau|}\right)\sqrt{\tau_2}\eta^2(\tau).
\end{equation}
The normalization $N(\tau)$ in \eqref{app:complete-wfn},
is given by $\mathcal N(\tau)=\mathcal N_0  \left(\sqrt{\tau_2}\eta^2(\tau)\right)^{\frac12\sum_i \mbf q_i^2}$,
where $\mathcal N_0$ is a constant that depends on $\tau$ only via the area $A$,
which is kept constant.
From the above relations we have 
$\mathcal N(\tau)\to \left(\frac{\tau}{\rmi |\tau|}\right)^{\frac12\sum_i\mbf q_i^2}\mathcal N(\tau)$.
The odd Jacobi theta functions transforms as
\begin{equation}
  \elliptic 1z{-1/\tau} = e^{-\rmi \pi/ 2}\sqrt{-\rmi \tau}
  e^{\rmi \pi\tau z^2}\elliptic 1 {\tau z} \tau
\end{equation}
implying
\begin{equation}
  \prod_{i<j}\left(\frac{\elliptic 1{z_{ij}/L}\tau}{\eta(\tau)}\right)^{\mbf q_i \mbf q_j}
  \to e^{-\rmi \frac{\pi}{4}(\mbf Q^2 -\sum_i \mbf q_i^2)}
  e^{\rmi \frac{\pi N_\Phi}\tau \sum_i (z_i/L)^2}
  e^{-\rmi \frac\pi\tau \mbf Z^{2}}
  \prod_{i<j}\left(\frac{\elliptic 1{z_{ij}^\prime/L}\tau}{\eta(\tau)}\right)^{\mbf q_i \mbf q_j}
\end{equation}
for the Jastrow factors.
Again the coordinates transform as $z\to\frac{\bar\tau}{|\tau|}z$ under $\mathcal S$.
For the Gaussian factors we have
\begin{equation}
  e^{-\frac{\tau}{2\tau_2}\sum_i(\im z_i)^2} \to 
  e^{\frac{\bar\tau}{2\tau_2|\tau|}\sum_i(\im \bar\tau z_i)^2}
  =e^{-\frac{\tau}{2\tau_2}\sum_i(\im z_i)^2}e^{-\rmi\frac{\tau_2}{2\tau}\sum_iz_i^2}\prod_i\US^{(i)},
\end{equation}
where $\US^{(i)}=\exp(\rmi2\pi N_\Phi x_iy_i)$ is the associated gauge transformation  given in \eqref{eq:US_UT_gauges} in the main text. 
The transformation of the CM functions can be worked out by using a multidimensional version of the Poisson resumation formula:
\begin{equation}
  \sum_{\mbf {q}\in\G}\exp\left(-\pi a\mbf {q}\cdot\mbf {q}+\mbf {q}\cdot\mbf {b}\right)=
  \frac{1}{\mathrm{vol}\left(\G\right)}
  \frac{1}{a^{\frac{n}{2}}}\sum_{\mbf {l}\in\G^{\star}}
  \exp\left[-\frac{\pi}{a}\left(\mbf {l}+\frac{\mbf {b}}{2\pi\rmi}\right)^{2}\right] \,, 
\end{equation}
where $\G^\star$ is the lattice dual to $\G$. The result is
\begin{eqnarray*}
  \mathcal F^{(\frac{-1}\tau)}_{\mbf h,\mbf t}(\mbf Z)
  &=&e^{\rmi\pi\frac{-1}{\tau}\mbf {h}^{2}}
  e^{2\pi\rmi\mbf {h}\cdot\frac{\mbf {Z}}{\left|\tau\right|}}
  \sum_{\mbf {q}\in\G}e^{\rmi\pi\frac{-1}{\tau}\mbf {q}^{2}}
  e^{2\pi\rmi\mbf {q}\cdot\left(\frac{\mbf {Z}}{\left|\tau\right|}+\mbf {t}-\frac 1\tau\mbf {h}\right)}\\
  &=&(\det K)^{-1/2}
  e^{\rmi\pi\frac{1}{\tau}\left(\frac{\tau}{\left|\tau\right|}\mbf Z\right)^{2}}
  \frac1{\eta(\tau)^n}
  \sum_{\mbf {l}\in\G^{\star}}
  e^{\rmi\pi\tau\left(\mbf {l}+\mbf {t}\right)^{2}}
  e^{\rmi2\pi\left(\mbf {l}+\mbf {t}\right)\left(\frac{\tau}{\left|\tau\right|}\mbf Z-\mbf {h}\right)} \, .
\end{eqnarray*} 
By \eqref{eq:GstarG}, we write $\mbf l=\mbf q+\mbf h'$,
where $\mbf q\in\G$ and $\mbf h'\in\G^\star/\G$.
The CM function is then rewritten as 
\begin{equation}
  \mathcal F_{\mbf h,\mbf t}(\mbf Z) \to e^{\rmi \frac\pi\tau \mbf Z^2}
  \sum_{\mbf h'\in \G^{\star}/\G}
  S_{\mbf h,\mbf h'+\mbf t} \mathcal F_{\mbf h'+\mbf t,-\mbf h}(\mbf Z),
\end{equation}
where $S_{\mbf h,\mbf h'} = (\det K)^{-1/2}e^{-2\pi\rmi \mbf h'\cdot \mbf h}$ is the modular $S$-matrix for the CFT. 
Putting the above results together, the chiral blocks transform under $\mathcal S$ as:
\begin{equation}
  \psi_{\mbf h,\mbf t} \to \left(\frac{\tau}{|\tau|}\right)^{\frac12\sum_i \mbf q_i^2}
  e^{-\rmi \pi (\mbf Q/2)^2} e^{-\rmi2\pi\mbf t\cdot\mbf h} \prod_i\US^{(i)}
  \sum_{\mbf h\in \G^*/\G}S_{\mbf h,\mbf h'+\mbf t} \psi_{\mbf h'+\mbf t,-\mbf h}  
\end{equation}
where the gauge factor is $\US=\exp\left(\rmi2\pi N_\Phi xy\right)$.
Using \eqref{eq_s_r_t_param} and the parametrization of $\G/\G^\star$ from section \ref{app:G_star_G},
we get the relations \eqref{modtrans}, \eqref{eq:US_UT_gauges} and \eqref{eq:B_ST} in the main text.

\subsection{Conformal blocks -- the $\mathcal T$ transformation}\label{app:T-transform}

The modular $\mathcal T$-transform is to be thought of as realigning the fundamental domain of the torus.
We will use that $\eta(\tau+1)=e^{\frac{\rmi\pi}{12}}\eta(\tau)$ and 
$\elliptic 1z{\tau+1} = e^{\frac{\rmi\pi}{4}}\elliptic 1z{\tau}$.
The normalization will transform as 
\begin{equation}
  \mathcal N(\tau+1)=e^{\rmi\pi\frac{1}{12}\sum_{i}\mbf {q}_{i}^{2}}\mathcal N(\tau)
\end{equation}

We also have
\begin{equation}
  \left(\frac{\elliptic 1{\frac{z_{ij}}{L_{x}}}{\tau+1}}{\eta\left(\tau+1\right)}\right)^{\mbf {q}_{i}\cdot\mbf {q}_{j}}
  =e^{\rmi\pi\frac16\mbf {q}_{i}\cdot\mbf {q}_{j}}
  \left(\frac{\elliptic 1{\frac{z_{ij}}{L_{x}}}{\tau}}{\eta\left(\tau\right)}\right)^{\mbf {q}_{i}\cdot\mbf {q}_{j}}.
\end{equation}
so the full Jastrow factor will pick up the phase 
\begin{equation}
  e^{\rmi\pi\frac{1}{6}\sum_{i<j}\left(\mbf {q}_{i}\cdot\mbf {q}_{j}\right)}
  =e^{\rmi\pi\frac{1}{12}\mbf {Q}^{2}}e^{-\rmi\pi\frac{1}{12}\sum_{i}\mbf q_i^2} \, .
\end{equation}

The Gaussian part contributes a coordinate dependent phase,
which is the gauge transformation $\UT(z) = \exp(\rmi\pi N_\Phi y^2)$,
given in \eqref{eq:US_UT_gauges} in the main text.

The CM part is trickier, and needs some care.
The difficulty is that $\mathcal F_{\mbf h,\mbf t}$
will change its boundary conditions.
When $\tau\rightarrow\tau+1$,
the sum over $\mbf q$ in the CM function \eqref{cmwf-full},
will pick up an extra factor $e^{\rmi\pi\left(\mbf {h}+\mbf {q}\right)^{2}}
=e^{\rmi\pi\mbf q\cdot\mbf q}e^{\rmi2\pi\mbf h\cdot\mbf q}e^{\rmi\pi\mbf h\cdot\mbf h}$, 
and there will also be an extra overall phase $e^{-2\pi\rmi\frac n{24}}$.
While the piece $e^{\rmi\pi\mbf {h}\cdot\mbf {h}}$ can be taken out of the sum, 
and $e^{\rmi2\pi\mbf h\cdot\mbf q}$ can be absorbed in the argument $\mbf Z$,
the factor $e^{\rmi\pi\mbf {q}\cdot\mbf {q}}$ requires some rewriting.
Using $\mbf {q}=\sum_{\beta}n_{\beta}\mbf {q}_{\beta}$ we get
\begin{eqnarray}
  \mbf q^2=\sum_{\alpha,\beta}n_\alpha n_\beta \mbf q_\alpha\cdot \mbf q_\beta=\sum_{\alpha}n_\alpha\,\mod 2 \, ,
\end{eqnarray}
and the same also holds true for $\mbf q\cdot\mbf h_0$, since
\begin{eqnarray}
  \mbf q\cdot\mbf h_0=\sum_\alpha n_\alpha\mbf q_\alpha\cdot\frac{\mbf Q}{N_\Phi}=\sum_\alpha n_\alpha \, .
\end{eqnarray}
From this follows
\begin{eqnarray}
  \mathcal{F}_{\mbf h,\mbf t}(\mbf Z|\tau+1) 
  & = & e^{\rmi\pi\mbf {h}\cdot\mbf {h}}\frac{e^{-2\pi\rmi\frac n{24}}e^{-\rmi2\pi\mbf h\cdot\mbf t}}{\eta(\tau)^n}
  \sum_{\mbf q \in \G} e^{\rmi\pi\tau (\mbf q+\mbf h)^2}
  e^{2\pi\rmi (\mbf q +\mbf h)\cdot(\mbf Z+\mbf t)}
  e^{2\pi\rmi \mbf q\cdot(\frac{\mbf h_0}2+\mbf h)}\nonumber\\
  & = & e^{\rmi\pi\mbf {h}\cdot\mbf {h}} \frac{e^{-2\pi\rmi\frac n{24}}e^{-\rmi2\pi\mbf h\cdot(\mbf t+\mbf h+\frac{\mbf h_0}2)}}{\eta(\tau)^n}
  \sum_{\mbf q \in \G} e^{\rmi\pi\tau (\mbf q+\mbf h)^2}
  e^{2\pi\rmi (\mbf q +\mbf h)\cdot(\mbf Z+\mbf t+\mbf h+\frac{\mbf h_0}2)}\nonumber\\
  & = & e^{\rmi\pi\mbf {h}\cdot\mbf {h}}e^{-2\pi\rmi\frac n{24}}
  \mathcal{F}_{\mbf h,\mbf t+\mbf h+\half {\mbf h}_0} (\mbf Z|\tau),
\end{eqnarray}
where the equality is obtained using \eqref{cmwf-full}.

Putting the above results together, the chiral blocks transform under $\mathcal T$ as
\begin{equation}\label{eq:T-on-psi}
  \psi_{\mbf h,\mbf t} \stackrel{\mathcal T}{\rightarrow}
  e^{\rmi2\pi\frac{1}{24}\mbf {Q}^{2}}
  e^{\rmi2\pi\left(\frac{\mbf {h}\cdot\mbf {h}}{2}-\frac{n}{24}\right)}
  \prod_i \UT^{(i)}
  \psi_{\mbf h,\mbf t+\mbf h+\frac{\mbf h_0}2} \, ,
\end{equation}
where $n$ is the level of the hierarchy, which equals the central charge of the underlying CFT.
Finally using the parametrization of $\G^\star/\G$ from section \ref{app:G_star_G}, we have
\begin{equation}\label{eq:T-on-psi2}
  \psi_s^{(r,t)}(z)\stackrel{\mathcal{T}}{\to}
  \prod_{i=1}^{N_e}\UT(z_i)\,\,
  B_{\mathcal T} \sum_{s^\prime=1}^qT_{s+\Delta_r,s^\prime+\Delta_r}\psi_{s^\prime}^{(r,t+r+N_\Phi)}(z)\,,
\end{equation}
where $\Delta_q(N_\Phi+q+r)/2$.
This summarizes the relations \eqref{modtrans} \eqref{eq:US_UT_gauges} and \eqref{eq:B_ST} in the main text.

\subsection{Modular transformations of the operators $t_{m,n}^{(\tau)}$ }\label{app:t_transform}

The translation operators that do not change the boundary conditions are
\begin{equation}
  t_{m,n}=\exp(m\eps\partial_x+n\eps\partial_y+\rmi2\pi nx),
\end{equation}
where $\eps=1/N_\Phi$.
Although the parameter $\tau$ does not explicitly enter these expressions, 
it is implicitly understood that $x$ and $y$ should be combined to the coordinate $z=L(x+\tau y)$.

Under $\mathcal T$, the coordinate $z$ changes as $z\to L(x+y+\tau y)$.
This induces the mapping $x\to x-y$ and $y\to y$,
with derivatives $\partial_x\to\partial_x$ 
and $\partial_y=\partial_x+\partial_y$.
It follows that 
\begin{equation*}
  t_{m,n}\to\exp((m+n)\eps\partial_x+n\eps\partial_y+\rmi2\pi n(x-y)).
\end{equation*}
We also know that the Gaussian factor changes as
$e^{\rmi\pi\tau N_\Phi y^2}\to e^{\rmi\pi\tau N_\Phi y^2}e^{\rmi\pi N_\Phi y^2}$.
The extra phase can be extracted as a gauge transformation 
$\UT=\exp(\rmi\pi N_\Phi y^2)$,
which applied to $t_{m,n}$, gives
\begin{equation}
  t_{m,n}\to \UT t_{m+n,n}\UT^{-1}.
\end{equation}

In a similar fashion we now consider $\mathcal S$,
under which  $\tau\to-1/\tau=-\bar\tau/|\tau|^2$,
which implies $L\to|\tau|L$,
such that $z/L=x+\tau y\to\frac{\bar\tau}{|\tau|^2}z/L=x-y/\tau$.
This induces $x\to y$ and $y\to-x$, such that
\begin{equation}\label{app_t_mn}
  t_{m,n}\to\exp(m\eps\partial_y-n\eps\partial_x+\rmi2\pi ny).
\end{equation}

For the gauge factor, we have $e^{\rmi\pi\tau N_\Phi y^2}\to e^{-\rmi\pi N_\Phi x^2/\tau}$,
which can be rewritten as $e^{\rmi\pi\tau N_\Phi y^2}\to e^{\rmi\pi\tau N_\Phi y^2}e^{-\rmi\pi N_\Phi z^2/(L^2\tau)}e^{\rmi2\pi N_\Phi xy}$, showing that the gauge transform is $\US=\exp(\rmi2\pi N_\Phi xy)$.
The holomorphic factor $e^{-\rmi\frac{\tau_2}{2\tau}z^2}$ can be combined with the rest of the wave function,
where it cancels pieces from the Jastrow factors.
We can finally write \eqref{app_t_mn} as 
\begin{equation}
  t_{m,n}=\US t_{-n,m}\US^{-1},
\end{equation}
which is equation \eqref{ttrans} in the text.

\section{The operator $\mathbb{D}_{\left(\alpha\right)} $}\label{app:D_operator}

Here we prove that $\mathbb D_{(\a)}$ defined by \eqref{micked} preserves the boundary conditions,
and quantum numbers of the primary correlator,
and satisfies the relations \eqref{mickedcom} and \eqref{eq:mod_ST_D}.
In doing this we also give the relevant expressions for general boundary conditions $(r,t)$. 
The outline of this rather long appendix is as follows:
In section \ref{app:D_qnums} we 
construct an operator that, when acting on a state $\psi$, leaves the $K_a$ quantum numbers unchanged, and 
in section Appendix \ref{ap:Modtrans_D} we give the modular transformations of the coefficients $\lambda_{m,n}$, and their implications for the operator $\mathbb D$.
Next, in section \ref{ap:Ferm_Bos_D} we consider general boundary conditions, and give formulas for the pertinent operator
$\mathbb D$ for both bosons and fermions. 
Section \ref{app:Sum_m_n}, 
specifies the limits of the sums in appearing in $\mathbb D$ , 
and finally, in section \ref{app:hierarchy_op},
we prove Eq. \eqref{mickedcom}.

\subsection{Quantum numbers}\label{app:D_qnums}

In section \ref{sec:Hierachy_for_lambda} we gave a physics argument that suggested that the operator $\mathbb D$,
that would be the torus version of the holomorphic derivative $\partial_z$ should involve a sum
\be{shift}
\sum_{m,n}\lambda_{m,n} T_{m,n}^{\left(\alpha\right)}
\ee
Note, however,
\begin{eqnarray*}
  T_1T_{m,n}^{\left(\alpha\right)}&=&
  e^{\rmi2\pi n\frac {p_\a}q}T_{m,n}^{\left(\alpha\right)}T_1\\
  T_2T_{m,n}^{\left(\alpha\right)}&=&
  e^{-\rmi2\pi m\frac {p_\a}q}T_{m,n}^{\left(\alpha\right)}T_2,
\end{eqnarray*}
which implies that when applied to a state with good $K_a$ quantum numbers,
the terms in the sum \eqref{shift} induces the shifts
$K_1\to K_1+n\frac{p_a}q$ and $K_2\to K_2-m\frac{p_a}q$. 
An obvious way to correct for this is to introduce an operator $T_{m^\prime,n^\prime}$,
which as special cases has the generators of the $q$-fold degenerate subspace, $T_1$ and $T_2$.
This leads to the ansatz,
\begin{equation*}
  \mathbb{D}_{\left(\alpha\right)} =\sum_{m,n}e^{N_\alpha K(\delta_{m,n})}T_{m,n}^{\left(\alpha\right)}T_{m',n'},
\end{equation*}
where the integers $m',n'$ should be chosen such that the quantum numbers are unchanged by $\mathbb D$, 
\ie so that $[\mathbb D_{(\alpha)},T_{m'',n''}]=0$ for all $m'',n''$.
We achieve this by imposing  the condition term by term,
\begin{equation}\label{dcond}
  \forall \ m'',n'' : \ \ [T_{m,n}^{\left(\alpha\right)}T_{m^\prime,n^\prime},T_{m'',n''}]=0,
\end{equation}
and requiring, $(m,n)$ and $(m',n')$ to be parallel,
\ie $(m',n')=(r_\a m,r_a,n)$.
The integer $r_\a$ is determined by using
\begin{equation*}
  T_{m^\prime,n^\prime}T_{m'',n''}=e^{\rmi2\pi \frac pq(m'n''-n'm'')}T_{m'',n''}T_{m^\prime,n^\prime},
\end{equation*}
in \eqref{dcond}, which leads to $\exp(\frac{\rmi2\pi}{q}(p_\alpha+r_\alpha p))=1$.
This is a modulo equation with formal solution $r_\alpha=-p^{-1}p_{\alpha}\mod q$.
As $p$ and $q$ are relatively prime, this equation always has a solution,
and the inverse of $p$ is $p^{-1}=p^{\varphi\left(q\right)-1}\mod q$, where, 
according to Euler's theorem, $\varphi\left(q\right)$, is the
number of integers smaller than or equal to $q$ that are relatively
prime to $q$. Thus we find the modular operator to be
\begin{equation}
  \mathbb{D}_{(\alpha)} =\sum_{m,n}e^{N_\alpha K(\delta_{m,n})}T_{m,n}^{(\alpha)}T_{-p^{-1}p_{\alpha}m,-p^{-1}p_{\alpha}n}.
\end{equation}
Next we fix the square root of $K(\delta_{m,n})$,
such that is has simple modular properties.

\subsection{Modular transformations } \label{ap:Modtrans_D}
In Section \ref{sec:Hierachy_for_lambda} we gave a physics based argument that determined $\lambda_{m,n}$ up to the phase $\zeta_{m,n}$.
To fix this phase we use the properties of $\eta$, $\tau_2$ and $\vartheta_1$,
given in Appendix \ref{app:S-transform} and \ref{app:T-transform},
to compute that $K(\delta_{m,n})$ transforms as
\begin{eqnarray*}
  e^{K(\delta_{m,n})}&\stackrel{\mathcal S} {\to}& 
  \left(\frac\tau{|\tau|}\right)
  \sqrt{\tau_2}\eta^3(\tau)
  \frac{e^{-\i\pi\tau m^2 \eps^2}e^{\rmi2\pi mn\eps^2}}{\elliptic 1{\eps(-n+\tau m)}\tau}\\
  &&=\left(\frac\tau{|\tau|}\right)
  e^{K(\delta_{-n,m})}e^{\rmi2\pi mn\eps^2}\\ 
  e^{K(\delta_{m,n})}&\stackrel{\mathcal T}{\to}&
  \sqrt{\tau_2}\eta^3(\tau)
  \frac{e^{-\i\pi\tau n^2 \eps^2}e^{-\i\pi n^2 \eps^2}}{\elliptic 1{\eps(m+n+\tau n)}\tau}\\
  && = e^{K(\delta_{m+n,m})}e^{-\rmi \pi n^2\eps^2}.
\end{eqnarray*}

We now fix $\zeta_{m,n}$ so that $\lambda_{m,n}$ has the simple transformation properties
\begin{eqnarray}
  \lambda_{m,n}  &\stackrel{\mathcal S} {\to}& \lambda_{-n,m}^{(\tau)}\left(\frac\tau{|\tau|}\right)  \nonumber\\
  \lambda_{m,n}  &\stackrel{\mathcal T} {\to}&  \lambda_{m+n,n}^{(\tau)} \, .
\end{eqnarray}
This implies the relations $\zeta_{-n,m}=\zeta_{m,n}e^{\rmi2\pi mn\eps^2}$ and 
$\zeta_{m+n,n}=\zeta_{m,n}e^{-\rmi \pi n^2\eps^2}$,
which are solved by taking $\zeta_{m,n}=e^{-\rmi \pi nm\eps^2}$. There is, however, one additional freedom 
in picking the phase.
The sign $(-1)^{l(m + n +mn)}$, where $l = 0,1$ is invariant under the modular transformations,
so we get the two solutions,
\begin{equation}\label{weight_app}
  \lambda_{m,n}=\sqrt{\tau_2}\eta^3(\tau)
  \frac{e^{-\rmi\pi\tau n^2\epsilon^2}e^{-\rmi\pi nm\epsilon^2}}
       {\elliptic 1{m\epsilon+n\epsilon\tau}{\tau}}(-1)^{l(m + n +mn)} \ \ \ \ l=0,1 \, .
\end{equation}
In the main text we will choose to put $(-1)^{l(m + n +mn)}$ in the overall sign factor $\xi$.
Using these $\lambda_{m,n}$ to define
\begin{equation}\label{app:D_N_even}
  \mathbb D_{(\a)} = \sum_{m,n}\lambda_{m,n}^{N_\alpha}T_{m,n}^{(\a)}T_{r_\a m,r_\a n},
\end{equation}
it is easy to derive the relations,
\begin{eqnarray}\label{app:transform_D}
  \mathbb D_{(\a)} &\stackrel{\mathcal S}{\to}& 
  \sum_{m,n}\left(\frac\tau{|\tau|}\right)^{N_\a}\lambda_{-n,m}^{N_\alpha}\US T_{-n,m}^{(\a)}T_{-r_\a n,r_\a m}\US^\dagger\nonumber\\
  &&=\left(\frac\tau{|\tau|}\right)^{N_\a}\US\mathbb D_{(\a)}\US^\dagger\nonumber\\
  \mathbb D_{(\a)}& \stackrel{\mathcal T}{\to}& 
  \sum_{m,n}\lambda_{m+n,n}^{N_\alpha}\US T_{m+n,n}^{(\a)}T_{r_\a (m+n),r_\a n}\US^\dagger\nonumber\\
  &&=\UT\mathbb D_{(\a)}\UT^\dagger,
\end{eqnarray}
assuming that the sums are over all integers. As discussed in the main text, these transformations
ensure that $ \mathbb D_{(\a)} \psi_s$ transforms covariantly under $\mathcal S$ and $\mathcal T$. 
In the next section we show that the particular $ \mathbb D_{(\a)} $ given here is appropriate only for
an even $N_\Phi$ and periodic boundary conditions, and give the relevant formulae for the general case,
including that of bosons.

\subsection{General boundary conditions } \label{ap:Ferm_Bos_D}

In this section explore the consequences of requiring that $\tilde\psi$ should transform the same way as $\psi$.
For simplicity, we consider here only states with odd $q$ in the filling fraction $\nu=p/q$.
This covers all fermion states and a large class of boson states.
When $q$ is odd, we can define the translation operator 
\begin{equation*}
  H_{t,r}=T_{\frac{qt}2,\frac{qr}2},
\end{equation*}
that changes the boundary conditions of $\psi_s$ as 
\[
\psi_s^{(r,t)}=H_{t,r}\psi_s^{(0,0)}.
\]
In this particular section, for convenience,
a different $r,t$-dependent phase is used in the definition of $\psi_s^{(r,t)}$.
As a consequence, the index relations of $\psi^{(r,t)}$ are different from \eqref{eq:s,t,r_relations}, as is the precise form of \eqref{modtrans}, but the $K_a$ quantum number are still the same.
We now assume a generic operator $\mathcal{O}$, such that 
\begin{equation}\label{app:eq:psi_tilde_O_psi}
  \tilde\psi_s^{\left(r,t\right)}=\mathcal O_{r,t}\psi_s^{\left(r,t\right)},
\end{equation}
is a state with the same boundary conditions as $\psi$.
Since $\tilde\psi$ and $\psi$ should obey the same equations,
$\tilde\psi_s^{(r,t)}=H_{t,r}\tilde\psi_s^{(0,0)}$ must hold for $\tilde\psi$,
which leads to the relation 
\begin{equation}
  \mathcal{O}_{r,t}=H_{t,r}\mathcal{O}H_{t,r}^{\dagger},
\end{equation}
where $\mathcal{O}=\mathcal{O}_{0,0}$.
Thus, for different boundary conditions,
different operators $\mathcal{O}_{r,t}$ are needed.

We now turn to the modular transformations \eqref{modtrans}, 
for $\tilde\psi$ and $\psi$.
As neither $qK_1$ nor $qK_2$, and therefore neither $H_{t,r}$ nor $\mathcal O_{r,t}$,
depends on $s$, we can disregard many of the details in \eqref{modtrans}
when considering modular transformations.
In fact, for our purposes,
all details except the changes in boundary conditions $(r,t)$,
can be suppressed for a cleaner notation.
Thus, under $\mathcal S $ and $\mathcal T$-transformations, $\psi$ transforms as
\begin{eqnarray}\label{app:red_S_T}
  \psi^{(r,t)}&\stackrel{\mathcal S}{\to}&\psi^{(t,r)}\nonumber\\
  \psi^{(r,t)}&\stackrel{\mathcal T}{\to}&\psi^{(r,t+r+N_\Phi)}.
\end{eqnarray}

For convenience, we also supress any constant phases $\tau/|\tau|$ as well as gauge transformations,
and just write $\mathcal O\to\mathcal O^{\mathcal S\,(\mathcal T)}$ under $\mathcal S$ ($\mathcal T$).
Requiring equal transformations under $\mathcal S$, for $\tilde\psi$ and $\psi$,
leads to 
\[
H_{r,t}\mathcal{O}H_{r,t}^{\dagger}\psi^{\left(t,r\right)}=H_{r,t}\mathcal{O}^{\mathcal{S}}H_{r,t}^{\dagger}\psi^{\left(t,r\right)}.
\]
Comparing the left and right hand sides, gives the condition $\mathcal{O}\rightarrow\mathcal{O}^{\mathcal{S}}=\mathcal{O}$,
\ie $\mathcal O$ should be invariant under $\mathcal S$-transformations.
The analogous condition on $\mathcal{T}$ gives
\[
H_{t+r+N_\Phi,r}\mathcal{O}H_{t+r+N_\Phi,r}^{\dagger}\psi^{\left(r,t+r+N_\Phi\right)}= H_{t+r,r}\mathcal{O}^{\mathcal{T}}H_{t+r,r}^{\dagger}\psi^{\left(r,t+r+N_\Phi\right)},
\]
which can be simplified to $\mathcal{O}\rightarrow\mathcal{O}^{\mathcal{T}}=H_{N_\Phi,0}\mathcal{O}H_{N_\Phi,0}^{\dagger}$,
by splitting $H_{t+r+N_\Phi,r}=H_{t+r,r}H_{N_\Phi,r}e^{\i\phi}$.
Thus, $\mathcal O$ must satisfy the two transformations
\begin{eqnarray}\label{app:O_tarns_S_T}
  \mathcal{O}&\stackrel{\mathcal S}{\to}&\mathcal{O}\nonumber\\
  \mathcal{O}&\stackrel{\mathcal T}{\to}&H_{N_\Phi,0}\mathcal{O}H_{N_\Phi,0}^{\dagger}.
\end{eqnarray}
Note that if $N_\phi$ is an even number,
$H_{N_\Phi,0}\mathcal{O}H_{N_\Phi,0}^{\dagger}=\mathcal{O}$ by virtue of \eqref{dcond}.
Thus, for an even number of fluxes,
$\mathcal O$ should be invariant under both $\mathcal S$ and $\mathcal T$.
An operator that has these transformation properties is $\mathbb D_{(\a)}$, defined in \eqref{app:D_N_even}. 

To find an operators that satisfies \eqref{app:O_tarns_S_T},
we first study $H_{N_\Phi,0}\mathbb T_{m,n}H_{N_\Phi,0}^{\dagger}$,
for the individual terms $\mathbb T_{m,n}=T_{m,n}^{\left(\alpha\right)}T_{r_{\alpha}m,r_{\alpha}n}$, in $\mathbb D$.
We have 
\begin{equation}
  H_{N_\Phi,0}\mathbb T_{m,n}H^\dagger_{N_\Phi,0}
  =\mathbb T_{m,n}e^{\rmi\pi nN_\Phi(p_\a+pr_\a)}
  =\mathbb T_{m,n}e^{\rmi\pi N_\Phi np_\a(1-pp^{-1})}
  =\mathbb T_{m,n}(-1)^{\Lambda N_\Phi n},
\end{equation}
where $pp^{-1}=1+\Lambda q \mod 2$.
When $p$ is an even number, then $\Lambda=1$,
such as for the fermionic $\nu=2/5$, and all bosonic states with odd denominator.
In a similar way, we also obtain
\begin{eqnarray*}
  H_{t,0}\mathbb T_{m,n}H^\dagger_{t,0}&=&\mathbb T_{m,n}(-1)^{\Lambda t n}\\
  H_{0,r}\mathbb T_{m,n}H^\dagger_{0,r}&=&\mathbb T_{m,n}(-1)^{\Lambda r m}.
\end{eqnarray*}

Clearly $\mathbb T_{m,n}$ in \eqref{app:D_N_even} has to be augmented by some phase $\xi_{m,n}$,
such that \eqref{app:O_tarns_S_T} is satisfied.
Since the phase to be accounted for is only a minus sign,
$\xi_{m,n}$ can be on the form $\xi_{m,n}=(-1)^{am+bn+cmn}$.
Repeating the calculation in \eqref{app:transform_D},
including $\xi_{m,n}$ leads to the requirements $\xi_{m,n}=\xi_{n,m}$ and $\xi_{m+n,n}=\xi_{m,n}(-1)^{\Lambda N_\Phi n}$.
The solution is $\xi_{m,n}=(-1)^{\Lambda N_\Phi)mn}(-1)^{l(m+n+mn)}$,
where again, as in \ref{ap:Modtrans_D}, $l=0,1$ is an unspecified integer parameter.

The $\mathbb D$ operator that has correct modular properties is
\begin{equation}\label{app:D_general}
  \mathbb D^{(r,t)}_{(\a)} = \sum_{m,n}\xi^{(r,t)}_{m,n}\lambda_{m,n}^{N_\alpha}T_{m,n}^{(\a)}T_{r_\a m,r_\a n},
\end{equation}
where
\begin{equation}\label{app:xi_general}
  \xi^{(r,t)}_{m,n}=(-1)^{l(m+n+mn)}(-1)^{\Lambda(N_\Phi mn+tn+rm)}.
\end{equation}
$l$ is not determined by this argument, which leaves two possible wave functions for any given choice of 
flux $N_{\Phi}$ and boundary condition $(r,t)$. 

\subsection{ Limits on the $(m,n)$ sums } \label{app:Sum_m_n}
The terms in the sums in \eqref{app:D_general} are periodic in $n$ and $m$, so to get a finite result we must
restrict the range of the summations. 
Recalling
\begin{equation*}
  D^{(\a)}_{m,n}=\xi_{m,n}^{(r,t)}\lambda_{m,n}^{N_{\alpha}}\mathbb{T}_{m,n}^{(\alpha)}
\end{equation*}
and noting the following index relations,
\begin{eqnarray}\label{eq_index_xi}
  \xi_{m+N_{\Phi},n}^{(r,t)} & = & (-1)^{l(1+n)+N_{\Phi}\Lambda(n+r)}\xi_{m,n}^{(r,t)}\nonumber\\
  \xi_{m,n+N_{\Phi}}^{(r,t)} & = & (-1)^{l(1+m)+N_{\Phi}\Lambda(m+t)}\xi_{m,n}^{(r,t)},
\end{eqnarray}
where $\Lambda=p^{-1}p+1\mod 2$, and $l$ is undefined.
The weight transforms as
\begin{eqnarray}\label{eq_index_lambda}
  \lambda_{m+N_{\Phi},n}^{N_{\alpha}} & = & (-1)^{N_\a}e^{-\i2\pi n\frac{p_{\alpha}}{2q}}\lambda_{m,n}^{N_{\alpha}}\nonumber\\
  \lambda_{m,n+N_{\Phi}}^{N_{\alpha}} & = & (-1)^{N_\a}e^{\i2\pi m\frac{p_{\alpha}}{2q}}\lambda_{m,n}^{N_{\alpha}}.
\end{eqnarray}
The index relations for $T_{m,n}$ and $T_{m,n}^{(\a)}$ depends on the boundary conditions on $\psi$ and are:
\begin{eqnarray}\label{eq_index_T}
  T_{m+q,n}\psi & = & (-1)^{np+R_{1}}T_{m,n}\psi\nonumber\\
  T_{m,n+q}\psi & = & (-1)^{mp+R_{2}}T_{m,n}\psi,
\end{eqnarray}
where $R_{j}=2qK_{j}$ is an integer.
Similarly 
\begin{eqnarray}\label{eq_index_T_alpha}
  T_{m+N_{\Phi},n}^{(\alpha)}\psi & = & \left(-1\right)^{N_{\a}(n+r)}T_{m,n}^{(\alpha)}\psi\nonumber\\
  T_{m,n+N_{\Phi}}^{(\alpha)}\psi & = & \left(-1\right)^{N_{\a}(m+t)}T_{m,n}^{(\alpha)}\psi.
\end{eqnarray}
where $t_{m+N_{\Phi},n}\psi=(-1)^{n+r}t_{m,n}\psi$ and $t_{m,n+N_{\Phi}}\psi=(-1)^{m+t}t_{m,n}\psi$ was used.
Combining \eqref{eq_index_T} and \eqref{eq_index_T_alpha},
together with $\left[T_{N_{\Phi},0}^{\left(\alpha\right)},T_{m,n}\right]=\left[T_{0,N_{\Phi}}^{\left(\alpha\right)},T_{m,n}\right]=0$, leads to 
\begin{eqnarray}\label{eq_index_TT}
  \mathbb{T}_{m+N_{\Phi},n}^{\left(\alpha\right)}\psi & = & 
  \left(-1\right)^{N_{\Phi}\left(r_{\alpha}np+r_{\alpha}R_{1}+n+r\right)}\mathbb{T}_{m,n}^{\left(\a\right)}\psi\nonumber\\
  \mathbb{T}_{m,n+N_{\Phi}}^{\left(\alpha\right)}\psi & = &
  \left(-1\right)^{N_{\Phi}\left(r_{\alpha}mp+r_{\alpha}R_{2}+m+t\right)}\mathbb{T}_{m,n}^{\left(\a\right)}\psi.
\end{eqnarray}
Thus, using \eqref{eq_index_xi}, \eqref{eq_index_lambda},
\eqref{eq_index_TT} and $pr_{\alpha}=\Lambda+1\mod 2$ gives
\begin{eqnarray}\label{eq_index_D}
  D_{m+N_{\Phi},n}\psi & = & (-1)^{l(1+n)+N_\a}
  e^{-\i2\pi n\frac{p_{\alpha}}{2q}}D_{m,n}\psi\nonumber\\
  D_{m,n+N_{\Phi}}\psi & = & (-1)^{l(1+m)+N_\a}
  e^{\i2\pi m\frac{p_{\alpha}}{2q}}D_{m,n}\psi.
\end{eqnarray}
From this equation follows that $D_{m,n}\psi$ is periodic in shifts of $2qN_\Phi$ in the indexes.
Thus, in order to account for \eqref{eq_index_D},
the lattice of summation in \eqref{app:D_general} needs to be extended to $\{m,n\}\in\mathbb Z_{2qN_\Phi}^2$.

\subsection{Proof that $[\mathbb{D}_{(\a)},\mathbb{D}_{(\b)}]=0$}\label{app:hierarchy_op}
For the construction given in \eqref{finwf} to be unambiguous,
with respect to the ordering of the operators $\mathbb{D}$,
it is imperative that $[\mathbb{D}_{(\a)},\mathbb{D}_{(\b)}]=0$,
for all $\a$ and $\beta$.
In this section, we will prove this.
Also here, we assume $q$ to be odd, such that $N_{\a}=N_{\Phi}\mod 2$.

The modular covariant $\mathbb{D}$-operator is 
\begin{equation}
  \mathbb{D}_{(\alpha)}^{(r,t)}=\sum_{m,n=0}^{2qN_\Phi}\xi_{m,n}^{(r,t)}\lambda_{m,n}^{N_{\alpha}}\mathbb{T}_{m,n}^{(\alpha)}=\sum_{m,n=0}^{2qN_\Phi}D_{m,n}
\end{equation}
where $r_{\alpha}=-p_{\alpha}p^{-1}$,
$\mathbb{T}_{m,n}^{(\alpha)}=T_{m,n}^{\left(\alpha\right)}T_{r_{\alpha}m,r_{\alpha}n}$ and 
$\xi_{m,n}$ is given by \eqref{app:xi_general}.
Now we consider the commutator $\left[\mathbb{D}_{\left(\alpha\right)},\mathbb{D}_{\left(\beta\right)}\right]$,
and write 
\begin{eqnarray*}
  \mathbb{D}_{(\alpha)}^{(r,t)}\mathbb{D}_{(\beta)}^{(r,t)} & = & \sum_{m,n=0}^{2qN_\Phi}\xi_{m,n}^{(r,t)}\lambda_{m,n}^{N_{\alpha}}\mathbb{T}_{m,n}^{(\alpha)}\times\sum_{m',n'=0}^{2qN_\Phi}\xi_{m',n'}^{(r,t)}\lambda_{m',n'}^{N_{\beta}}\mathbb{T}_{m',n'}^{(\beta)}\\
  & = & \sum_{m',n'=0}^{2qN_\Phi}\xi_{m',n'}^{(r,t)}\lambda_{m',n'}^{N_{\beta}}\times\sum_{m,n=0}^{2qN_\Phi}\xi_{m,n}^{(r,t)}\lambda_{m,n}^{N_{\alpha}}\cdot\mathbb{T}_{m,n}^{(\alpha)}\mathbb{T}_{m',n'}^{(\beta)}.
\end{eqnarray*}

By the definition of $r_{\alpha}$,
the full many-body translation operator $T_{r_{\beta}m^{\prime},r_{\beta}n^{\prime}}$ can be commuted through $\mathbb{T}_{m,n}^{(\alpha)}$.
Left is to commute $T_{m^{\prime},n^{\prime}}^{(\beta)}$
through $\mathbb{T}_{m,n}^{(\alpha)}$.
As $\alpha\neq\beta$, ($[\mathbb{D}_{(\alpha)},\mathbb{D}_{(\alpha)}]=0$ trivially),
a phase is only picked up between $T_{r_{\alpha}m,r_{\alpha}n}$ and $T_{m^{\prime},n^{\prime}}^{\left(\beta\right)}$.
More precisely, it is only the $T_{r_{\alpha}m,r_{\alpha}n}^{(\beta)}$ part of $T_{r_{\alpha}m,r_{\alpha}n}$ that will contribute.
This gives 
\begin{equation*}
  T_{r_{\alpha}m,r_{\alpha}n}^{(\beta)}T_{m^{\prime},n^{\prime}}^{\left(\beta\right)}=e^{\rmi2\pi r_{\alpha}\left(mn^{\prime}-nm^{\prime}\right)\frac{p_{\beta}}{q}}T_{m^{\prime},n^{\prime}}^{\left(\beta\right)}T_{r_{\alpha}m,r_{\alpha}n}^{\left(\beta\right)},
\end{equation*}
such that a phase of $\Upsilon=e^{-\rmi2\pi\left(mn^{\prime}-nm^{\prime}\right)\frac{p^{-1}p_{\alpha}p_{\beta}}{q}}$ is acquired.
The commutator $\mathbb{D}_{(\alpha)}^{(r,t)}\mathbb{D}_{(\beta)}^{(r,t)}$ can thus we written as
\begin{equation*}
  \mathbb{D}_{(\alpha)}^{(r,t)}\mathbb{D}_{(\beta)}^{(r,t)} 
  = \sum_{m,n=0}^{2qN_\Phi}D_{m,n}^{(\a)}
  \times\sum_{m',n'=0}^{2qN_\Phi} D_{m',n'}^{(\b)} \times \Upsilon.
\end{equation*}

We note that $\Upsilon$ is invariant under a shift of $N_\Phi$ in any of $n,m,n',m'$.
From \eqref{eq_index_D} it follows that 
\[
D^{(\a)}_{m,n}\Upsilon=D^{(\a)}_{m+2N_{\Phi}m^{\prime}p^{-1}p_{\beta},n+2N_{\Phi}n^{\prime}p^{-1}p_{\beta}},
\]
such that 
\begin{eqnarray*}
  \mathbb{D}_{(\alpha)}^{(r,t)}\mathbb{D}_{(\beta)}^{(r,t)} & = & 
  \sum_{m',n'=0}^{2qN_\Phi}D_{m',n'}^{(\beta)}\times
  \sum_{m,n=0}^{2qN_\Phi}D_{m+2N_{\Phi}m^{\prime}p^{-1}p_{\beta},n+2N_{\Phi}n^{\prime}p^{-1}p_{\beta}}^{\left(\alpha\right)}\\
  & = & \mathbb{D}_{(\alpha)}^{(r,t)}\times\mathbb{D}_{(\beta)}^{(r,t)},
\end{eqnarray*}
by shifting the boundaries of the summation.
Thus $\left[\mathbb{D}_{\left(\alpha\right)},\mathbb{D}_{\left(\beta\right)}\right]=0$,
and \eqref{finwf} is well defined.

\section{The thermodynamic limit}   \label{sec:thermo_limit}
In order to ensure that in the thermodynamic, \ie the large $N_e$ or large $L$ limit,
the wave functions \eqref{finwf} do agree with the corresponding ones on the plane, 
we must show that effectively $\mathbb D_{(\alpha)} \to \prod_j\partial_{z_j} $. 

By expanding $\mathbb D_{(\alpha)}$ in powers of $\epsilon = 1/N_\Phi$,
and rewriting $\partial_x$ and $\partial_y$ in in terms of $\partial_z$ and $\partial_{\bar z}$,
and also writing $x =(z + \bar z)/2$,
we get a sum of polynomials in $\partial_z$, $\partial_\zb$, $z$ and $\bar z$.
Although not obvious, we know that when acting on the correlators of primary operators $\psi_s$,
this will produce, up to a gauge transformation and the ubiquitous Gaussian, 
a polynomial in the coordinates $z_i$.
This follows since the full wave function is by construction in the LLL, 
and furthermore, it will hold term by term in the expansion in $\epsilon$. 
Thus, the final outcome will be a power series in $\epsilon$,
consisting of polynomials in $\partial_{z_i}$ and $z_i$,
acting on the polynomial in $\psi_s$. 
Since this polynomial is symmetric when exchanging any two coordinates belonging different groups,
it has to be multiplied by a polynomial from the expansion of $\mathbb D_{(\alpha)}$ that distinguish all the coordinates
in the different groups.
If this is to be achieved by acting on one group,
all particle have to be affected. 
Since the particles are identical,
the lowest order operator that will do this for the group $\a$ is 
\be{limit}
\prod_{i\in I_\a} (\partial_i - c {z_i} /{\ell_B^2} ),
\ee
where $c$ is some constant.
The term corresponding to $c=0$ exactly gives the planar wave functions,
while the terms also containing powers of $z_i$ will have higher angular momentum,
and physically most likely correspond to edge excitations.

\section{Special functions}\label{app:special-functions}

In this appendix we collect the main properties of these functions that will be used throughout the main text.
The generalized Jacobi $\vartheta$-function is defined as 
\begin{equation}
  \genelliptic abz{\tau}=\sum_{k=-\infty}^{\infty}e^{\rmi\pi\tau\left(k+a\right)^{2}}e^{\rmi2\pi\left(k+a\right)\left(z+b\right)}\label{eq:gen_theta_def}
\end{equation}
where $\Im\left(\tau\right)>0$ for convergence. 
The two main periodic properties are
\begin{equation}
  \genelliptic ab{z+n}{\tau}=e^{\rmi2\pi an}\genelliptic abz{\tau}\label{eq:gen_theta_z+n}
\end{equation}
where $n\in\mathbb{Z}$ and 
\begin{equation}
  \genelliptic ab{z+c\tau}{\tau}=e^{-\rmi2\pi c\left(z+b\right)}e^{-\rmi\pi\tau c^{2}}\genelliptic{a+c}bz{\tau}\label{eq:gen_theta_z+n_tau}
\end{equation}
where $c\in\mathbb{R}$.
Under transformations of the lattice parameter
$\tau$ the relations are
\begin{equation}
  \genelliptic abz{\tau+n}=e^{-\rmi\pi a\left(1+a\right)n}\genelliptic a{an+\frac{n}{2}+b}z{\tau}\label{eq:gen_theta_z+tau+n}
\end{equation}
where $n\in\mathbb{Z}$. 
Under inversion of the lattice parameter $\tau\rightarrow-\frac{1}{\tau}$,
the transformation is
\begin{eqnarray}
  \genelliptic abz{-\frac{1}{\tau}} & = & \sqrt{-\rmi\tau}e^{\rmi\tau\pi z^{2}}e^{\rmi2\pi ba}\genelliptic b{-a}{\tau z}{\tau}\label{eq:gen_theta_taun_inverse}
\end{eqnarray}
The first elliptic theta function $\elliptic 1z\tau$ is a special case with 
\begin{eqnarray}
  \elliptic 1z{\tau} & = & \genelliptic{\frac{1}{2}}{\frac{1}{2}}z{\tau}\label{eq:theta_1}\\
\end{eqnarray}
and is odd under $z\rightarrow-z$.
Of importance is also the Dedekind $\eta$-function
\begin{equation}
  \eta(\tau)=e^{\frac{\rmi\pi\tau}{12}}\prod_{n=1}^{\infty}\left(1-e^{2\pi\rmi n\tau}\right),
\end{equation}
with transformation properties
\begin{eqnarray}
  \eta(-\frac1\tau) &=& \sqrt{-\rmi \tau}\eta(\tau)\\
  \eta(\tau+1) &=& e^{\frac{\rmi\pi}{12}}\eta(\tau)
\end{eqnarray}

\bibliographystyle{unsrt} 
\bibliography{Torus_Ref}

\end{document}